\documentclass[%
reprint,
superscriptaddress,
nobibnotes,
amsmath,amssymb,
aps,
prl,
floatfix
]{revtex4-2}

\makeatother
\usepackage{graphicx}  
\usepackage{dcolumn}   
\usepackage{bm}        
\usepackage{amssymb}   
\usepackage{subfigure}
\usepackage{xcolor}
\usepackage{footmisc}
\usepackage[normalem]{ulem}
\usepackage{filecontents}
\usepackage[colorlinks=true]{hyperref}
\usepackage{pdfpages}
\usepackage{etoolbox} 

\makeatletter
\patchcmd{\@outputpage@head}{\@ifx{\LS@rot\@undefined}{}{\LS@rot}}{}{}{}

\begin{document}
\preprint{APS/123-QED}

\title{Phase analysis of Ising machines and their implications on optimization}
\author{Shu Zhou}
\author{K. Y. Michael Wong}
\email{phkywong@ust.hk}
\affiliation{Department of Physics, The Hong Kong University of Science and Technology, Hong Kong SAR, China.}
\author{Juntao Wang}
\affiliation{Department of Physics, The Hong Kong University of Science and Technology, Hong Kong SAR, China.}
\affiliation{Theory Lab, Central Research Institute, 2012 Labs, Huawei Technologies Co. Ltd., Hong Kong SAR, China.}
\author{David Shui Wing Hui}
\author{Daniel Ebler}
\email{ebler.daniel1@huawei.com}
\author{Jie Sun}
\email{j.sun@huawei.com}
\affiliation{Theory Lab, Central Research Institute, 2012 Labs, Huawei Technologies Co. Ltd., Hong Kong SAR, China.}
\date{\today}
\begin{abstract}

Ising machines, which are dynamical systems designed to operate in a parallel and iterative manner, have emerged as a new paradigm for solving combinatorial optimization problems. Despite computational advantages, the quality of solutions depends heavily on the form of dynamics and tuning of parameters, which are in general set heuristically due to the lack of systematic insights. Here, we focus on optimal Ising machine design by analyzing phase diagrams of spin distributions in the Sherrington-Kirkpatrick model. We find that that the ground state can be achieved in the phase where the spin distribution becomes binary, and optimal solutions are produced where the binary phase and gapless phase coexist. Our analysis shows that such coexistence phase region can be expanded by carefully placing a digitization operation, giving rise to a family of superior Ising machines, as illustrated by the proposed algorithm digCIM.
\end{abstract}

\maketitle

Discrete optimization problems are ubiquitous in science and engineering, encompassing protein folding \cite{Ooka2023}, route planning \cite{Bao9564593}, information processing \cite{LeVinee17052895}, and machine learning \cite{Bryngelson1987-yp}, among many others \cite{Lucas_2014}. The Ising model has become a powerful bridge for modeling and analyzing these problems from the viewpoint of statistical physics \cite{ RevModPhys.39.883}. However, finding the system states corresponding to the optimal solution of the target problems remains computationally hard~\cite{FBarahona_1982,arora2005non}.

An Ising problem, which is equivalent to the class of quadratic unconstrained binary optimization problems (QUBO), generally takes the form of
\begin{equation}
\min_{\sigma\in\{-1,+1\}^N}-\frac{1}{2}\sum_{ij} \sigma_iJ_{ij}\sigma_j,
\end{equation}
where $\sigma\in\{\pm1\}^N$ is the set of $N$ binary spins coupled through the matrix $[J_{ij}]_{N\times N}$.
Analog Ising machines (AIMs) are non-von-Neumann computing paradigms designed to find ground state configurations of Ising models more efficiently than conventional search-based algorithms. 
In abstract terms, AIMs compute solution states through time-evolution of a coupled dynamical system
\begin{equation}
\frac{d x_i}{d t}= \mathcal{F}_i(\bm{x};\bm{\mu}) + \zeta_i, i\in\{1,\dots,N\}.
\label{eq:dynamics}
\end{equation}
Here $\bm{x} = (x_1,x_2,\dots,x_N),  x_i \in \mathbb{R}$ denotes analog spin values, $\bm{\mu}$ denotes control parameters
and $\zeta_i$ represents noise.
The system's dynamics is governed by a driving term $\mathcal{F}_i(\bm{x},\bm{\mu})$ that is distinct for each AIM variant. Once the system settles into its stationary state, binary spin values $\sigma_i$ are assigned via the decoding rule $\sigma_i = {\rm sgn}(x_i)$\footnote{When $x_i = 0$, the decoding rules are not well-defined. In such cases, an arbitrary sign is assigned.}, producing a candidate solution to the Ising problem. A number of AIMs have been proposed~\cite{Mohseni2022}, such as quantum annealers \cite{PhysRevE.58.5355, Johnson2011QuantumAW, Boixo_2014, Hauke_2020}, optical Ising machines based on pulse lasers \cite{Marandi_2014, Yamamoto2017, PhysRevA.88.063853}, coupled oscillator networks \cite{Chou2019, mi13071016}, as well as special purpose chip-implementations on conventional hardware \cite{PhysRevX.5.021027, 7350099, Aramon_2019, Yamamoto2020, Leleu2021-cz, inagaki2016coherent, doi:10.1126/sciadv.abh0952}. The performance of AIMs, however, can vary significantly, preventing them from large-scale adoption in practical applications.

Considerable efforts have been made to enhance the accuracy by modifying the dynamics or constraining the distribution of analog spins. A prominent example is the Coherent Ising Machine (CIM) which is characterized by the driving term 
\begin{equation}\label{eq:CIM}
\mathcal{F}^{\rm{CIM}}_i (\bm{x}, a ,\xi) = -x_{i}^{3} + a x_{i}+ \xi \sum_{j=1, j \neq i}^{N} J_{ij } x_{j} .
\end{equation}
The terms correspond (from left to right) to cubic saturation, driving with net gain $a$, and coupling with interaction strength $\xi$~\cite{PhysRevA.88.063853}. Variants of CIM include clipCIM~\cite{Order-of-magnitude} as an optoelectronic version which bounds the amplitude of the analog spins, simCIM \cite{Tiunov:19} which further removes the saturation term from $\mathcal{F}^{\rm{CIM}}$ \cite{Tiunov:19}, and discrete Simulated Bifurcation (dSB) which is a momentum-based machine discretizing the analog spins in the interaction term~\cite{Goto2021} to mitigate errors. However, while promising, accurate solutions require manual fine-tuning of control parameters due to the lack of theoretical guidance.

In this letter, we elucidate how the choice of system dynamics and parameters shapes the spin distribution of Ising machines and determines the solution accuracy. Concretely, we study the static spin distribution of AIMs by applying the replica method to AIMs with the Sherrington-Kirkpatrick (SK) model of spin glasses~\cite{Sherrington1975, Kirkpatrick1977}, revealing a rich palette of phases that emerge in AIMs. 
In particular, we discover that a version of AIM, referred to as digCIM, can realize the microscopic equation describing the steady-state solutions of the Ising problem, namely, the TAP equation~\cite{Thouless_1977}, and find that such solutions can be accessed where the binary phase coexists with the gapless phase. Based on these findings, we propose a superior family of AIMs characterized by digitized driving terms, as illustrated by digCIM which produces optimal solutions in a wide range of the gain parameter.

To reveal distinct phases that commonly appear in AIMs, 
we first consider the example of CIM with driving term $\mathcal{F}_{\rm CIM}$ and white noise $\zeta_i$ with temperature $T$, where $\langle\zeta_i(t)\rangle  = 0$ and $\langle\zeta_i(t)\zeta_j(t')\rangle = 2T\delta_{ij}\delta(t-t')$. 
Equation~(\ref{eq:CIM}) can then be written as a Langevin equation describing the gradient descent of the 
Hamiltonian $H = \frac{1}{4}\sum_i x_i^4 - \frac{a}{2} \sum_i x_i^2 - \frac{\xi}{2} \sum_{i\ne j} J_{ij}x_i x_j.$

We consider Ising problems defined by the SK model
in which the couplings $J_{ij}$ follow a Gaussian distribution with zero mean and variance $J^2/N$~\cite{Sherrington1975, Kirkpatrick1977}. The SK model is suitable for accuracy analyses as in the thermodynamic limit the free energy of the system can be derived analytically using the replica method (see SM Sec.1 and 3), 
and its complex energy landscape is typical of combinatorial optimization problems \cite{Mezard1987}. The method has been applied to study one of the AIMs \cite{Yamamura2023}. Solutions of the replica method depend on the level of replica symmetry-breaking (RSB) \cite{Mezard1987}. The replica symmetric (RS) solution is stable in the paramagnetic phase. In the spin glass phases we will quote results of full RSB (FRSB) where possible, and one-step RSB (1RSB) approximation for some numerical results, which produce rather accurate predictions except phase transition points.

We introduce tools to analyze the phases of AIMs. First, the (analog) spin distribution is the instantaneous distribution of the individual spins in equilibrium at zero temperature of an AIM, given by $P(x)=[\langle \delta(x_i-x)\rangle]_{J_{ij}}$. Here, the angular and square brackets denote thermal and disordered averages. In the framework of the replica method the value of a representative spin $x$ is determined by a mean-field effective free energy. For CIM, it is given by (setting $J\xi=1$)
\begin{equation}
   g(x) = \frac{1}{4} x^4 - \frac{a_{\rm eff}}{2} x^2 - wx,
\label{eq:g}
\end{equation}
where $w$ represents the cavity field, which is the field experienced by the spin $x$ when the system excluding $x$ is in thermodynamic equilibrium. The distribution of the cavity field depends on the level of RSB. Parameter $a_{\rm eff} = a + \chi$ is the effective gain where $\chi$ is the susceptibility which is equal to the thermodynamic average of $\partial x/\partial h$ when an infinitesimal external field $h$ is applied to the system. Hence, $\chi$ represents the mean-field influence of the spin on its neighbors which is fed back to the spin itself \footnote{This is commonly referred to as the Onsager reaction \cite{Opper2001}.}. The 1RSB spin distribution of CIM is derived in  Sec. 4 of SM and reads
\begin{equation}
    P(x) = \int Dv \frac{\int Du[\int dx' e^{-\beta g(x')}]^{m-1}  e^{-\beta g(x)}}
    {\int Du[\int dx' e^{-\beta g(x')}]^m}.
\label{eq:distribution}
\end{equation}
Here, $Du \equiv e^{-u^2/2} du/\sqrt{2\pi}$ is the Gaussian measure, and $u$ and $v$ are Gaussian variables forming the cavity field $w$ 
according to $w = \sqrt{q_0} v + \sqrt{q_1 - q_0}u$. In the 1RSB framework, the solution space consists of a collection of clusters, with $q_1$ and $q_0$ representing the correlation of solutions in the same and different clusters, respectively. 
The parameter $m$ characterizes the relative weights of the same and different clusters, and $m\beta$ approaches a constant when $T \to 0$. The inner expression is the Gibbs distribution shaped by $g(x)$ in Eq.~\eqref{eq:g}. 
At $T=0$, the spin distribution is dominated by the values $x^*$,  which are the minimizers of $g(x)$.

Next we introduce the decoded energy $E_{\rm dec}=-[\langle \sum_{ij} J_{ij} {\rm sgn}x_i {\rm sgn}x_j \rangle]_{J_{ij}}/(2N)$,
which represents the thermodynamic average of the Ising energy after binarization of the analog spin values of AIM at equilibrium. The decoded energy at $T=0$ is a lower bound on the accessible energy of the AIM (see Appendix \ref{A1} for details).

\begin{figure}[htb]
\centering
\includegraphics[width=0.47\textwidth]{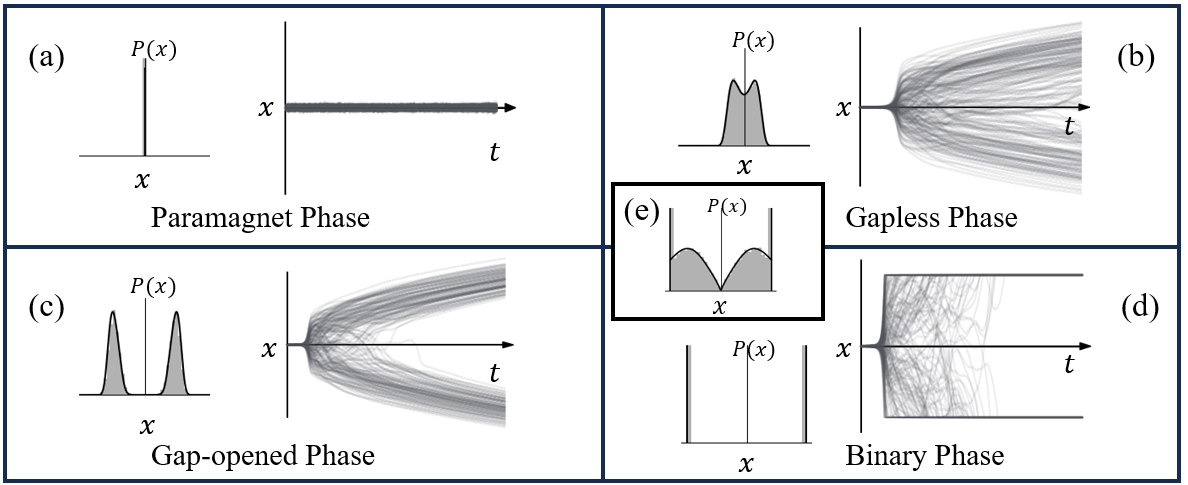}
\caption{AIM phases as characterized by distinct modes of spin distribution $P(x)$ together with typical AIM dynamics of $x$ vs $t$ ($T=0$). (a) Paramagnet phase.  (b) Gapless phase.  (c) Gap-opened phase. (d) Binary phase. (e) Gapless-binary coexistence region.
}
\label{fig:distributions}
\end{figure}

Phases of AIM are characterized by spin distribution modes as depicted in Fig.~\ref{fig:distributions} which includes {\it paramagnetic phase} (spin distribution is a delta function at $T=0$), {\it gapless phase} (single-band distribution), {\it gap-opened phase} (double-band distribution), {\it binary phase} (dual delta distribution), and a gapless-binary coexistence scenario associated with the ground state of the SK model~\cite{Horner_2007, Sommers1984}.
Note that the phases characterize the qualitative properties of AIMs at fixed parameters.

To explore these phases and their impact on optimization, we start with CIM for which Figs.~\ref{fig:distributions}(a)-(c) appear in order as the control parameter $a$ changes from negative to positive values in Eq.~(\ref{eq:CIM}).

{\it Paramagnetic phase (Fig.~\ref{fig:distributions}(a)).}
For CIM dynamics with $T=0$ at low gains $a\leq-2$, the
spin distribution is a delta peak centered at zero and has vanishing thermodynamic average $[\langle x_i\rangle^2]_{J_{ij}} = 0$. 
The individual spins do not have a preferred direction to be positive or negative. For $T>0$, the paramagnetic phase still exists even for higher gains. The decoded energy is already negative, indicating locally optimal clusters in the spin configurations. However, due to the inversion symmetry of the Hamiltonian, clusters favoring either orientation remain equally probable and the decoded energy is suboptimal.

\begin{figure}[ht]
\centering
\hspace{-0.4cm}
    \includegraphics[scale=0.52]{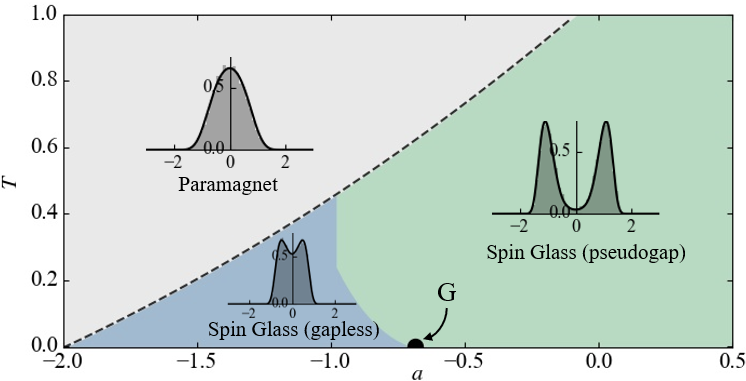}\\
\caption{Phase diagram of CIM in the space of gain $a$ and temperature $T$ with typical analog spin distributions shown in each region, respectively located at $(a, T) = (-1.5, 0.5)$ (paramagnetic phase, defined by $q_1= 0$), $(-1.5, 0.03)$ (gapless spin glass phase, defined by $q_1 > 0$), and $(0.5, 0.03)$ (pseudogap spin glass phase, defined by $q_1 >0$ and $a_{\rm{eff}}>0$). In these distributions, the solid black lines correspond to 1RSB theory, while the grey shadings indicate simulation results. The dashed phase line is obtained in RS. G is the 1RSB gap-opening point at $T = 0$. The paramagnetic phase also consists of a pseudogap region but for clarity it is not shown in the figure (see Sec. 4 of SM).}
\label{fig:phase_diagram}
\end{figure}

{\it Gapless phase (Fig.~\ref{fig:distributions}(b))}. 
At fixed temperature $T$, when the gain $a$ exceeds certain threshold the thermodynamic average $[\langle x_i\rangle^2]_{J_{ij}}$ of the spin population  becomes nonzero. This is due to the dominance of one of the two inversion-symmetric states at the phase transition line between paramagnet and spin glass, breaking inversion symmetry. Compared with the paramagnetic phase, the spin distribution is broadened, and bimodal for sufficiently large gain. The two peaks remain connected in a single energy band.

At $T=0$, at the bifurcation threshold $a=-2$ the decoded energy is continuous with that of the paramagnetic phase but with a kink. The decoded energy of $-2/\pi$ at this point corresponds to the minimum eigenvalue of the coupling matrix with elements $J_{ij}\xi$ \cite{potters_bouchaud_2020}\footnote{The eigenvalue distribution of the matrix $J_{ij}\xi$ obeys Wigner’s semicircular law with center 0 and radius $2J\xi$.}. Although 
the spin values at this point is somehow indicative of the optimal configuration \cite{Juntao2023}, they may still deviate considerably from the optimal value. For the SK model, 
the decoded energy of $-0.64$ remains higher than the theoretical optimal energy $-0.76$. 

\begin{figure}[ht] 
    \centering \hspace{-0.2cm}
    \includegraphics[width=0.5\textwidth]{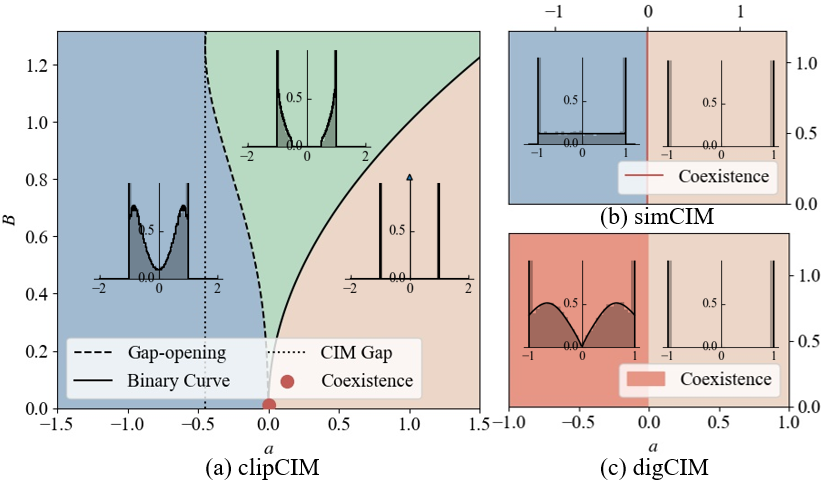}
\caption{\label{fig:clip_sim_phase} Phase diagram and the spin distribution for clipCIM ($B = 1$), simCIM and digCIM in the space of gain $a$ and bound $B$ at zero temperature. In Fig. 3(a), the solid and dashed phase lines are obtained by FRSB and simulation, respectively. Distributions are annotated  similarly to those in Fig. \ref{fig:phase_diagram}.}
\label{fig:a-b_phases}
\end{figure}

{\it Gap-opened phase (Fig.~\ref{fig:distributions}(c)).} 
When the gain is further increased above a larger threshold (e.g., $G$ in Fig.~\ref{fig:phase_diagram} for $T=0$),
the bimodal distribution becomes separated by a gap. 
At $T = 0$, the gap-opening point $a = -\chi$ (i.e., $a_{\rm eff} = 0$) can be derived by analyzing how $g(x^*)$ evolves with $a_{\rm eff}$ (see Appendix~\ref{A2}). Since $\chi=0.45$  in FRSB, the gap-opening point becomes  $a=-0.45$ \cite{Yamamura2023}. 
At finite temperatures, thermally excited spins are located in the gap states, and the gap becomes a pseudogap (green region in Fig.~\ref{fig:phase_diagram}).

\begin{figure}[ht] 
\centering \hspace{-0.2cm}
\includegraphics[width=0.48\textwidth]{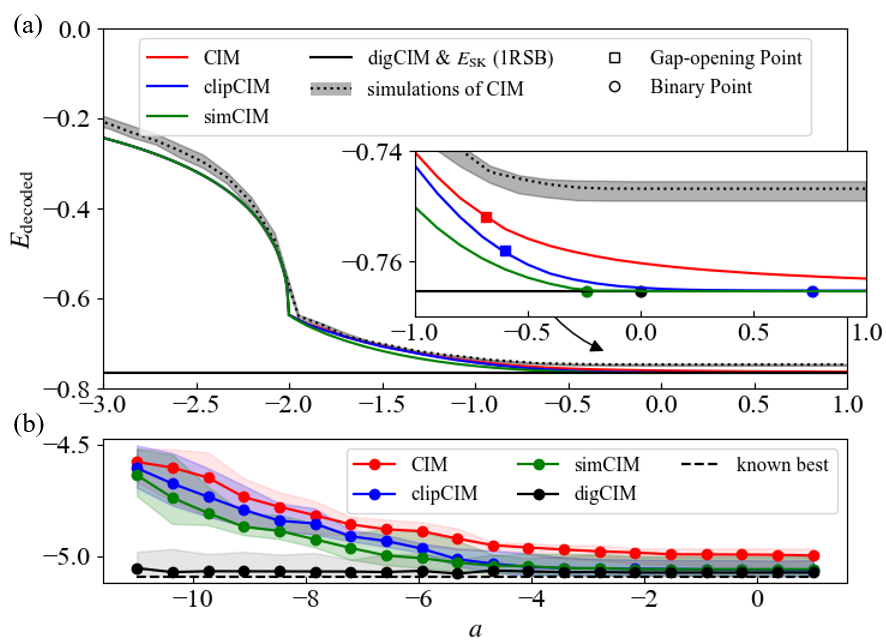}
\caption{(a) Theoretical dependence of CIM, clipCIM ($B = 1$), simCIM, and digCIM's decoded energy on the gain $a$ at zero temperature. 
Dotted line shows the simulation results for 10,000 steps, $N = 10,000$ and $T=10^{-5}$, bounded by extremal values in the shaded areas.
(b) The same experiments applied to G1 in Gset \cite{Ye_2003}, which exhibits similar characteristics.}
\label{fig:CIM}
\end{figure}

The 1RSB decoded energy of the CIM reaches $-0.753$ at this point, which is significantly closer to the optimal energy compared with the energy in the gapless phase (see the red curve of Fig.~\ref{fig:CIM}). Beyond the gap-opening point, the decoded energy slowly decreases towards the optimal value with increasing gain $a$, but does not reach it. 

The double peaks with equal magnitudes of the distribution $P(x)$, shown in the green region of Fig.~\ref{fig:a-b_phases}, becomes increasingly prominent as $a$ increases, implying that the system becomes more Ising-like. Although it was recognized that the spin magnitudes need to be uniform for the analog system to be an accurate optimizer of Ising problems~\cite{Order-of-magnitude, PhysRevE.95.022118}, 
the spin amplitudes in CIM still exhibit a spread even for large values of $a$. This leads to underperformance of the Ising machine, which has previously motivated the introduction of a clipping function~\cite{Tiunov:19, Bohm2019-pi}. 
We will show next that the binary phase therein can attain the ground state.

{\it Binary phase (Fig.~\ref{fig:distributions}(d))} is characterized by a spin distribution that assumes 
two delta-peaks (at values $x=\pm 1$), fully emulating the binarized states of an Ising model.
This means each spin is frozen at $\pm1$
and both the order parameters and the decoded energy are fixed. 
However, for CIM the spin values approach the binary distribution
only for infinitely large systems and gain, i.e.  $N, a \to \infty$.  In clipCIM, a clipping function $\phi(x) = \max(-B, \min(B, x))$ on the amplitudes of $x$ is imposed at each iteration. Here, with $T=0$, the distribution $P(x)$ is bounded by $\pm B$ with a continuous distribution in between (blue region of Fig.~\ref{fig:clip_sim_phase}(a)). The gap-opening curve starts at $(a,B)=(0,0)$ and ends at the gap-opening point of CIM when $B$ is very large, i.e. $a=-0.45$ in FRSB \cite{Yamamura2023}. When $a$ increases further at zero temperature, the gap of the spin distribution widens, and the continuous component vanishes eventually, leaving behind only the delta components. 
The boundary between the gap-opening phase and binary phase is theoretically given by $a_{\rm eff} = B^2$~\footnote{This can be derived by comparing the half-gap width $\sqrt{a_{\rm eff}}$ with the clipping threshold $B$.}. In FRSB, $\chi = 0$ along the boundary \cite{Thouless_1977, Schmidt2008}, implying that $a = B^2$. Furthermore, we derive that the decoded energy in the binary phase is the ground state energy of the SK model, independent of the level of RSB (see Sec. 5 of SM) and equals $-0.763$.~\cite{GParisi_1980, Crisanti2002, Palassini_2008, Kim2007, Boettcher2004}.  Below we will explore how the driving term in clipCIM can be modified to broaden the range of gain for the binary phase and squeeze the suboptimal gap-opened phase.
 
{\it Gapless-binary transition of distribution.}  The profile of the spin distribution is influenced by the presence of the cubic saturation term (see \cite{PhysRevE.95.022118}). The AIM without cubic saturation, referred to as simCIM~\cite{Tiunov:19}, follows the dynamics ${\mathcal F}_i^{\rm simCIM} = ax_i + \sum_j J_{ij}x_j$ with clipping threshold $B$~\footnote{A variant of simCIM with additional momentum, known as ballistic Simulated Bifurcation (bSB), was found to have an excellent accuracy and speed performance~\cite{Goto2021}.}.
For simCIM, the spin distribution is much broader due to the removal of the saturation term, and remains gapless until it becomes fully binary, see Fig.~\ref{fig:a-b_phases}(b). 
Hence, the transition line of the binary points coincides with the gap-opening points 
with $a_{\rm eff} = 0$ at $T = 0$.
Bypassing the gap-opened phase causes the binary point to set in at a lower gain (Fig.~\ref{fig:CIM}), and the decoded energy is lower than that of clipCIM at the same setting.

Towards further improving the optimization performance, we note that the contributions of the spins to the local field are digitized in the binary phase, where the ground state is reached. Below, we show that the digitization operation is indeed closely related to the ground state and can lead to superior performance when implemented appropriately, leading to a new type of AIM.

{\it Digitization.} To see the above relation, we resort to the TAP equations \cite{Thouless_1977} which microscopically describe the steady states of mean-field models such as the SK model.
At $T = 0$, the local field $H_i$ at node $i$ is given by the TAP equation 
$H_i = \sum_j J_{ij} {\rm sgn} H_j - \chi {\rm sgn} H_i$. Since $\chi=0$ at $T=0$ in FRSB \cite{Thouless_1977, Schmidt2008}, a necessary ground state condition is $H_i = \sum_j J_{ij} {\rm sgn} H_j$. Motivated by this result, we propose a type of AIM which we call digCIM that naturally satisfies the TAP equation throughout the entire range of $a$, covering both regimes with and without continuous components. The digCIM dynamics is similar to the simCIM dynamics, with a clipping threshold $B$ but digitized components of interactions $\mathcal{F}_i^{\rm digCIM} = ax_i + \sum_j J_{ij}{\rm sgn}x_j$. We note that dSB \cite{Goto2021}, which uses the same driving term but employed momentum dynamics, thus belongs to the same family as digCIM with the same steady-state behavior (see Appendix~\ref{A3}). Specifically, a steady-state solution of the SK model can be mapped to that of the digCIM family by clipping the fields $H_i$ to within the bounds $\pm B|a|$, and the converse is valid if one notes that ${\rm sgn} x$ in the digCIM solution is independent of the magnitude of $x$.

For negative $a$, the continuous component of the digCIM spin distribution vanishes linearly at $x = 0$ as shown in Fig.~\ref{fig:clip_sim_phase}(c), with positive and negative bands touching only in a single point at origin, resembling the local field distribution of the SK model obtained from solving the TAP equations \cite{Thouless_1977, Horner_2007, Sommers1984}. Above the binary point at $a = 0$, the binary spins have the same sign as the TAP solutions. Hence, digCIM is able to attain the ground states of the SK model at all values of the gain $a$, irrespective of $B$ (see Fig. ~\ref{fig:CIM}).

{\it Implications to dynamics.} In the following, we elucidate optimal regimes of the dynamics for attaining ground state solutions. The static analysis of the phases predicts that the decoded energy attains the ground state of the binary system {\it beyond} the binary point, but simulation dynamics yield decoded energies higher than the minimum. The situation is analogous to the geometric landscape at high gain of the CIM studied in \cite{Yamamura2023}, where the authors computed the distribution of local minima, showing that local minima prevent the convergence to the ground state. Besides, to prevent the dynamics from freezing prematurely into suboptimal states, it is favorable to work in regimes where the positive and negative bands of the spin distribution are not separated by gaps. For instance, for clipCIM, at zero temperature this regime is located at the triple point $(a_{\rm eff} ,B) = (0, 0)$ where gapless, gapped, and binary phases coexist. For simCIM, this regime is the gapless-binary coexistence curve. For digCIM, the coexistence curve is identical to the line of binary points. However, since the steady states of digCIM at negative values of $a$ are isomorphic through rescaling $x$ and $B$ by $|a|$, the region of optimal decoding expands to cover the entire region of $a<0$.

In practice, the operation of AIMs requires a good annealing path with appropriate control of the system parameters. Including temperature annealing, we found that annealing terminating in the coexistence regions leads to much higher decoding accuracy than otherwise (see SM Sec. 12). For digCIM, the system depends on the temperature alone as the sole parameter, resulting in a path with strongly reduced complexity. Together with the 2-dimensional coexistence region, the accessibility of the ground state is significantly enhanced.

Gapped distributions at $T=0$ degenerate into distributions with pseudogaps at finite temperatures. This changes effectively the phase diagram: for instance in clipCIM the binary curve becomes the coexisting curve, see Fig. \ref{fig:a-b_phases}(a). Such changes can guide the trajectory in parameter space to find optimal decoding schemes. In Sec. 13-14 of SM, we illustrate for various AIMs how good trajectories can lead to superior performance.

{\it Generalization to practical problems.}
To validate our results beyond the SK model we used the Gset \cite{Ye_2003}, which includes random graphs of varying sizes and topologies, as a benchmark. The decoded energies of different systems are in strong agreement with our theoretical predictions (see Fig.~\ref{fig:CIM}(b)), with digCIM yielding the most accurate results.
We found that optimal decoding in the coexistence region is also applicable in these models. 
The parameter search for optimal performance can be implemented 
solely in the space of $a$, $B$, and $T$ as the coexistence regions of clipCIM and simCIM are $a = B^2$ and $a = 0$, respectively \footnote{We assume $\chi = 0$ in practical problems}. Experimental evidence suggests that these settings extend across various models (see SM Sec. 13). Remarkably, digCIM delivers the best performance for the QUBO-QLIB suite \cite{Furini2019}. The performance surpasses not only the other Ising machine systems discussed in this paper, but also solves 19 out of 23 problems in the suite, significantly exceeding all current state-of-the-art mainstream commercial solvers (see Appendix~\ref{A4}).

{\it Conclusion.} We have studied the spin distribution and the decoded energy in a family of AIMs optimizing the energy of the SK model. Their behaviors can be classified into several phases: paramagnetic, gapless, gapped, and binary. We observe that the gapless-binary coexistence region is the most favorable working region for solution accessibility and accuracy. Based on this insight and digCIM's ability to solve the TAP equations, we propose that AIMs with the digitized driving term, such as digCIM and dSB, are the preferred choice. 

This theoretical framework readily generalizes to the steady-state behavior of momentum-based systems, including aSB, bSB, and dSB~\cite{Goto2021}. Optimal AIM performance requires further understanding non-equilibrium spin distribution dynamics to design parameter schedules that reach coexistence regions, ultimately guiding future system development.

\begin{acknowledgments}
{\it Acknowledgments}—This work is partially supported by a Huawei CSTT project.
\end{acknowledgments}


\bibliographystyle{apsrev4-2}
\bibliography{\jobname.bib}

\quad
\appendix

\section{The 1RSB Decoded Energy}
\label{A1}
\quad
Using the 1RSB framework, we find the decoded energy of CIM at zero temperature
\begin{widetext}
\begin{equation}
    \begin{aligned}
        E_{\rm dec} = &-2Je^{\frac{1}{4}m\beta a_{\rm eff}^2\Theta(a_{\rm eff})} \left[[|x^* |]_u \right]_v \left[[\delta(w)]_u \right]_v - \frac{1}{2} m\beta J\left\{\left[[|x^*|]_u\right]_v^2 - \left(\left[[x^*]_u\right]_v\left[[{\rm sgn}x^*]_u\right]_v\right)^2\right\}.
    \end{aligned}
\end{equation}
\end{widetext}
Here, for a quantity $O(x)$ in terms of of $x$, we use the notations $[O(x)]_u = \int Du\ e^{-m\beta g(x)} O(x)$ $/\int Du\ e^{-m\beta g(x)}$, and $[[O(x)]_u]_v = \int Dv [O(x)]_u$, with $\Theta$ being the Heaviside step function. The decoded energy expression for other AIM systems is similar and detailed in the SM.

\section{The Gap-opening Point}
\label{A2}
\quad
At $T = 0$, the minimizer $x^*$ of $g(x)$ dominates the distribution $P(x)$, i.e., $x^{*3} - a_{\rm eff}x^* = w$. This shows that when $a_{\rm eff} = -1$ (i.e., $a = -2$) increases to $a_{\rm eff} = 0$ (i.e., $a = -0.45$), $x^*$ is a single-valued function of the cavity field $w$, and the distribution $P(x)$ is continuous. On the other hand, when $a_{\rm eff}$ is positive, there exists a range of $w$ in which multiple values of $x^*$ for each $w$ are possible. In this case, the value of $x^*$ that yields the minimum $g(x)$ is dominant. This leads to the emergence of a gap in the distribution $P(x)$. Hence, the Gap-opening Point is given by $a_{\rm eff} = 0$. 

\section{The Family of DigCIM}
\label{A3}
Previous efforts, such as a variant of HTNN \cite{Haribara_2016} and the momentum-based algorithm dSB \cite{Goto2021}, have experimentally explored related concepts. Some of these approaches have achieved excellent performance. However, these studies did not identify the theoretical optimality of the digCIM family. In this work, we demonstrate that this optimality emerges from the superior performance observed in the coexistence region of the phase diagram, as well as from the steady-state properties characterized by ground-state equations such as the TAP equations. 
The success probability of digCIM and dSB scale similarly with problem size (Sec. 14 of SM). For the GSET benchmark, the time to solution of digCIM is comparable to dSB on smaller datasets with advantages, and longer for large instances (Sec. 15 of SM). 

\section{QPLib-QUBO Result of DigCIM}
\label{A4}
Performance comparison between DigCIM and conventional solvers on QUBO problems from the QPLIB benchmark \cite{Furini2019, QUBO-QPLIB}.

\begin{table}[ht]
\begin{ruledtabular}
\begin{tabular}{cc}
Solver Name & Number of Solved Problems\cite{QUBO-QPLIB} \\ \hline
BARON & 12 \\ 
SCIP & 7 \\ 
MCSPARSE & 12 \\ 
GUROBI & 13 \\ 
QUBOWL & 15 \\ 
BIQBIN & 9 \\ 
SHOT & 12 \\ 
DigCIM & 19 \\ 
\end{tabular}
\end{ruledtabular}
\end{table}
 DigCIM consistently outperforms state-of-the-art results, solving 19  out of 23 problems under the same hardware and time constraints. Details on implementation time, hardware specifications, and the performance of other Ising machines are provided in SM Sec.14.

\clearpage
\newpage
\newpage
\pagestyle{empty}


\section*{Supplementary material}

# Supplementary Materials

July 11, 2025

# Contents

# 1 The RS solution of the coherent Ising machine (CIM)

Table 1 summarizes the dynamical equations of the family of Analog Ising Machines (AIMs) discussed in this paper, which takes the general form

$$\frac{dx_i(t)}{dt} = \mathcal{F}_i(t) + \zeta_i(t), \tag{1}$$

followed by a mapping from $x_i$ to $\phi(x_i)$, which denotes the post-processing function applied after each step of operation. In this context, $\phi$ is generally a clipping function which prevents the amplitude of a soft spin from becoming too large. In the above equation, $\mathcal{F}_i$ is the driving term in the dynamics of an AIM variant, and $\zeta_i(t)$ is the noise term satisfying $\langle \zeta_i(t) \rangle = 0$ and $\langle \zeta_i(t) \zeta_j(t') \rangle = 2T\delta_{ij}\delta(t-t')$, and $T$ is the noise temperature.

Table 1: The Analog Ising Machines discussed in this paper.

| AIMs | CIM | clipCIM | simCIM | digCIM |
|---|---|---|---|---|
| $\mathcal{F}_i$ | $-x_i^3 + ax_i + \sum_j J_{ij} x_j$ | $-x_i^3 + ax_i + \sum_j J_{ij} x_j$ | $ax_i + \sum_j J_{ij} x_j$ | $ax_i + \sum_j J_{ij}\,\mathrm{sgn}\,x_j$ |
| $\phi(x)$ | $x$ | $\max(-B, \min(B, x))$ | $\max(-B, \min(B, x))$ | $\max(-B, \min(B, x))$ |

To begin with, consider the CIM for optimizing the Hamiltonian of the Sherrington-Kirkpatrick (SK) model,

$$H = -\frac{1}{2}\sum_{i \neq j} J_{ij}\sigma_i\sigma_j. \tag{2}$$

It is a fully-connected Ising spin glass model with the randomly quenched couplings $J_{ij}$ following independent Gaussian distributions with mean $\langle J_{ij} \rangle = 0$ and variance $\langle J_{ij}^2 \rangle = \delta_{ij} J^2/N$. The corresponding analog Ising machine replaces the Ising spins $\{\sigma_i\}$ by continuous variables $\{x_i\}$ and adds two extra terms. The first extra term is quadratic in $x_i$ with the gain $a$ tunable by the pump rate of the CIM. The second one is a fourth order term representing the nonlinearity that restricts the magnitude of $\{x_i\}$. This yields the Hamiltonian given by

$$H = -\frac{a}{2}\sum_i x_i^2 + \frac{1}{4}\sum_i x_i^4 - \frac{\xi}{2}\sum_{i \neq j} J_{ij} x_i x_j. \tag{3}$$

Let $Z(T) = \int \exp(-\beta H(\vec{x'}))\mathbf{d}(\vec{x'})$ be the partition function where $\beta \equiv 1/T$ is the inverse temperature. After replicating the system $n$ times,

$$Z^n = \prod_{\alpha=1}^n \left[\prod_{i=1}^N \int_{-\infty}^{\infty} dx_i^\alpha \exp\left(-\frac{\beta}{4}\sum_{i,\alpha}(x_i^\alpha)^4 + \frac{\beta a}{2}\sum_{i,\alpha}(x_i^\alpha)^2 + \frac{\beta \xi}{2}\sum_{i \neq j,\alpha} x_i^\alpha J_{ij} x_j^\alpha\right)\right]. \tag{4}$$

Consider the disorder average $[\cdots]_J$ of the interactions,

$$\left[\exp\left(\frac{\beta\xi}{2}\sum_{i \neq j,\alpha} x_i^\alpha J_{ij} x_j^\alpha\right)\right]_J = \exp\left(\sum_{i<j}\frac{\beta^2 J^2 \xi^2}{2N}\sum_{\alpha,\beta} x_i^\alpha x_j^\alpha x_i^\beta x_j^\beta\right). \tag{5}$$

Completing the squares of the terms dependent on a node index, and noting that terms with $i = j$ are negligible,

$$\left[\exp\left(\frac{\beta\xi}{2}\sum_{i\neq j,\alpha} x_i^\alpha J_{ij} x_j^\alpha\right)\right]_J = \exp\left[\frac{\beta^2 J^2 \xi^2}{2N}\sum_{\alpha<\beta}\left(\sum_i x_i^\alpha x_i^\beta\right)^2 + \frac{\beta^2 J^2 \xi^2}{4N}\sum_\alpha\left(\sum_i (x_i^\alpha)^2\right)^2\right]. \quad (6)$$

Applying the Hubbard-Stratonovich identity to the square terms,

$$\left[\exp\left(\frac{\beta\xi}{2}\sum_{i\neq j,\alpha} x_i^\alpha J_{ij} x_j^\alpha\right)\right]_J = \prod_\alpha \left(\int \frac{dq_{\alpha\alpha}}{\sqrt{4\pi/N\beta^2 J^2\xi^2}}\right) \prod_{\alpha<\beta}\left(\int \frac{dq_{\alpha\beta}}{\sqrt{2\pi/N\beta^2 J^2\xi^2}}\right) \\
\times \exp\left[-\frac{N\beta^2 J^2 \xi^2}{4}\sum_\alpha q_{\alpha\alpha}^2 - \frac{N\beta^2 J^2 \xi^2}{2}\sum_{\alpha<\beta} q_{\alpha\beta}^2 + \frac{\beta^2 J^2 \xi^2}{2}\sum_{\alpha,i} q_{\alpha\alpha}(x_i^\alpha)^2 + \beta^2 J^2\xi^2 \sum_{\alpha<\beta,i} q_{\alpha\beta} x_i^\alpha x_i^\beta\right]. \quad (7)$$

Note that the interactions between nodes have been decoupled. We thus obtain an expression with the integrals of the nodes factorized,

$$[Z^n]_J = \prod_\alpha \left(\int \frac{dq_{\alpha\alpha}}{\sqrt{4\pi/N\beta^2 J^2\xi^2}}\right) \prod_{\alpha<\beta}\left(\int \frac{dq_{\alpha\beta}}{\sqrt{2\pi/N\beta^2 J^2\xi^2}}\right) \exp\left\{-\frac{N\beta^2 J^2 \xi^2}{4}\sum_\alpha q_{\alpha\alpha}^2 - \frac{N\beta^2 J^2 \xi^2}{2}\sum_{\alpha<\beta} q_{\alpha\beta}^2\right.\\
\left.+ N\log \prod_\alpha\left(\int dx^\alpha\right)\exp\left[-\frac{\beta}{4}\sum_\alpha (x^\alpha)^4 + \frac{\beta a}{2}\sum_\alpha (x^\alpha)^2 + \frac{\beta^2 J^2\xi^2}{2}\sum_\alpha q_{\alpha\alpha}(x^\alpha)^2 + \beta^2 J^2\xi^2 \sum_{\alpha<\beta} q_{\alpha\beta} x^\alpha x^\beta\right]\right\}. \quad (8)$$

In the thermodynamic limit, the partition function is dominated by the saddle point of the parameters $q_{\alpha\alpha}$ and $q_{\alpha\beta}$. We first find the solution in the replica symmetric ansatz, which is applicable to the low gain regime. In this ansatz, $q_{\alpha\alpha} = Q$ and $q_{\alpha\beta} = q$ for $\alpha \neq \beta$. The partition function reduces to

$$[Z^n]_J = \exp\left(N\left\{-\frac{1}{4}\beta^2 J^2\xi^2\left[n^2 q^2 + n(Q^2 - q_1^2)\right] + \log\prod_\alpha\left(\int dx_\alpha\right)\exp\left[-\frac{\beta}{4}\sum_\alpha x_\alpha^4\right.\right.\right.\\
\left.\left.\left.+ \frac{\beta a}{2}\sum_\alpha x_\alpha^2 + \frac{1}{2}\beta^2 J^2\xi^2 q\left(\sum_\alpha x_\alpha\right)^2 + \frac{1}{2}\beta^2 J^2\xi^2(Q-q)\sum_\alpha x_\alpha^2\right]\right\}\right). \quad (9)$$

Then, using the Hubbard-Stratonovich identity, the replica indices in the log term can be decoupled, yielding

$$[Z^n]_J = \exp\left(N\left\{-\frac{1}{4}\beta^2 J^2\xi^2\left[n^2 q^2 + n(Q^2 - q^2)\right] + \log\int Dv\left(\prod_\alpha \int dx_\alpha\right)\right.\right.\\
\left.\left.\times \exp\left[-\frac{\beta}{4}\sum_\alpha x_\alpha^4 + \frac{\beta a}{2}\sum_\alpha x_\alpha^2 + \frac{1}{2}\beta^2 J^2\xi^2(Q-q)\sum_\alpha x_\alpha^2 + \beta J\xi\sqrt{q}v\sum_\alpha x_\alpha\right]\right\}\right), \quad (10)$$

where $Dv \equiv dv \exp(-u^2/2)/\sqrt{2\pi}$ is the Gaussian measure. In the limit $n \to 0$, we can obtain the expression of the average free energy per node $f$ using the replica formula

$$-\beta f = \frac{1}{N}[\log Z]_J = \lim_{n\to 0}\frac{[Z^n]_J - 1}{Nn}. \quad (11)$$

Factorizing the product of integrals,

$$-\beta f = -\frac{1}{4}\beta^2 J^2 \xi^2 (Q^2 - q^2) + \int Dv \log\left(\int \mathrm{d}x e^{-\beta g}\right), \tag{12}$$

where the effective free energy $g$ is given by

$$g = \frac{1}{4}x^4 - \frac{1}{2}\left[a + J^2 \xi^2 \beta (Q - q)\right] x^2 - J\xi\sqrt{q}vx. \tag{13}$$

The expressions of the order parameters are given by the saddle point equations,

$$\begin{aligned} q &= \int Dv \left(\frac{\int \mathrm{d}x\, e^{-\beta g} x}{\int \mathrm{d}x\, e^{-\beta g}}\right)^2, \\ Q &= \int Dv \frac{\int \mathrm{d}x\, e^{-\beta g} x^2}{\int \mathrm{d}x\, e^{-\beta g}}. \end{aligned} \tag{14}$$

In the above equations, the term $\beta(Q - q)$ in the expression of the effective free energy $g$ can be interpreted in the light of the fluctuation-response relation. To see this, we write

$$\chi \equiv \beta(Q - q) = \frac{\partial}{\partial h} \int Dv \frac{\int \mathrm{d}x\, e^{-\beta g + \beta hx} x}{\int \mathrm{d}x\, e^{-\beta g + \beta hx}}\bigg|_{h=0}. \tag{15}$$

Hence, $\chi$ can be interpreted as the susceptibility of $x$ in the presence of an external field.

## 2  The paramagnetic phase

We first consider the paramagnetic phase at nonzero temperature. This phase is characterized by the order parameter $q = 0$. In this case, the parameter $Q$ can be obtained from the self-consistent equation for given values of $a$ and $T$,

$$Q = \frac{\int dx e^{-\beta[\frac{1}{4}x^4 - \frac{1}{2}(a + \beta J^2 \xi^2 Q)x^2]} x^2}{\int dx e^{-\beta[\frac{1}{4}x^4 - \frac{1}{2}(a + \beta J^2 \xi^2 Q)x^2]}}. \tag{16}$$

To ensure that $q = 0$ is a stable solution in this phase, we consider the first order Taylor expansion of $q$ in Eq. (14), yielding

$$q = \int Dv \left\{\frac{\int dx e^{-\beta[\frac{1}{4}x^4 - \frac{1}{2}(a + \beta J^2 \xi^2 Q)x^2]} \beta J\xi\sqrt{q}vx^2}{\int dx e^{-\beta[\frac{1}{4}x^4 - \frac{1}{2}(a + \beta J^2 \xi^2 Q)x^2]}}\right\}^2, \tag{17}$$

As the term inside the curly bracket becomes $Q$ after extracting extra terms, this simplifies to

$$q = \beta^2 J^2 \xi^2 Q^2 q. \tag{18}$$

Therefore, the paramagnetic-spin glass phase transition is given by

$$T = J\xi Q. \tag{19}$$

Besides the order parameter $q$, the second moment $Q$ also undergoes a continuous transition from 0 to $q$ at the paramagnet-spin glass transition in the zero temperature limit (different from the nonzero temperature case in which $Q$ takes finite values). This simplifies the numerical detection of the transition point as $Q$ is

independent of thermal averaging. To see this, we note that in the zero temperature limit in the paramagnetic phase, Eq. (16) reads

$$Q = \frac{\int dx e^{\frac{\beta}{2}(a+\beta J^2\xi^2 Q)x^2} x^2}{\int dx e^{\frac{\beta}{2}(a+\beta J^2\xi^2 Q)x^2}}, \tag{20}$$

leading to

$$Q = \frac{T}{2J^2\xi^2}(-a - \sqrt{a^2 - 4J^2\xi^2}). \tag{21}$$

This paramagnetic solution breaks down at the transition point $a = -2J\xi$. Taylor expansion of Eq. (16) to the next order yields the phase transition line in the space of $T$ versus $a$ in the low temperature limit

$$T \approx \frac{1}{3}J\xi(a + 2J\xi). \tag{22}$$

In the limit of zero temperature, the integrals of $x$ are dominated by the minimizer $x^* := \arg\min_x(g)$, that is,

$$x^{*3} - a_{\text{eff}} x^* = J\xi\sqrt{q}v. \tag{23}$$

In the above expression, $a_{\text{eff}} \equiv a + J^2\xi^2\chi$ is the effective gain. Note that the dynamics of a node is no longer merely driven by the external term $a$. Additionally, the term $J^2\xi^2\chi$ represents the influence of the node on its neighbors which is fed back to the node itself, and is commonly referred to as the Onsager reaction.

It can be observed that $x^*$ is a function of the Gaussian variable $v$. In the cavity approach to the SK model [1], $J\xi\sqrt{q}v$ represents the *cavity field*, which is the field experienced by a node when the system excluding the node is in thermodynamic equilibrium. In the zero temperature limit, $Q$ approaches $q$. The saddle point equation of $q$ reduces to

$$q = \int Dv \, x^{*2}. \tag{24}$$

Note that in the zero temperature limit, the susceptibility is the product of $\beta$ (which approaches infinity) and $Q - q$ (which approaches zero). In this case, its value can be determined from the fluctuation-response relation. As the external field $h$ plays the same role as the cavity field, taking the derivative with respect to $h$ is equivalent to taking the derivative with respect to $J\xi\sqrt{q}v$. Therefore,

$$\frac{\partial x^*}{\partial h} = (3x^{*2} - a_{\text{eff}})^{-1}, \tag{25}$$

leading to

$$\chi = \int Dv \, (3x^{*2} - a_{\text{eff}})^{-1}. \tag{26}$$

In the low gain regime, $x^* = 0$, resulting in $q = 0$ and $\chi = -1/(a + J^2\xi^2\chi)$. This is the paramagnetic phase in which each variable has no preference to be positive or negative. From Eq. (23), the perturbation of $x^*$ due to a perturbation $\delta q$ is

$$(\delta x^*)^2 = \frac{J^2\xi^2 v^2}{a_{\text{eff}}^2}\delta q. \tag{27}$$

In view of Eq. (24), this implies that the paramagnetic phase becomes unstable at $a_{\text{eff}} = -J\xi$. Hence, the zero solution of the Ising machine bifurcates at $a = -2J\xi$. Note that $a = -2J\xi$ is the minimum eigenvalue

5

of the coupling matrix $J_{ij}\xi$, whose eigenvalue distribution obeys Wigner's semicircular law with center 0 and radius $2J\xi$ [2]. This agrees with the prediction and simulation results of previous studies based on bifurcation analysis [3].

## 3 The 1RSB solution of the coherent Ising machine

When the system enters the spin glass phase, the replica symmetric solution becomes unstable, as verified in Sec. 7. The first-step replica symmetry-breaking (1RSB) ansatz provides a more appropriate framework [1]. In the 1RSB ansatz, the structure of the matrix of $q_{\alpha\beta}$ is characterized by having the $n$ replicas divided into $n/m$ blocks of size $m$. The diagonals take the value $Q$, while other elements in the diagonal blocks take the value $q_1$, and the elements in the off-diagonal blocks are $q_0$. Thus, the matrix take the shape

$$q = \begin{pmatrix} Q & q_1 & q_1 & & & \\ q_1 & Q & q_1 & & q_0 & \\ q_1 & q_1 & Q & & & \\ \hline & & & Q & q_1 & q_1 \\ & q_0 & & q_1 & Q & q_1 \\ & & & q_1 & q_1 & Q \end{pmatrix}. \tag{28}$$

The cross-replica terms in the partition function can then be simplified to

$$\sum_{\alpha<\beta} q_{\alpha\beta} x^\alpha x^\beta = \frac{1}{2} \left\{ q_0 \left( \sum_\alpha^n x^\alpha \right)^2 + (q_1 - q_0) \sum_{\text{block}=1}^{n/m} \left( \sum_{\alpha \in \text{block}}^m x^\alpha \right)^2 - q_1 \sum_\alpha (x^\alpha)^2 \right\}, \tag{29}$$

$$\sum_{\alpha<\beta} q_{\alpha\beta}^2 = \frac{1}{2} \left[ n^2 q_0^2 + nm(q_1^2 - q_0^2) - nq_1^2 \right]. \tag{30}$$

The partition function reduces to

$$[Z^n]_J = \exp\left[ N \left( -\frac{1}{4}\beta^2 J^2 \xi^2 \left[ n^2 q_0^2 + nm\left(q_1^2 - q_0^2\right) + n(Q^2 - q_1^2) \right] + \log \prod_\alpha \left( \int dx_\alpha \right) \exp\left\{ -\frac{\beta}{4} \sum_\alpha x_\alpha^4 \right. \right. \right.$$
$$\left. \left. \left. + \frac{\beta}{2} a \sum_\alpha x_\alpha^2 + \frac{1}{2}\beta^2 J^2 \xi^2 \left[ q_0 \left( \sum_\alpha x_\alpha \right)^2 + (q_1 - q_0) \sum_{\text{block}=1}^{n/m} \left( \sum_{\alpha \in \text{ block}}^m x_\alpha \right)^2 + (Q - q_1) \sum_\alpha x_\alpha^2 \right] \right\} \right) \right]. \tag{31}$$

Then, using the Hubbard-Stratonovich identity, the replica indices in the log term can be decoupled, yielding

$$[Z^n]_J = \exp\left[ N \left( -\frac{1}{4}\beta^2 J^2 \xi^2 \left[ n^2 q_0^2 + nm\left(q_1^2 - q_0^2\right) + n(Q^2 - q_1^2) \right] + \log \int Dv \left( \prod_\alpha \int dx_\alpha \right) \right. \right.$$
$$\left. \times \exp\left[ -\frac{\beta}{4} \sum_\alpha x_\alpha^4 + \frac{\beta a}{2} \sum_\alpha x_\alpha^2 + \beta J \xi v \sqrt{q_0} \sum_\alpha x_\alpha + \frac{\beta^2 J^2}{2} \xi^2 (Q - q_1) \sum_\alpha x_\alpha^2 \right] \right. \tag{32}$$
$$\left. \left. \times \prod_{\text{block}=1}^{n/m} \left\{ \int Du \exp\left[ \beta J \xi u \sqrt{q_1 - q_0} \sum_{\alpha \in \text{ block}}^m x_\alpha \right] \right\} \right) \right].$$

In the limit $n \to 0$, we can obtain the expression of the average free energy per node $f$ using the replica formula. Due to the symmetry between the blocks, products of terms involving individual replicas become identical, and all replica indices of $x$ can be omitted,

$$-\beta f = -\frac{1}{4}\beta^2 J^2 \xi^2 [m(q_1^2 - q_0^2) + Q^2 - q_1^2] + \frac{1}{m} \int Dv \log \int Du \left( \int \mathrm{d}x e^{-\beta g} \right)^m, \tag{33}$$

where the effective free energy $g$ is given by

$$g = \frac{1}{4}x^4 - \frac{1}{2}\left[a + J^2 \xi^2 \beta (Q - q_1)\right] x^2 - J\xi \left(\sqrt{q_0}v + \sqrt{q_1 - q_0}u\right) x. \tag{34}$$

The expressions of the order parameters are given by the saddle point equations,

$$\begin{aligned}
q_0 &= \int Dv \left[ \frac{\int Du \left( \int \mathrm{d}x\, e^{-\beta g} \right)^{m-1} \left( \int \mathrm{d}x\, e^{-\beta g} x \right)}{\int Du \left( \int \mathrm{d}x\, e^{-\beta g} \right)^m} \right]^2, \\
q_1 &= \int Dv \frac{\int Du \left( \int \mathrm{d}x\, e^{-\beta g} \right)^{m-2} \left( \int \mathrm{d}x\, e^{-\beta g} x \right)^2}{\int Du \left( \int \mathrm{d}x\, e^{-\beta g} \right)^m}, \\
Q &= \int Dv \frac{\int Du \left( \int \mathrm{d}x\, e^{-\beta g} \right)^{m-1} \left( \int \mathrm{d}x\, e^{-\beta g} x^2 \right)}{\int Du \left( \int \mathrm{d}x\, e^{-\beta g} \right)^m}.
\end{aligned} \tag{35}$$

The saddle point equation for the 1RSB order parameter $m$ is given by

$$\begin{aligned}
&\frac{1}{4}J^2\xi^2(q_1^2 - q_0^2) + \frac{1}{(m\beta)^2} \int Dv \log \left[ \int Du \left( \int \mathrm{d}x\, e^{-\beta g} \right)^m \right] \\
&+ \frac{1}{m\beta} \int Dv \frac{\int Du \left( \int \mathrm{d}x\, e^{-\beta g} \right)^m [-\beta^{-1} \log(\int \mathrm{d}x\, e^{-\beta g})]}{\int Du \left( \int \mathrm{d}x\, e^{-\beta g} \right)^m} = 0.
\end{aligned} \tag{36}$$

In the above equations, the term $\beta(Q - q_1)$ can be interpreted as the susceptibility of $x$ in the presence of an external field. In the light of the fluctuation-response relation,

$$\chi \equiv \beta(Q - q_1) = \frac{\partial}{\partial h} \int Dv \frac{\int Du \left( \int \mathrm{d}x\, e^{-\beta g} \right)^m \left( \int \mathrm{d}x\, e^{-\beta g + \beta h x} x \right) / \left( \int \mathrm{d}x\, e^{-\beta g + \beta h x} \right)}{\int Du \left( \int \mathrm{d}x\, e^{-\beta g} \right)^m} \bigg|_{h=0}. \tag{37}$$

To conserve space in the main text, we adopt the notation $[O(x)]_u = \frac{\int Du\, e^{-m\beta g(x)} O(x)}{\int Du\, e^{-m\beta g(x)}}$ and $[[O(x)]_u]_v = \int Dv [O(x)]_u$ for a quantity $O(x)$ dependent on $x$. For clarity in our calculations in the Supplementary Materials, we still keep the complete expression of above averages in the subsequent sections.

## 4 Zero temperature limit of the 1RSB solution, distributions and phases

In the limit of zero temperature, the integral of $x$ is dominated by the minimizer $x^* := \arg\min_x(g)$, that is,

$$x^{*3} - a_{\text{eff}} x^* = J\xi w. \tag{38}$$

7

In the above expression, $a_{\text{eff}} \equiv a + J^2 \xi^2 \chi$ is the effective gain, and $w \equiv \sqrt{q_0} v + \sqrt{q_1 - q_0} u$. Analogous to the RS case, $J\xi w$ represents the cavity field. Yet, beyond the RS solution, the 1RSB solution takes into account the rough energy landscape of the solution space, which consists of a collection of subspace commonly referred to as "pure states", with $q_1$ and $q_0$ representing the correlation of solutions in the same and different pure states, respectively. The saddle point equations reduce to

$$q_0 = \int Dv \left[ \frac{\int Du \, e^{-m\beta g^*} x^*}{\int Du \, e^{-m\beta g^*}} \right]^2,$$
$$q_1 = \int Dv \frac{\int Du \, e^{-m\beta g^*} x^{*2}}{\int Du \, e^{-m\beta g^*}},$$
(39)

where $g^*$ is given by

$$g^* = \frac{1}{4} x^{*4} - \frac{1}{2} a_{\text{eff}} x^{*2} - J\xi w x^*. \tag{40}$$

The saddle point equation for the 1RSB order parameter $m$ is given by

$$\frac{1}{4} J^2 \xi^2 (q_1^2 - q_0^2) + \frac{1}{(m\beta)^2} \int Dv \log \left( \int Du \, e^{-m\beta g^*} \right) + \frac{1}{m\beta} \int Dv \frac{\int Du \, e^{-m\beta g^*} g^*}{\int Du \, e^{-m\beta g^*}} = 0. \tag{41}$$

One notices that when $T \to 0$, $m\beta$ converges to a constant, while $m$ approaches zero. Hereafter, the combination $m\beta$ will continue to be used in the zero temperature limit, as it remains a constant.

In the zero temperature limit, the susceptibility can be determined from the fluctuation-response relation,

$$\chi = \int Dv \frac{\int Du \, e^{-m\beta g^*} \left( 3x^{*2} - a_{\text{eff}} \right)^{-1}}{\int Du \, e^{-m\beta g^*}}. \tag{42}$$

At the equilibrium state at finite temperature, there are two distinct probability densities for $x_i$, which can be expressed as:

$$P_{\text{inst}}(c) = \int Dv \frac{\int Du [\int dx \, e^{-\beta g(x)}]^{m-1} e^{-\beta g(c)}}{\int Du [\int dx \, e^{-\beta g(x)}]^m},$$
$$P_{\text{aver}}(\langle c \rangle) = \int Dv \frac{\int Du [\int dx \, e^{-\beta g(x)}]^m \delta \left[ \frac{\int dx \, e^{-\beta g(x)} x}{\int dx \, e^{-\beta g(x)}} - \langle c \rangle \right]}{\int Du [\int dx \, e^{-\beta g(x)}]^m}.$$
(43)

The first density characterizes the instantaneous distribution of spins, whereas the second density represents the thermal-averaged amplitude of spins. In the paramagnetic phase, the variance of $P_{\text{aver}}$ is $q = 0$ according to the replica symmetric solution, implying that it is a delta function at $\langle c \rangle = 0$. This corresponds to the delta function (the brown and red curves) in Fig. 1(c). Decoding the ground state energy using the thermal-averaged spin values is not possible. On the other hand, $P_{\text{inst}}$ has a nonzero distribution. As we shall see, the nonzero distribution endows the system the ability, albeit limited, to decode the ground state energy even in the paramagnetic phase.

In the spin glass phase, $P_{\text{inst}}$ is gapless at low gains (the blue line in Fig. 1(a)). At higher gains and zero temperature, a gap opens in $P_{\text{inst}}$. At finite temperature, the gap becomes a pseudogap, but $P_{\text{inst}}$ remains to have two maxima, facilitating efficient decoding of the minimum energy (Fig. 1(b)), The pseudogap region

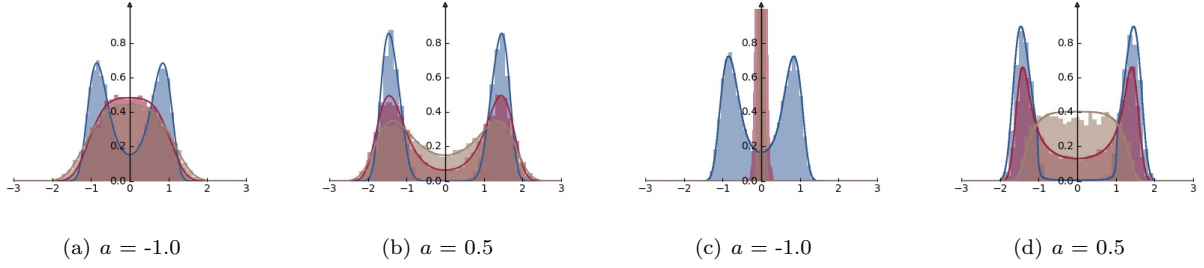

(a) $a = -1.0$  (b) $a = 0.5$  (c) $a = -1.0$  (d) $a = 0.5$

Figure 1: Probability distribution functions for analog spins. The solid line is the theoretical value based on Eq. (43) and the shaded area is the numerical simulation of the CIM dynamics. The brown, red and blue colors correspond to the values at temperatures of 1, 0.5 and 0.03 respectively. Fig. 1(a) and 1(b) are the instantaneous distribution $P_{\text{inst}}$, while Fig. 1(c) and 1(d) are the thermal average distribution $P_{\text{aver}}$.

is also present in the paramagnetic phase. Furthermore, we found that $P''_{\text{inst}}(0) > \beta a_{\text{eff}} P_{\text{inst}}(0)$. Hence, in the pseudogap region where $a_{\text{eff}} > 0$, $P_{\text{inst}}$ always has two maxima.

The CIM's phase diagram in Fig. 2 summarizes the different types of distribution. The paramagnetic phase with a delta function of $P_{\text{aver}}$ is indicated by the blue area. The spin glass phase with a broadened $P_{\text{inst}}$ is indicated by the orange area. The pseudogap region, spanning across both the paramagnetic and spin glass phases, is indicated by the green area.

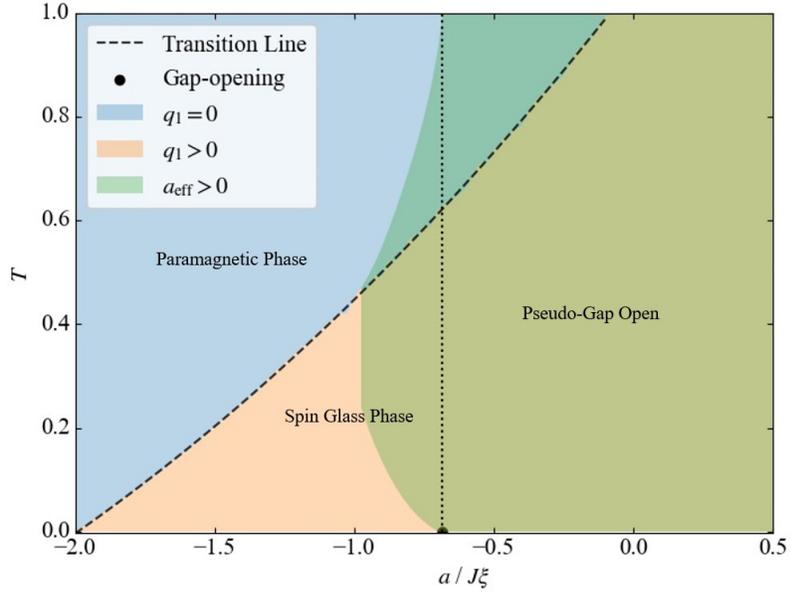

Figure 2: The phase diagram of the coherent Ising machine

Below, we focus on the instantaneous distribution, and $P_{\text{inst}}(x)$ is simply denoted as $P(x)$. At zero temper-

ature, the distribution of the $x$-components is given by

$$P(x) = \int Dv \frac{\int Du\ e^{-m\beta g^*} \delta(x - x^*)}{\int Du\ e^{-m\beta g^*}}, \tag{44}$$

Two regimes of the $P(x)$ can be identified. When the effective gain $a_\text{eff}$ is negative, $x^*$ is a single-valued function of the cavity field $w$. This implies that $P(x)$ is a continuous distribution.

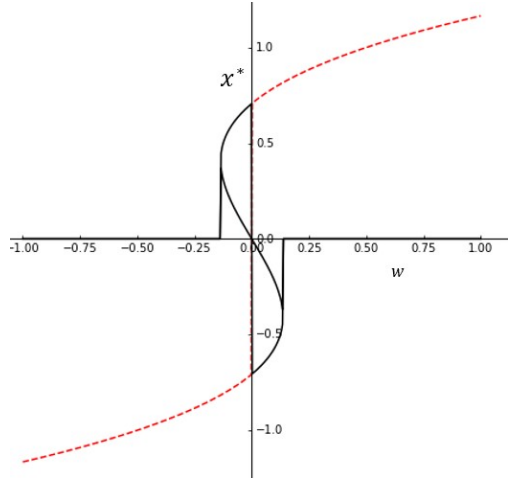

Figure 3: When $a_\text{eff}$ is positive, the stationary condition of the effective free energy provides three roots. The optimal root is the red dashed line, which is an odd and discontinuous function of $w$.

On the other hand, when $a_\text{eff}$ is positive, there exists a range of $w$ in which multiple values of $x^*$ for each $w$ are possible. In this case, the value of $x^*$ that yields the minimum $g^*$ is dominant. As shown in Fig. 3, this leads to the emergence of a gap in the two-band distribution $P(x)$. Hence, the gap-opening point is given by

$$a_\text{eff} = 0. \tag{45}$$

The distribution $P(x)$ for increasing values of gain $a$ is shown in Fig. 4(a).

In the two-band regime, most saddle point equations remain the same. However, the susceptibility $\chi$ depends on $\partial x^*/\partial h$, which diverges at $w = 0$. This irregularity is resolved by explicitly writing the limits of the integration variable $u$ in the numerator of Eq. (37), which reads

$$\chi = \frac{\partial}{\partial h}\left[\int Dv \left(\int Du\ e^{-m\beta g^*}\right)^{-1} \left(\int_{u_0(h)}^{\infty} Du\ e^{-m\beta g^*} x^*(h) + \int_{-\infty}^{u_0(h)} Du\ e^{-m\beta g^*} x^*(h)\right)\right]\bigg|_{h=0}, \tag{46}$$

where $x^*(h)$ is the solution of $[x^*(h)]^3 - a_\text{eff} x^*(h) = J\xi w + h$ minimizing $g^*$, and $J\xi[\sqrt{q_0}v + \sqrt{q_1 - q_0}u_0(h)] = h$. Considering both the one-band and two-band regimes, the result is

$$\chi = \int Dv \frac{\int Du\ e^{-m\beta g^*}(3x^{*2} - a_\text{eff})^{-1}}{\int Du\ e^{-m\beta g^*}} + \Theta(a_\text{eff}) \frac{2\sqrt{a_\text{eff}}}{\sqrt{2\pi(q_1 - q_0)}} \int Dv \frac{e^{-\frac{q_0 v^2}{2(q_1 - q_0)} + \frac{1}{4}m\beta a_\text{eff}^2}}{\int Du\ e^{-m\beta g^*}}. \tag{47}$$

where $\Theta(x) = 1$ for $x > 0$ and 0 otherwise. Note that the second term in the numerator originates from the discontinuous change of the $x$-component across the gap of the distribution when the external drive $h$ is turned on, and is present only in the two-band regime.

10

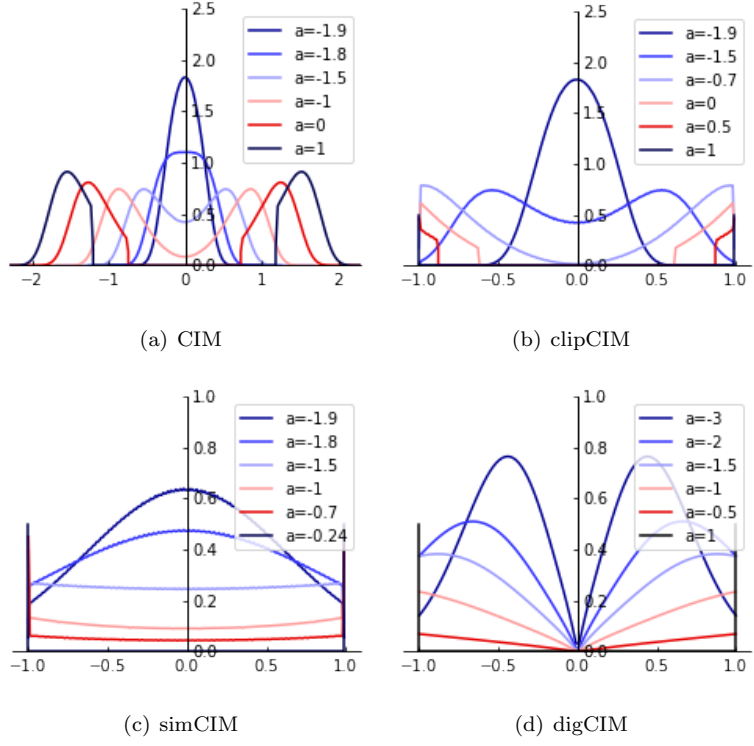

(a) CIM

(b) clipCIM

(c) simCIM

(d) digCIM

Figure 4: The distributions $P(x)$ of (a) CIM, (b) clipCIM, (c) simCIM (bSB) , (d) digCIM (dSB) at different values of the gain $a$. $B = 1$ for the clipCIM, simCIM and digCIM. (a) to (c) are obtained in the 1RSB ansatz, whereas (d) is obtained from the simulated field distribution of the SK model, which will be shown to be equivalent to the full RSB ansatz.

## 5 Decoded Energy

In an Ising machine, the decoded energy corresponds to the energy of the target Ising model that the machine provides, which is obtained by taking the sign of the variables $x_i$. The expression for the decoded energy (density) of an analog Ising machine is given by

$$E_{\text{decoded}} = \left[ \frac{-\frac{1}{2N} \int d\boldsymbol{x} \exp(-\beta H(\boldsymbol{x})) \operatorname{sgn} \boldsymbol{x}^T \boldsymbol{J} \operatorname{sgn} \boldsymbol{x}}{\int \exp\left(-\beta H\left(\boldsymbol{x}'\right)\right) d\boldsymbol{x}'} \right]_J. \tag{48}$$

Applying the replica method, we focus on the decoded nodes in replica 1 as the number of replicas, $n$, approaches zero. Furthermore, since the couplings are identically distributed, it is convenient to consider only interactions involving node 1,

$$[\langle E_{\text{decoded}} \rangle]_J = -\frac{1}{2} \left[ \prod_{\alpha,i} \int dx_i^\alpha \exp(-\sum_\alpha \beta H(\boldsymbol{x}^\alpha)) \sum_{l \neq 1}^N \operatorname{sgn} x_l^1 J_{1l} \operatorname{sgn} x_1^1 \right]_J. \tag{49}$$

11

Separating the averaging of the couplings feeding node 1 and the rest, we obtain

$$[\langle E_{\text{decoded}}\rangle]_J = -\frac{1}{2}\prod_{\alpha,i}\int dx_i^\alpha e^{-\frac{\beta}{4}(x_i^\alpha)^4 + \frac{\beta a}{2}(x_i^\alpha)^2} \sum_{l\neq 1}\int \frac{dJ_{1l}}{\sqrt{2\pi J^2/N}}$$
$$\times \exp\left[-\frac{NJ_{1l}^2}{2J^2} + \sum_\alpha \beta\xi J_{1l}x_1^\alpha x_l^\alpha\right] J_{1l}\,\text{sgn}\,x_1^1\,\text{sgn}\,x_l^1 \qquad (50)$$
$$\times \prod_{(ij)\neq(1l)}\int \frac{dJ_{ij}}{\sqrt{2\pi J^2/N}}\exp\left[-\frac{NJ_{ij}^2}{2J^2} + \sum_\alpha \beta\xi J_{ij}x_i^\alpha x_j^\alpha\right].$$

The distribution of $J_{1l}$ is Gaussian with mean $J^2\beta\xi\sum_\alpha x_1^\alpha x_l^\alpha/N$. Hence, the average over $J_{1l}$ can be replaced by its mean. The resultant expression is thus the disorder average over the system of the mean, yielding

$$[\langle E_{\text{decoded}}\rangle]_J = -\frac{1}{2N}\sum_{\alpha,l\neq 1}\beta\xi J^2\langle x_1^\alpha x_l^\alpha \text{sgn}x_1^1\text{sgn}x_l^1\rangle. \qquad (51)$$

In the 1RSB ansatz, this expression becomes

$$[\langle E_{\text{decoded}}\rangle]_J = -\frac{1}{2}\beta\xi J^2[\langle x^1\text{sgn}x^1\rangle^2 + (m-1)\langle x^1\text{sgn}x^2\rangle^2 + (n-m)\langle x^1\text{sgn}x^n\rangle^2]. \qquad (52)$$

Hence, we get

$$[\langle E_{\text{decoded}}\rangle]_J = -\frac{1}{2}\beta\xi J^2\left[\int Dv\frac{\int Du\left(\int dx\,e^{-\beta g}x\,\text{sgn}\,x\right)\left(\int dx\,e^{-\beta g}\right)^{m-1}}{\int Du\left(\int dx\,e^{-\beta g}\right)^m}\right]^2$$
$$-\frac{1}{2}(m-1)\beta\xi J^2\left[\int Dv\frac{\int Du\left(\int dx\,e^{-\beta g}x\right)\left(\int dx\,e^{-\beta g}\,\text{sgn}\,x\right)\left(\int dx\,e^{-\beta g}\right)^{m-2}}{\int Du\left(\int dx\,e^{-\beta g}\right)^m}\right]^2$$
$$-\frac{1}{2}(n-m)\beta\xi J^2\left[\int Dv\left(\frac{\int Du\left(\int dx\,e^{-\beta g}x\right)\left(\int dx\,e^{-\beta g}\right)^{m-1}}{\int Du\left(\int dx\,e^{-\beta g}\right)^m}\right)\right. \qquad (53)$$
$$\left.\times \left(\frac{\int Du\left(\int dx\,e^{-\beta g}\,\text{sgn}\,x\right)\left(\int dx\,e^{-\beta g}\right)^{m-1}}{\int Du\left(\int dx\,e^{-\beta g}\right)^m}\right)\right]^2.$$

In the limit $\beta \to \infty$, the expression contains an indeterminate term that is the product of $\beta$ and a vanishing factor. Expressing the term as a response to an external field, we find

$$\int Du e^{-m\beta g}\frac{\int dx\,e^{-\beta g}x\,\text{sgn}\,x}{\int dx\,e^{-\beta g}} - \int Du e^{-m\beta g}\frac{\left(\int dx\,e^{-\beta g}x\right)\left(\int dx\,e^{-\beta g}\,\text{sgn}\,x\right)}{\left(\int dx\,e^{-\beta g}\right)^2}$$
$$= \frac{1}{\beta}\frac{\partial}{\partial h}\left(\int Du e^{-m\beta g}\frac{\int dx\,e^{-\beta g+\beta h x}\,\text{sgn}\,x}{\int dx\,e^{-m\beta g+\beta h x}}\right). \qquad (54)$$

Separating the integrals for positive and negative $x$,

$$\int Du e^{-m\beta g}\frac{\int dx\,e^{-\beta g}x\,\text{sgn}\,x}{\int dx\,e^{-\beta g}} - \int Du e^{-m\beta g}\frac{\left(\int dx\,e^{-\beta g}x\right)\left(\int dx\,e^{-\beta g}\,\text{sgn}\,x\right)}{\left(\int dx\,e^{-\beta g}\right)^2}$$
$$= \frac{1}{\beta}\frac{\partial}{\partial h}\left(\int_{-\frac{\sqrt{q_0}v+h}{\sqrt{q_1-q_0}}}^{\infty} Du e^{-m\beta g} - \int_{-\infty}^{-\frac{\sqrt{q_0}v+h}{\sqrt{q_1-q_0}}} Du e^{-m\beta g}\right) = \frac{1}{\beta}\frac{2}{\sqrt{2\pi(q_1-q_0)}}\exp\left(-\frac{q_0 v^2}{2(q_1-q_0)} + \frac{1}{4}m\beta a_{\text{eff}}^2\right),$$
$$\qquad (55)$$

We arrive at the expression for the decoded energy at zero temperature

$$[\langle E_{\text{decoded}} \rangle]_J = -\frac{2J}{\sqrt{2\pi(q_1-q_0)}} \left( \int Dv \frac{\int Du\, e^{-m\beta g^*} |x^*|}{\int Du\, e^{-m\beta g^*}} \right) \int Dv \frac{\exp\left(-\frac{q_0 v^2}{2(q_1-q_0)} + \frac{1}{4}m\beta a_{\text{eff}}^2 \Theta(a_{\text{eff}})\right)}{\int Du\, e^{-m\beta g^*}}$$
$$-\frac{1}{2}m\beta J \left\{ \left[ \int Dv \frac{\int Du\, e^{-m\beta g^*} |x^*|}{\int Du\, e^{-m\beta g^*}} \right]^2 - \left[ \int Dv \left( \frac{\int Du\, e^{-m\beta g^*} x^*}{\int Du\, e^{-m\beta g^*}} \right) \left( \frac{\int Du\, e^{-m\beta g^*} \operatorname{sgn} x^*}{\int Du\, e^{-m\beta g^*}} \right) \right]^2 \right\}. \quad (56)$$

## 6 Rescaling and summary of CIM results

In our analysis so far, we observe that the energy has the dimension of the fourth order of the current $x$, and the dimensions of other quantities are determined accordingly. Using $J\xi$ as the basic unit, which scales as $x^2$, we note that $a$, $Q$, $q_1$ and $q_0$ scale as $J\xi$, $\chi$ scales as $(J\xi)^{-1}$, and the decoded energy scales as $J$. Hereafter, we adopt the convention that $J\xi = 1$ and summarize the equations below in terms of the dimensionless variables.

$$g \equiv \frac{1}{4}x^4 - \frac{1}{2}a_{\text{eff}} x^2 - \left(\sqrt{q_0} v + \sqrt{q_1 - q_0} u\right) x. \quad (57)$$

$$g^* \equiv \min_x(g) \qquad x^* \equiv \arg\min(g). \quad (58)$$

$$q_1 = \int Dv \frac{\int Du\, e^{-m\beta g^*} x^{*2}}{\int Du\, e^{-m\beta g^*}}. \quad (59)$$

$$q_0 = \int Dv \left( \frac{\int Du\, e^{-m\beta g^*} x^*}{\int Du\, e^{-m\beta g^*}} \right)^2. \quad (60)$$

$$\chi = \int Dv \frac{\int Du\, e^{-m\beta g^*} (3x^{*2} - a_{\text{eff}})^{-1}}{\int Du\, e^{-m\beta g^*}} + \frac{\Theta(a_{\text{eff}}) 2\sqrt{a_{\text{eff}}}}{\sqrt{2\pi(q_1-q_0)}} \int Dv \frac{e^{-\frac{q_0 v^2}{2(q_1-q_0)} + \frac{1}{4}m\beta a_{\text{eff}}^2}}{\int Du\, e^{-m\beta g^*}}. \quad (61)$$

$$\frac{1}{4}(q_1^2 - q_0^2) + \frac{1}{(m\beta)^2} \int Dv \ln \left( \int Du\, e^{-m\beta g^*} \right) + \frac{1}{m\beta} \frac{\int Du\, e^{-m\beta g^*} g^*}{\int Du\, e^{-m\beta g^*}} = 0. \quad (62)$$

$$E_{\text{decoded}} = -\frac{2}{\sqrt{2\pi(q_1-q_0)}} \left( \int Dv \frac{\int Du\, e^{-m\beta g^*} |x^*|}{\int Du\, e^{-m\beta g^*}} \right) \int Dv \frac{e^{-\frac{q_0 v^2}{2(q_1-q_0)} + \Theta(a_{\text{eff}}) \frac{1}{4}m\beta a_{\text{eff}}^2}}{\int Du\, e^{-m\beta g^*}}$$
$$-\frac{1}{2}m\beta \left\{ \left[ \int Dv \frac{\int Du\, e^{-m\beta g^*} |x^*|}{\int Du\, e^{-m\beta g^*}} \right]^2 - \left[ \int Dv \left( \frac{\int Du\, e^{-m\beta g^*} x^*}{\int Du\, e^{-m\beta g^*}} \right) \left( \frac{\int Du\, e^{-m\beta g^*} \operatorname{sgn} x^*}{\int Du\, e^{-m\beta g^*}} \right) \right]^2 \right\}. \quad (63)$$

The pseudo-code for solving the above saddle point equations in the single-band regime is shown in algorithm 1.

**Algorithm 1** The numerical method for determining the order parameters

**Input:** Constants $a, m\beta$, Initial values $\chi, q_0, q_1$.

**Output:** Order parameters $q_1, q_0$, and $\chi$.

1: Initialize $q_1, q_0$, and $\chi$;
2: **for** i = 0, iter_num, i++ **do**
3:     Set $q_1', q_0', \chi'$ as zero;
4:     **for** each Gaussian variable $v$ **do**
5:         Set dnmntr, nmrtr_$(q_1, q_0, m\beta)$ as zero;
6:         **for** each Gaussian variable $u$ **do**
7:             Get $x^*$ from $-x^3 + (a + \chi)x + (\sqrt{q_0}v + \sqrt{q_1 - q_0}u) = 0$;
8:             Get $\partial_h x^* = (3x^{*2} - a_{\text{eff}})^{-1}$;
9:             Sum over dnmntr $+= e^{-m\beta g^*}$;
10:            Sum over nmrtr_$q_1$ $+= e^{-m\beta g^*} x^{*2}$;
11:            Sum over nmrtr_$q_0$ $+= e^{-m\beta g^*} x^*$;
12:            Sum over nmrtr_$\chi$ $+= e^{-m\beta g^*} \partial_h x^*$;
13:         **end for**
14:         $q_1' += $ nmrtr_$q_1$/dnmntr;
15:         $q_0' += $ (nmrtr_$q_0$/dnmntr)$^2$;
16:         $\chi' += $ nmrtr_$\chi$/dnmntr;
17:     **end for**
18:     Update $q_1, q_0, \chi = \frac{q_1'}{\#v}, \frac{q_0'}{\#v}, \frac{\chi'}{\#v}$;
19: **end for**
20: Repeat the calculations for a range of values of $m\beta$ and find the root of its saddle point equation by line search.

We observe that the saddle point equation for $m\beta$ exhibits instability when it is iterated. Consequently, this algorithm refrains from updating this parameter during each iteration. Instead, line search techniques are further employed to determine the final $m\beta$. Upon obtaining the order parameters $q_1, q_0$ and the limiting values $\chi, m\beta$, one can proceed to compute the free energy $f$, energy $e$, and decoded energy $E_{\text{decoded}}$.

Figure 5 displays the decoded energy contours for different parameter pairs using the decoded energy expression. It is apparent that when the energy is low, the rate of energy decrease with the gain and the temperature becomes increasingly slow. Moreover, the effect of $a$ on the decoded energy is akin to that of the inverse temperature in this system. Lowering the temperature (noise level) or raising $a$ helps to attain states at lower energy levels. As both $a$ and the inverse temperature increase, the energy reaches approximately $-0.765$, which corresponds to the ground state of the SK model under the 1RSB assumption. In practical applications of CIMs, achieving lower energy states quickly may be prioritized over lowest energy

14

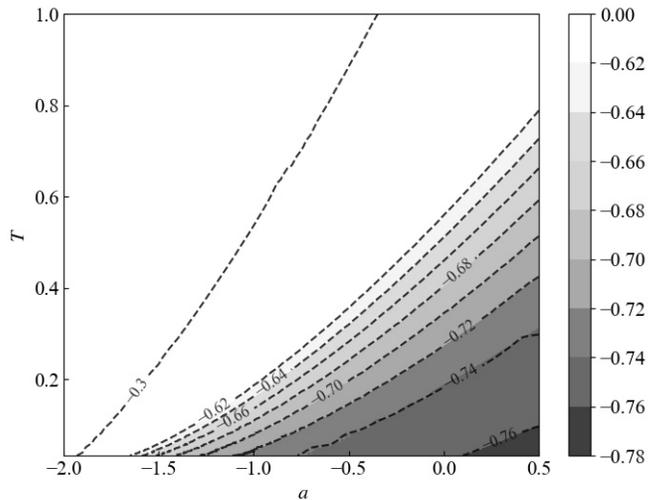

Figure 5: Decoded energy contour line in the space of $a$ and $T$.

states. However, directly setting low $T$ or high $a$ risks premature local minima trapping. Based on Fig. 5, researchers can choose optimal parameters or annealing schedules for the Ising machine that yield the target decoded energy. Note that the contour presented has been constructed with the parameter $\beta = 1/T$ ranging from 1 to 30, which implies that the temperature does not approach absolute zero. Consequently, the contour plot in Fig. 5 does not capture the behavior of the energy feature for temperatures in the vicinity of zero. This limitation is further demonstrated in Fig. 6(b).

Figure 6 presents the results of numerical experiments with 10,000 spins ($T = 10^{-4}$) by linearly increasing $a$ from $-2$ to 1 with simulation time. These results are compared with the theoretical value at zero temperature. The simulation and analytical results are found to be in good agreement. As shown in Fig. 4(a), when the gain is low ($a < -2$), the order parameter is in the paramagnetic phase with $Q = 0$. At $a = -2$, $Q$ starts to undergo a continuous transition. This agrees excellently with the prediction that transition sets in when the gain $a$ becomes equal to the minimum eigenvalue of the coupling matrix [3]. When $a > -2$, $Q$ increases and becomes linear asymptotically. This is expected since for $a \gg 1$, $x^*$ approaches $\text{sgn} w \sqrt{a_\text{eff}} \pm O(1)$, such that $Q$ approaches $a_\text{eff}$. As the magnitude of the variables $x_i$ becomes uniformly distributed, the system becomes Ising-like. It is recognized that the uniform distribution of the variable magnitudes is a prerequisite for the analog system to serve as an accurate optimizer of Ising problems [4, 5]. Thus, we expect that accurate optimization can be achieved for sufficiently strong gains.

Figure 6(b) shows the dependence of the decoded energy on the gain. It is interesting to note that the decoded energy is already negative even in the paramagnetic phase. This indicates that locally optimal clusters are already present in the spin configurations. However, due to the inversion symmetry of the Hamiltonian, clusters favoring either state of inversion symmetry remain equally probable. The exact expression at $T = 0$

shows that below the bifurcation threshold,

$$E_{\text{decoded}} = -\frac{2J\chi}{\pi} = \frac{J}{\pi}\left(a + \sqrt{a^2 - 4}\right). \tag{64}$$

We remark that the above expression is applicable in the limit of vanishing temperature. At exactly zero temperature in the paramagnetic phase, the dynamical behavior is dominated by other spurious states.

At the bifurcation threshold $a = -2$, one of the two inversion-symmetric states becomes dominant and the inversion symmetry is broken. The decoded energy changes continuously with $a$, but a kink exists at the bifurcation point as the square root behavior below the threshold (see Eq. (64)) changes to a linear behavior above,

$$[\langle E_{\text{decode}}\rangle]_J = -\frac{2J}{\pi}\left(1 + \frac{a+2}{3}\right) \quad a \gtrsim -2. \tag{65}$$

There were hints that the spin configuration at this point is a precursor of the optimal configuration [3], but noting that the decoded energy of $-0.64$ remains higher than the optimal energy of $-0.76$, the configuration may still be considerably remote from the optimal configuration.

When $a$ increases above the bifurcation point, the decoded energy drops rapidly. When it reaches the gap-opening point at $a = -0.70$, the decoded energy already reaches -0.753, very close to the optimal energy. When $a$ increases further, the decoded energy decreases smoothly to the optimum. The asymptotic theoretical value of the 1RSB decoded energy extrapolated to infinite values of $a$ is $-0.766$, which is slightly below the value of $-0.765$ obtained from simulations around $a \sim O(1)$. For reference, the optimal ground state energy of the SK model is $-0.763 \sim -0.765$ [6, 7], and the 1RSB value is known to be slightly below. This shows that the agreement between the CIM and the SK model is excellent.

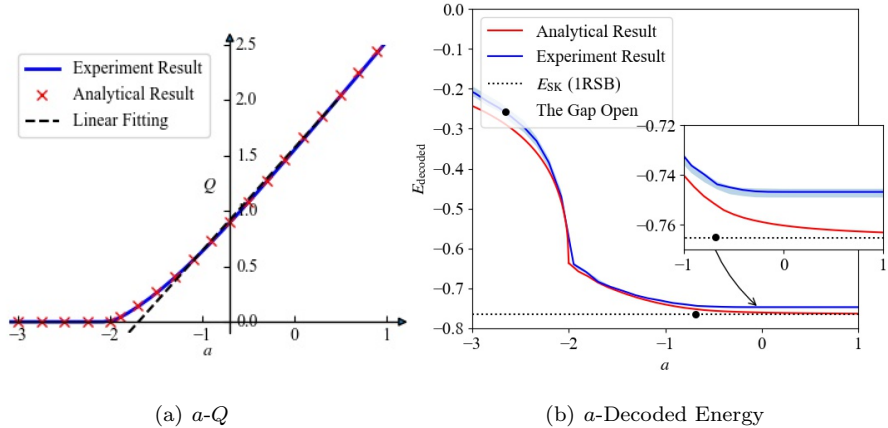

(a) $a$-$Q$  (b) $a$-Decoded Energy

Figure 6: (a) The dependence of the order parameter $Q$ on the gain $a$ in the zero temperature limit. In the simulations, the second moment of CIM's analog spins are obtained from one trial. When $a$ is large, the function approaches a linear relation. (b) The dependence of CIM's decoded energy on the gain $a$ at zero temperature. $E_{\text{SK}}(1RSB)$ is taken to be $-0.765$. In the simulations, the CIM's decoded energy is obtained from 20 trials, and the shaded areas are the maximum and minimum values for those experiments.

# 7 Stability of the RS solution

In the main text and the previous sections of this Supplementary Materials, we asserted that the replica symmetric (RS) solution becomes unstable when $q_1 > 0$. This instability prompted us to employ the 1-step replica symmetry-breaking (RSB) ansatz. In this section, we will substantiate this assertion by determining the de Almeida-Thouless (AT) line [8] and demonstrate that the Hessian matrix of the free energy is not positive definite when $q_1 > 0$. For convenience, we first rescale the variables using $\beta J \xi q_{\alpha\beta} = q_{\alpha\beta}$, which allows us to express the free energy as

$$f = -\lim_{n\to 0}\frac{1}{\beta n}\left[-\sum_{\alpha<\beta}\frac{(q_{\alpha\beta})^2}{2} - \sum_\alpha\frac{(q_{\alpha\alpha})^2}{4} + \log\prod_\alpha\int dx_\alpha \exp\left[-\frac{\beta}{4}\sum_\alpha x_\alpha^4\right.\right.$$
$$\left.\left.+\frac{\beta}{2}a\sum_\alpha x_\alpha^2 + \beta J\xi\sum_{\alpha<\beta}q_{\alpha\beta}x_\alpha x_\beta + \beta J\xi\sum_\alpha\frac{q_{\alpha\alpha}}{2}x_\alpha^2\right]\right]. \tag{66}$$

To assess the stability, we expand $[f]$ to second order of small deviations around the RS solution: $q_{\alpha\beta} = q + \epsilon_{\alpha\beta}$; $L_0 = -\frac{\beta}{4}\sum_\alpha x_\alpha^4 + \frac{\beta}{2}a\sum_\alpha x_\alpha^2 + \beta J\xi q\sum_{\alpha<\beta}x_\alpha x_\beta + \beta J\xi\frac{q}{2}\sum_\alpha x_\alpha^2$.

Substituting these values into the free energy expression, we obtain

$$\log\prod_\alpha\int dx_\alpha\exp\left[L_0 + \beta J\xi\sum_{\alpha<\beta}\epsilon_{\alpha\beta}x_\alpha x_\beta + \beta J\xi\sum_\alpha\frac{\epsilon_{\alpha\alpha}x_\alpha^2}{2}\right],$$
$$\approx \log\prod_\alpha\int dx_\alpha\exp(L_0) + \frac{\beta^2 J^2\xi^2}{2}\sum_{\alpha<\beta}\sum_{\gamma<\delta}\epsilon_{\alpha\beta}\epsilon_{\gamma\delta}\langle x_\alpha x_\beta x_\gamma x_\delta\rangle_{L_0} - \frac{\beta^2 J^2\xi^2}{2}\sum_{\alpha<\beta}\sum_{\gamma<\delta}\epsilon_{\alpha\beta}\epsilon_{\gamma\delta}\langle x_\alpha x_\beta\rangle_{L_0}\langle x_\gamma x_\delta\rangle_{L_0}$$
$$+\frac{\beta^2 J^2\xi^2}{4}\sum_\delta\sum_{\alpha<\beta}\epsilon_{\delta\delta}\epsilon_{\alpha\beta}\langle x_\delta^2 x_\alpha x_\beta\rangle_{L_0} - \frac{\beta^2 J^2\xi^2}{4}\sum_\delta\sum_{\alpha<\beta}\epsilon_{\delta\delta}\epsilon_{\alpha\beta}\langle x_\delta^2\rangle_{L_0}\langle x_\alpha x_\beta\rangle_{L_0}$$
$$+\frac{\beta^2 J^2\xi^2}{8}\sum_\alpha\sum_\beta\epsilon_{\alpha\alpha}\epsilon_{\beta\beta}\langle x_\alpha^2 x_\beta^2\rangle_{L_0} - \frac{\beta^2 J^2\xi^2}{8}\sum_\alpha\sum_\beta\epsilon_{\alpha\alpha}\epsilon_{\beta\beta}\langle x_\alpha^2\rangle_{L_0}\langle x_\beta^2\rangle_{L_0}.$$
(67)

Here, $\langle\cdots\rangle_{L_0}$ represents the average taken with the replica-symmetric weight $e^{L_0}$. Subsequently, the Hessian matrix of the free energy can be expressed as (with $x^2$ rescaled as $\beta J\xi x^2$ for calculation simplification):

$$\Delta \equiv \frac{1}{2}\sum_{\alpha<\beta}\sum_{\gamma<\delta}\epsilon_{\alpha\beta}\epsilon_{\gamma\delta}\left(\delta_{(\alpha\beta)(\gamma\delta)} - \langle x_\alpha x_\beta x_\gamma x_\delta\rangle_{L_0} + \langle x_\alpha x_\beta\rangle_{L_0}\langle x_\gamma x_\delta\rangle_{L_0}\right)$$
$$-\frac{1}{4}\sum_\delta\sum_{\alpha<\beta}\epsilon_{\delta\delta}\epsilon_{\alpha\beta}\left(\langle x_\delta^2 x_\alpha x_\beta\rangle_{L_0} - \langle x_\delta^2\rangle_{L_0}\langle x_\alpha x_\beta\rangle_{L_0}\right) \tag{68}$$
$$+\frac{1}{8}\sum_\alpha\sum_\beta\epsilon_{\alpha\alpha}\epsilon_{\beta\beta}\left(2\delta_{\alpha\beta} - \langle x_\alpha^2 x_\beta^2\rangle_{L_0} + \langle x_\alpha^2\rangle_{L_0}\langle x_\beta^2\rangle_{L_0}\right).$$

We denote the matrix of coefficients for this quadratic form in $q_{\alpha\beta}$ by $G$, that is:

$$
\begin{aligned}
G_{\alpha\alpha\alpha\alpha} &= \frac{1}{8}\left(2 - \langle x_\alpha^4 \rangle_{L_0} + \langle x_\alpha^2 \rangle_{L_0}^2\right) \equiv A; \\
G_{\alpha\alpha\beta\beta} &= \frac{1}{8}\left(\langle x_\alpha^2 \rangle_{L_0} \langle x_\beta^2 \rangle_{L_0} - \langle x_\alpha^2 x_\beta^2 \rangle_{L_0}\right) \equiv B; \\
G_{\alpha\beta\alpha\beta} &= \frac{1}{2}\left(1 - \langle x_\alpha^2 x_\beta^2 \rangle_{L_0} + \langle x_\alpha x_\beta \rangle_{L_0}^2\right) \equiv P; \\
G_{\alpha\beta\alpha\delta} &= \frac{1}{2}\left(\langle x_\alpha x_\beta \rangle_{L_0} \langle x_\alpha x_\delta \rangle_{L_0} - \langle x_\alpha^2 x_\beta x_\delta \rangle_{L_0}\right) \equiv Q; \\
G_{\alpha\beta\gamma\delta} &= \frac{1}{2}\left(\langle x_\alpha x_\beta \rangle_{L_0} \langle x_\gamma x_\delta \rangle_{L_0} - \langle x_\alpha x_\beta x_\gamma x_\delta \rangle_{L_0}\right) \equiv R; \\
G_{\alpha\alpha\alpha\beta} &= \frac{1}{4}\left(\langle x_\alpha x_\beta \rangle_{L_0} \langle x_\alpha^2 \rangle_{L_0} - \langle x_\alpha^3 x_\beta \rangle_{L_0}\right) \equiv C; \\
G_{\gamma\gamma\alpha\beta} &= \frac{1}{4}\left(\langle x_\alpha x_\beta \rangle_{L_0} \langle x_\gamma^2 \rangle_{L_0} - \langle x_\alpha x_\beta x_\gamma^2 \rangle_{L_0}\right) \equiv D.
\end{aligned}
\tag{69}
$$

From this, we can derive the positive definite conditions for $G$:

$$
\begin{aligned}
\lambda_1, \lambda_2 &= \frac{1}{2}\left\{A - B + P - 4Q + 3R \pm \sqrt{(A - B - P + 4Q - 3R)^2 - 8(C - D)^2}\right\} > 0, \\
\lambda_3 &= P - 2Q + R > 0.
\end{aligned}
\tag{70}
$$

In the RS ansatz, we get from the second relation that

$$
\beta^2 J^2 \xi^2 \int Dv \left[\frac{\int dx e^{-\beta g} x^2}{\int dx e^{-\beta g}} - \left(\frac{\int dx e^{-\beta g} x}{\int dx e^{-\beta g}}\right)^2\right]^2 < 1. \tag{71}
$$

Along the phase boundary between the paramagnetic and spin glass phases, this reduces to

$$
1 = \beta^2 J^2 \xi^2 Q^2, \tag{72}
$$

This implies that the boundary of AT stability is given by $T = J\xi Q$. This is identical to the phase boundary between paramagnet and spin glass, showing that the RS solution of the entire spin glass phase is unstable.

The stability analysis can be extended to study the stability of the 1RSB solution, and it can be shown that the 1RSB solution is also unstable. In fact, all finite-step RSB solutions of the SK model are unstable, and the model is described by the infinite-step RSB solution, referred to as the full RSB (FRSB) solution. Nevertheless, it was widely accepted that the 1RSB solution already represents a significant improvement from the RS solution [1]. Further steps of replica symmetry-breaking lead to less significant improvements, and hence 1RSB serves as a widely-used approximation. Compared with the RS ground-state energy of $-0.798$, 1RSB solution yields $-0.765$, much closer to the true value of $-0.763$.

## 8 The 1RSB solution of the clipCIM

Decimation is a common technique to boost the optimization performance of binary models, leading to enhanced accuracy [9–16] and robustness against hyperparameter uncertainties [17]. When applied to CIMs, a fraction of nodes with the strongest amplitudes of $c$-components are frozen according to their signs. The

dynamics of the rest of the nodes then continue, with the prospect of improved convergence under the influence of the frozen nodes. Recent experiments on CIMs with the trap-and-correct technique demonstrate the effectiveness of this method [3].

To theoretically explore the advantage of decimation, we consider the optimization performance of a clipped CIM. We analyze the Hamiltonian of a CIM in which the variables are restricted to within the range $[-B, B]$, thus forcing amplitudes greater than $B$ to be clipped. As we are considering the equilibrium state of the system, we assume that the decimated nodes have been optimally chosen. Thus, our analysis provides an upper bound to the performance of practical optimizers, and approximates asymptotically those with prudent choice of decimated nodes.

We consider the clipping function defined as

$$\phi(x) := \begin{cases} B & if \quad x > B; \\ x & if \quad x \in [-B, B]; \\ -B & if \quad x < -B, \end{cases} \quad (73)$$

This function imposes a boundary of $\pm B$ on the variable $x_i$. Hence, the partition function of the 1RSB analog Ising machine becomes

$$Z^n = \prod_{i,\alpha=1}^{n} \int_{x_i^\alpha \in [-B,B]} dx_i^\alpha \exp\left(-\frac{\beta}{4}\sum_{i,\alpha}(x_i^\alpha)^4 + \frac{\beta a}{2}\sum_{i,\alpha}(x_i^\alpha)^2 + \frac{\beta \xi}{2}\sum_{i\neq j,\alpha} x_i^\alpha J_{ij} x_j^\alpha\right). \quad (74)$$

Carrying out similar calculations as the CIM, we can derive the related quantities for the clipCIM. We present the zero-temperature case results here after rescaling:

$$g \equiv \frac{1}{4}x^4 - \frac{1}{2}a_{\text{eff}}x^2 - \left(\sqrt{q_0}v + \sqrt{q_1 - q_0}u\right)x. \quad (75)$$

$$g^* \equiv \min_x(g) \qquad x^* \equiv \arg\min(g). \quad (76)$$

$$q_1 = \int Dv \frac{\int Du\, e^{-m\beta g^*} x^{*2}}{\int Du\, e^{-m\beta g^*}}. \quad (77)$$

$$q_0 = \int Dv \left(\frac{\int Du\, e^{-m\beta g^*} x^*}{\int Du\, e^{-m\beta g^*}}\right)^2. \quad (78)$$

$$\chi = \int Dv \frac{\int Du\, e^{-m\beta g^*}(3x^{*2} - a_{\text{eff}})^{-1}\Theta(B - |x^*|)[\Theta(-a_{\text{eff}}) + \Theta(a_{\text{eff}})\Theta(|x^*| - \sqrt{a_{\text{eff}}})]}{\int Du\, e^{-m\beta g^*}}$$
$$+ \Theta(a_{\text{eff}})\frac{2\min(\sqrt{a_{\text{eff}}}, B)}{\sqrt{2\pi(q_1 - q_0)}} \int Dv \frac{e^{-\frac{q_0 v^2}{2(q_1 - q_0)} - m\beta[\frac{x^4}{4} - a_{\text{eff}}\frac{x^2}{2}]_{x=\min(\sqrt{a_{\text{eff}}}, B)}}}{\int Du\, e^{-m\beta g^*}}. \quad (79)$$

$$\frac{1}{4}(q_1^2 - q_0^2) + \frac{1}{(m\beta)^2}\int Dv \ln\left(\int Du\, e^{-m\beta g^*}\right) + \frac{1}{m\beta}\frac{\int Du\, e^{-m\beta g^*} g^*}{\int Du\, e^{-m\beta g^*}} = 0. \quad (80)$$

$$
\begin{aligned}
E_{\text{decoded}} &= -\frac{2}{\sqrt{2\pi(q_1-q_0)}} \left( \int Dv \frac{\int Du\ e^{-m\beta g^*} |x^*|}{\int Du\ e^{-m\beta g^*}} \right) \int Dv \frac{e^{-\frac{q_0 v^2}{2(q_1-q_0)} - \Theta(a_{\text{eff}})m\beta[\frac{x^4}{4} - a_{\text{eff}}\frac{x^2}{2}]_{x=\min(\sqrt{a_{\text{eff}}}, B)}}}{\int Du\ e^{-m\beta g^*}} \\
&\quad - \frac{1}{2} m\beta \left\{ \left[ \int Dv \frac{\int Du\ e^{-m\beta g^*} |x^*|}{\int Du\ e^{-m\beta g^*}} \right]^2 - \left[ \int Dv \left( \frac{\int Du\ e^{-m\beta g^*} x^*}{\int Du e^{-m\beta g^*}} \right) \left( \frac{\int Du\ e^{-m\beta g^*} \operatorname{sgn} x^*}{\int Du\ e^{-m\beta g^*}} \right) \right]^2 \right\}.
\end{aligned}
\tag{81}
$$

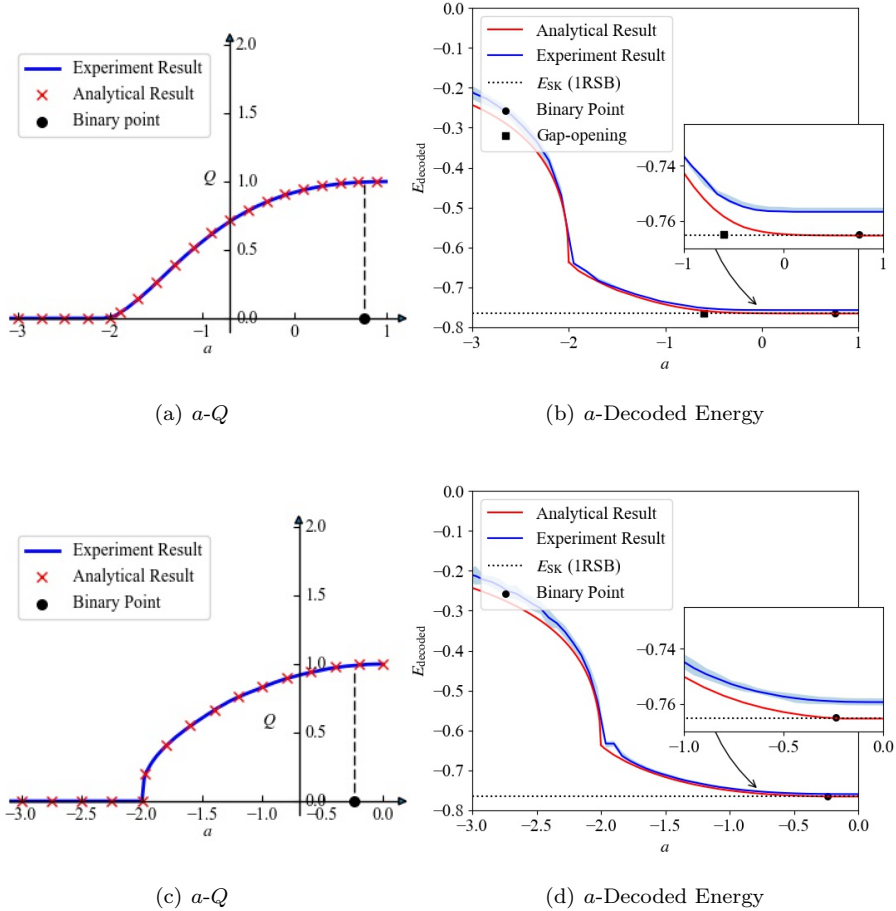

(a) $a$-$Q$

(b) $a$-Decoded Energy

(c) $a$-$Q$

(d) $a$-Decoded Energy

Figure 7: (a) The second moment ($Q$) of analog spins (from a single trial) for the clipCIM as a function of $a$ in the zero temperature limit and $B = 1$. (b) The mean value of the decoded energy of the clipCIM (20 trials), in comparison with the theoretical decoded energy value at zero temperature. One notices that in the gap-opening regime, the simulated decoded energy approaches the ground state of the SK model. (c) and (d) show the corresponding results for the simCIM. In the insets of (b) and (d), the analytical result does not fully attain the ground state until the binary point is reached.

The above equations allow us to plot the order parameter and the decoded energy as a function of the gain

$a$ as shown in Fig. 7(a) and Fig. 7(b). One notices that with the increase of $a$, the order parameters $Q$ or $q_1$ approach $B^2$. (The effect of decreasing $B$ is similar.) In these cases, the distribution $P(x)$ approaches one with two delta function peaks, as illustrated in Fig. 4(b), and the decoded energy at zero temperature is equivalent to SK model's ground state energy.

Thus, a natural question relevant to the search for the optimal solution is, what are the ranges of $a$ or $B$ in which $Q$ or $q_1$ equals to $B^2$. Using Eq. (76), we know that when $w = 0$, the minimizer of the effective free energy provides the smallest absolute value of $x^* = \sqrt{a_{\text{eff}}}$ as shown in Fig. 3. We call this value **the half gap width**, since it represents the half-width between the two peaks within the distribution $P(x)$, and is non-zero only when gap-opening occurs. If the half gap width exceeds $B$, then $P(x)$ transforms into a dual delta function (as shown in Fig. 4(b)). Thus, we can derive the following condition for the binary phase:

$$\min |x^*| = \sqrt{a + \chi} \geq B \tag{82}$$

This equation defines the region in the space of $a$ and $B$ where the ground state of the SK model can be attained by the clipCIM. We refer to the boundary of this region as the binary curve.

To derive the binary curve, we substitute the solution $c = B\text{sgn}w$ into the saddle point equations of the clipCIM, where $\sqrt{q_1}w = \sqrt{q_0}v + \sqrt{q_1 - q_0}u$. For $B = 1$, they become

$$q_1 = 1. \tag{83}$$

$$q_0 = \int Dv \left( \frac{\int Du\ e^{m\beta|w|}\text{sgn}w}{\int Du\ e^{m\beta|w|}} \right)^2. \tag{84}$$

$$\chi = \frac{2}{\sqrt{2\pi(1-q_0)}} \int Dv \frac{e^{-\frac{q_0 v^2}{2(1-q_0)}}}{\int Du\ e^{m\beta|w|}}. \tag{85}$$

$$\frac{1}{4}(1 - q_0^2) + \frac{1}{(m\beta)^2} \int Dv \ln \left( \int Du\ e^{m\beta|w|} \right) + \frac{1}{m\beta} \frac{\int Du\ e^{m\beta|w|}|w|}{\int Du e^{m\beta|w|}} = 0. \tag{86}$$

$$[\langle E_{\text{decoded}} \rangle]_J = -\chi - \frac{1}{2}m\beta(1 - q_0^2). \tag{87}$$

The solution is $q_0 = 0.477$, $m\beta = 1.361$, and $\chi = 0.239$. The decoded energy per node is $-0.765$, agreeing with the ground state value of the SK model. For general values of $B$, the solution is obtained by multiplying $q_0$ and $m\beta$ by $B^2$ and $B^{-2}$, respectively. Hence, remarkably, the parameters along the entire binary curve are constants after rescaling. Typically, at non-zero temperatures, small values of $a$ and large values of $B$ are preferable for obtaining a better solution. Therefore, the preferred parameters correspond to the case where the equality replaces the inequality in (82), which theoretically attains the ground state with minimal restrictions. Figure 8(a) shows the phase diagram in the space of $a$ and $B$ in which the binary curve separates the region with the decoded energy equal to the SK ground state energy (orange) and that above (blue).

An implication of the above result is that the ground state energy can be attained at a pump rate $a \sim O(1)$. A recent analysis showed that for the CIM solution to the ground state, it is sufficient that the pump

rate $a \sim O(\sqrt{N})$ [3]. Here, we demonstrate that in practice, the required pump rate is much lowered by introducing clipping.

## 9 The 1RSB solution of the simCIM

A related machine that gains advantage through clipping is the Simulated Coherent Ising Machine (simCIM) [18]. Its dynamics is obtained by removing the saturation term of clipCIM,

$$\mathcal{F}_i(t) = ax_i + \sum_j J_{ij}x_j \quad \text{for} |x_i| \leq B. \tag{88}$$

When its dynamics reaches equilibrium, the Hamiltonian of the equilibrium state is given by

$$H = -\frac{1}{2}\sum_i ax_i^2 - \sum_{(i,j)} J_{ij}x_ix_j, \tag{89}$$

where $x_i$ is restricted by $|x_i| \leq B$. The Hamiltonian of this dissipative system is equivalent to the potential energy of another popular algorithm, bSB (ballistic simulated bifuraction), which is momentum-based [19]. Therefore the results of the analysis can also be applied to it. Using the replica method, the saddle point equations are given by

$$g \equiv -\frac{1}{2}a_{\text{eff}}x^2 - \left(\sqrt{q_0}v + \sqrt{q_1-q_0}u\right)x. \tag{90}$$

$$g^* \equiv \min_{|x|\leq B}(g) \qquad x^* \equiv \arg\min(g). \tag{91}$$

$$q_1 = \int Dv \frac{\int Du\, e^{-m\beta g^*} x^{*2}}{\int Du\, e^{-m\beta g^*}}. \tag{92}$$

$$q_0 = \int Dv \left(\frac{\int Du\, e^{-m\beta g^*} x^*}{\int Du\, e^{-m\beta g^*}}\right)^2. \tag{93}$$

$$\chi = \Theta(-a_{\text{eff}})\frac{1}{|a_{\text{eff}}|}\int Dv\, \frac{\int Du\, e^{-m\beta g^*}\Theta(|a_{\text{eff}}|B - |\sqrt{q_0}v + \sqrt{q_1-q_0}u|)}{\int Du\, e^{-m\beta g^*}} \tag{94}$$
$$+ \Theta(a_{\text{eff}})\frac{2B}{\sqrt{2\pi(q_1-q_0)}}\int Dv\frac{e^{-\frac{q_0v^2}{2(q_1-q_0)}+\frac{1}{2}m\beta a_{\text{eff}}B^2}}{\int Du\, e^{-m\beta g^*}}.$$

$$\frac{1}{4}(q_1^2 - q_0^2) + \frac{1}{(m\beta)^2}\int Dv\ln\left(\int Du\, e^{-m\beta g^*}\right) + \frac{1}{m\beta}\frac{\int Du\, e^{-m\beta g^*}g^*}{\int Du\, e^{-m\beta g^*}} = 0. \tag{95}$$

$$[\langle E_{\text{decoded}}\rangle]_J = -\frac{2}{\sqrt{2\pi(q_1-q_0)}}\left(\int Dv\frac{\int Du\, e^{-m\beta g^*}|x^*|}{\int Du\, e^{-m\beta g^*}}\right)\int Dv\frac{e^{-\frac{q_0v^2}{2(q_1-q_0)}-\Theta(a_{\text{eff}})\frac{1}{2}m\beta_{\text{eff}}B^2}}{\int Du\, e^{-m\beta g^*}}$$
$$-\frac{1}{2}m\beta\left\{\left[\int Dv\frac{\int Du\, e^{-m\beta g^*}|x^*|}{\int Du\, e^{-m\beta g^*}}\right]^2 - \left[\int Dv\left(\frac{\int Du\, e^{-m\beta g^*}x^*}{\int Du\, e^{-m\beta g^*}}\right)\left(\frac{\int Du\, e^{-m\beta g^*}\text{sgn}x^*}{\int Du\, e^{-m\beta g^*}}\right)\right]^2\right\}. \tag{96}$$

Note that the saddle point equations are invariant if we rescale $x$ by $B$, and $q_1$, $q_0$, $g$,$[\langle E_\text{decoded}\rangle]_J$, $\beta^{-1}$ by $B^2$. Hence, it is sufficient to study the case for $B = 1$.

As shown in Fig. 4(c), the distribution $P(x)$ has a much broader distribution than CIM and clipCIM, owing to the absence of confinement due to the quartic term. It is remarkable to note that when the gain reaches the gap-opening point at $a = -0.24$, the distribution of $x$ already becomes a pair of delta functions at $x = \pm B$. The solution is again given by the binary curve equations (83) to (87), and it reaches the ground state of the SK model. Hence, the gap-opening point of simCIM is also the binary point and the gapless phase is absent in simCIM. The phase diagram in the space of $a$ and $B$ is shown in Fig. 8(b). Observe that the SK ground state is achieved at low values of $a$, demonstrating the superiority of simCIM over CIM and clipCIM in the search for the ground state. Such superiority was confirmed in the experimental comparison between the versions of CIM and simCIM with ballistic dynamics [19].

Furthermore, the Ising and gap-opening points of simCIM coincide. This ensures that the probability density at $x = 0$ remains nonzero until the entire system fully transitions to an Ising-like state at zero temperature. The gapless nature of the probability density at $x = 0$ indicates that it is easy for spins to change signs during the optimization dynamics. For other systems such as CIM and clipCIM, the existence of the gapped region prevents the spins to flip easily during the optimization dynamics, and noise (or finite temperature) is required to facilitate the spin update process. SimCIM, in contrast, does not require such assistance. Numerical experiments also support this assertion.

The phase diagrams of clipCIM and simCIM can be generalized to the case of FRSB. In FRSB, the susceptibility $\chi$ of the SK model vanishes [20, 21]. Hence, in Fig. 8(a) for clipCIM, the boundary between the gap-opening phase and binary phase becomes $a = B^2$. The boundary between the gapless and gap-opening phases is obtained by numerically detecting the gap of the clipCIM distribution. It starts from $(a, B) = (0, 0)$ and in the limit of large $B$, it approaches the gap-opening point of CIM, which is $a = -0.45$ as obtained by Yamamura *et al.* using 37-step RSB of the gap-opening point of CIM, and is expected to approximate well the FRSB value [22]. Remarkably, the shift of $a$ from 1RSB to FRSB is effectively independent of $B$ ($\approx 0.24$ at both $B = 0$ and $\infty$). In Fig. 8(b) for simCIM, the gapless-binary boundary is given by $a = 0$.

23

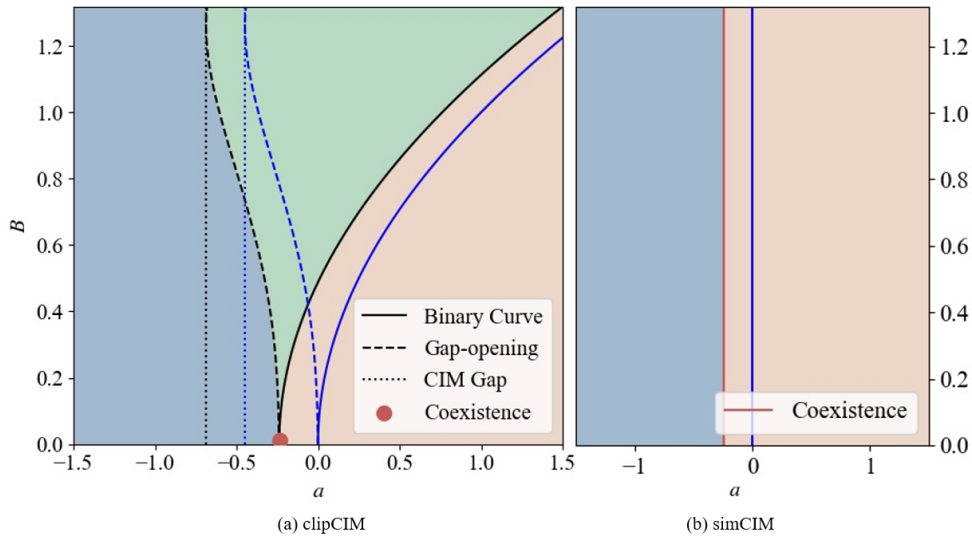

Figure 8: 1RSB phase diagrams for clipCIM and simCIM in the $(a, B)$ parameter space. For both systems, the additional blue lines indicate the corresponding full RSB phase transition positions. The binary/coexistence phase boundaries are given by $\sqrt{a_{\text{eff}}} = B$ for clipCIM and $a_{\text{eff}} = 0$ for simCIM. In (a), the full RSB gap-opening transition line is obtained by numerically detecting the gap of the clipCIM distribution for the SK model.

## 10 The stationary solutions of the digCIM

Here, we introduce the digitized CIM (digCIM) with gradient descent dynamics and binarized driving term, given by

$$\mathcal{F}_i(t) = ax_i + \sum_j J_{ij}\text{sgn}x_j \quad \text{for}|x_i| \leq B. \tag{97}$$

Similar forms have been introduced as variants of the HTNN (Hopfield-Tank Neural Network) in [23] and as a gradient-based version of dSB (discrete Simulated Bifurcation) in the Supplementary Materials of [19]. Both serve as benchmarks in those studies. However, these studies have not achieved the optimality of digCIM, as discovered in our work, both in theory and through simulation. The former study employed a soft digitization approach, while the latter lacked the phase diagram analysis that our work emphasizes, details of which we will present shortly.

In contrast to our system, [23] and [19]'s investigations overlooked the necessity of temperature annealing to initially access a gapless phase and subsequently enter its coexistence region. Instead, these earlier attempts primarily focused on adjusting the gain—$a$, a strategy that, according to our theoretical framework, cannot effectively alter the energy landscape. These systems either remained stuck in an almost frozen region ~~(coexistence phase)~~ with a constant low temperature while annealing a scaling parameter, or they surfed in

24

the gapless phase when setting a constant high temperature with a suboptimal decoded energy. One can immediately notice this after obtaining the digCIM's phase diagram. Consequently, the results obtained by digCIM are comparable to those of the momentum-based algorithm dSB. Moreover, the following theory can be extended to dSB due to their shared Hamiltonian structure, providing theoretical support for the advantages of dSB as well.

To study the digCIM, We first introduce the auxiliary variables $s_i$ defined by

$$s_i \equiv \text{sgn} x_i = \lim_{A \to \infty} \tanh(A x_i). \tag{98}$$

Then the dynamical equation is transformed to

$$\frac{ds_i}{dt} = A(1 - s_i^2) \left\{ \frac{a}{2A} \frac{\partial}{\partial s_i} [(1 + s_i) \ln(1 + s_i) + (1 - s_i) \ln(1 - s_i)] + \sum_j J_{ij} s_j \right\} \quad \text{for} |s_i| \leq 1. \tag{99}$$

Hence, the dynamics stops when $s_i$ reaches the bound $\pm 1$, or when the equilibrium state reaches a stationary point of the Hamiltonian

$$H = -\frac{a}{2A} [(1 + s_i) \ln(1 + s_i) + (1 - s_i) \ln(1 - s_i)] - \sum_j J_{ij} s_i s_j \quad \text{for} |s_i| \leq 1. \tag{100}$$

Here, we analyze the Hamiltonian as a candidate for describing the system behavior. Applying the replica method, we find that the effective free energy $g$ can be expressed in terms of the cavity field $\sqrt{q_0} v + \sqrt{q_1 - q_0} u$,

$$g = -\frac{a}{2A} [(1 + s) \ln(1 + s) + (1 - s) \ln(1 - s)] - \frac{1}{2} \chi s^2 - (\sqrt{q_0} v + \sqrt{q_1 - q_0} u) s \quad \text{for} |s| \leq 1. \tag{101}$$

At zero temperature and in the limit $A \to \infty$, $g$ is dominated by its minimum at the boundaries $s = \pm 1$,

$$s^* = \text{sgn}(\sqrt{q_0} v + \sqrt{q_1 - q_0} u). \tag{102}$$

Not surprisingly, the corresponding saddle point equations are identical to those of the binary curve in Eqs. (83) to (87), and the decoded energy becomes the ground state energy of the SK model. Remarkably, this energy is attained for all values of $a$ with no need of any transition from suboptimal states to the SK ground state at the binary point.

However, whereas the distribution of analog spins is well described by the double delta peak when $a$ is positive, it is very different from simulation results when $a$ is negative, and further analysis is required. In this respect, we consider the steady-state equations of the local fields in the SK model, commonly known as the TAP equations [20]. This set of equations is based on mean-field arguments of statistical physics and its validity is independent of the replica method [20]. It is applicable to extensively connected networks such as the SK model. As will be seen below, its only connection with the replica method is the value of the susceptibility. In the TAP approach, the local field $H_i$ of node $i$ is defined by $\langle \sigma_i \rangle \equiv \tanh(\beta H_i)$. At $T = 0$, the local field at node $i$ of the SK model is given by

$$H_i = \sum_j J_{ij} \text{sgn} h_j, \tag{103}$$

where $h_j$ is the cavity field at node $j$. The cavity field of a node is the field experienced by the node when its impact on the rest of the network is neglected. This impact, commonly known as the Onsager reaction, is expected to modify the cavity field into the true local field of the node. Due to the fully connected structure of the SK model, the Onsager reaction term can be calculated, yielding

$$H_i = \sum_j J_{ij}\mathrm{sgn}H_j - \chi\mathrm{sgn}H_i. \tag{104}$$

Comparing with the steady-state solution of Eq. (97), we note that the last term is extra in the TAP equation. However, it is important to note that in the full RSB solution, $\chi$ vanishes at $T = 0$ [20, 21]. Consequently, the TAP equation reduces to

$$H_i = \sum_j J_{ij}\mathrm{sgn}H_j. \tag{105}$$

We propose that given a solution $\{H_i\}$ of the TAP equation, we may map it to a stationary state of the digCIM dynamics with negative gain $a$ by setting

$$x_i = \min\left(\frac{|H_i|}{|a|}, B\right)\mathrm{sgn}\,H_i. \tag{106}$$

To see this, we consider the cases with $|x_i| \leq B$ and $|x_i| > B$ separately. In the former case, the steady-state equation of digCIM yields

$$|a|x_i = \sum_j J_{ij}\mathrm{sgn}x_j. \tag{107}$$

Substituting Eq. (106) into the right-hand side of the digCIM equation, we see that

$$\sum_j J_{ij}\mathrm{sgn}x_j = \sum_{|H_j|\leq|a|} J_{ij}\mathrm{sgn}\left(\frac{H_j}{a}\mathrm{sgn}H_j\right) + \sum_{|H_j|>|a|} J_{ij}\mathrm{sgn}(B\mathrm{sgn}H_j). \tag{108}$$

Thus, the expression in the above equation is equal to $H_i = |a|x_i$ according to the TAP equations, consistent with the steady state of digCIM.

For $|H_i| > |a|$, substituting Eq. (106) into the dynamical equation of digCIM, we have

$$\frac{dx_i}{dt} = -|a|B\,\mathrm{sgn}\,H_i + \sum_j J_{ij}\,\mathrm{sgn}\,H_j = (|H_i| - |a|B)\,\mathrm{sgn}\,H_i. \tag{109}$$

This shows that the rate of change of $|x_i|$ is directed along the direction of magnitude increase. However, since $|x_i| = B$ already lies on the boundary, it has also reached the steady state of the digCIM dynamics.

It is important to note that in practical applications, if $a < 0$, then the solution $x_i = \frac{H_i}{|a|}$ is both confined and stable. This implies that the clipping function, which prevents the analog spins from increasing indefinitely, is not required to constrain the magnitudes of the spin values in digCIM when $a < 0$.

In summary, we have verified that the existence of an SK stationary state is a sufficient condition for the existence of the stationary state of the digCIM dynamics described by Eq. (106).

The converse can be verified similarly, namely, if $\{x_i\}$ is a stationary state of the digCIM dynamics, then the corresponding stationary state of the SK model is given by

$$H_i = \sum_j J_{ij} \operatorname{sgn} x_j. \tag{110}$$

To verify this, we look for solutions of the TAP equation with $\operatorname{sgn} H_i = \operatorname{sgn} x_i$. Then, for $|\sum_j J_{ij} x_j| \leq B/|a|$, the steady state of $x_i$ lies in the the continuous range of the analog spin distribution, and $H_i = |a|x_i$ directly. On the other hand, for $|\sum_j J_{ij}\operatorname{sgn} x_j| > B/|a|$, the steady state of $x_i$ is constrained by $B$, but nevertheless, $H_i = \sum_j J_{ij}\operatorname{sgn} x_j$ remains valid.

Hence, we have verified the correspondence between the stationary states of the digCIM dynamics and the SK model. The distribution of the analog spins can then be determined by the field distribution $P(H)$ via

$$P(x) = \begin{cases} |a|P(H=|a|x) + \int dH \Theta(|H|-|a|)P(H)\left[\frac{1}{2}\delta(x-B) + \frac{1}{2}\delta(x+B)\right] & \text{for } a < 0 \\ \frac{1}{2}\delta(x-B) + \frac{1}{2}\delta(x+B) & \text{for } a \geq 0 \end{cases} \tag{111}$$

The local field distribution of the SK model, $P(H)$, has been extensively studied both analytically and numerically. It can be obtained through multi-step RSB or by generating an empirical distribution from samples [24,25]. A key finding from these studies is that the probability density at $H=0$ is nil and increases linearly as $|H|$ in the neighborhood, implying that the probability density of $x$ at zero also vanishes. Thus, the probability density has a profile intermediate between gap-opening and gapless distributions. Despite the absence of a gap in the distribution of $x$ (when $a < 0$), this feature effectively plays the role of gap-opening, thus providing the SK model's ground state energy across all ranges of $a$.

The vanishing of the gap in the spin distribution of digCIM is a consequence of the vanishing susceptibility $\chi$. For non-vanishing values of $\chi$, Eq. (104) becomes

$$H_i = F\left(\sum_j J_{ij} H_j\right), \tag{112}$$

where $F(x) = x + \chi \operatorname{sgn} x$, implying that a gap of width $2\chi$ exists in the spin distribution at $x=0$. As our digCIM simulation shows that the gap shrinks to 0, we have demonstrated that digCIM yield solutions consistent with the FRSB case.

The vanishing of the probability density at $x=0$ further confirms that the digCIM dynamics is stable at the interior points with $|x| < B$ when $a$ is negative. Consider the dynamics of perturbations $\delta x_i$ when the digCIM dynamics has reached a stationary state in Eq. (97). We see that the second term has contributions dependent on $\delta(x_j)$. Since $P(0) = 0$, the perturbation dynamics only depends on the first term, resulting in $d(\delta x_i)/dt = a\delta x_i$. Thus, stationary states with interior $x_i$ are possible. On the other hand, for $a > 0$, the interior points are not stable, and the analog spin distribution becomes a Ising like dual-delta function, as depicted in Fig. 4(d) and 9(a). Thus $a = 0$ is the binary point of digCIM system.

Contrary to simCIM, our simulations show that annealing the temperature is crucial to the quality of the decoding performance of digCIM. Although the ground state energy exists in the full range of $a$, its attainment is influenced by the choice of the dependence of $a$ on $T$ which affects the magnitude of the analog spin during temperature annealing. In general, a larger $a$ necessitates a larger $T$. This is consistent with the implementation of the dSB system, in which the momentum plays the role of noise.

Especially, one should not anneal $a$ for digCIM at low temperatures, since the probability distribution of $x = 0$ almost vanishes. Such annealing is merely a transition from one frozen system to another, with the sign of the soft spins remaining unchanged.

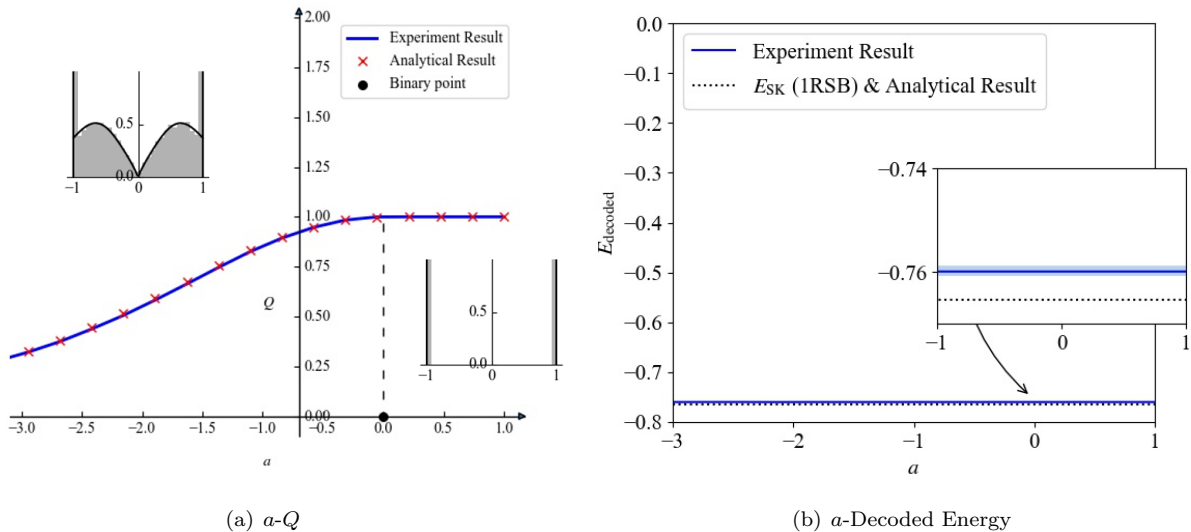

(a) $a$-$Q$ \hspace{4em} (b) $a$-Decoded Energy

Figure 9: (a) The second moment ($Q$ or $q_1$) and distributions of analog spins from a single trial for digCIM as a function of the gain $a$. Beyond $a > 0$, $Q$ stabilizes at a constant value. In these distributions, the solid black lines correspond to the theoretical values, while the grey shaded regions in the spin distribution indicate the results obtained from simulations. The left distribution is derived from $a = -2$, whereas the right one corresponds to $a = 1$. (b) Mean value of decoded energy of digCIM (20 Trials) compared with theoretical decoded energy. Notably, the decoded energy remains independent of $a$, consistently matching the energy landscape of the SK model.

We close this section by comparing the performances of the decoded energies of CIM, clipCIM, simCIM and digCIM. The original CIM manages to approach the ground state of the SK model asymptotically, but there is no binary point beyond which it collapses to the ground state. We found numerical evidence that the required value of $a$ to reach the ground state scales as $\sqrt{N}$, roughly of the same order of magnitude predicted by the sufficient condition proposed in [3]. On the other hand, by introducing clipping, the decoded energy of the clipCIM is much closer to the SK ground state, and it collapses to the ground state beyond the binary point. The simCIM, with the nonlinear decay term eliminated, has an even lower decoded energy, and the

SK ground state sets in at an even lower binary point. The digCIM, with the local fields computed using binarized variables, has the decoded energy identical to the SK ground state for all values of $a$, even the binary point of it is at $a = 0$. These results agree with the experimental observation that the Ballistic Simulated Bifurcation Machine outperforms the CIM [19].

## 11 Summary of the physical implications of gap-opening and binary point

In this work, we emphasized two phase transition points. The first is the 'gap-opening point', defined as **the value of $a$ above which there is a discontinues 'gap' (zero density region) appearing in the analog spins' distribution at zero temperature.** The second, referred to as the 'binary point', is characterized by **the value of $a$ above which the analog spin distribution becomes an Ising-like dual delta function at zero temperature.**

The gap-opening is a significant milestone in the process whereby target systems undergo a transition into an Ising system. Each analog spin adopts a meaningful configuration without ambiguity. During the network dynamics, the system typically starts with an initial gapless spin distribution and a decreasing decoded energy. At the instant the gap opens, further decrease of the decoded energy slows down significantly. Specifically, at zero temperature, after the gap opens, the system's decoded energy will no longer change even as $a$ continues to rise, since the gap prevents the spins from changing signs.

The binary point is the end point of the process through which target systems evolve into an Ising system during an annealing process in the space of $a$ and $T$ to search for the ground state. This implies the importance of ensuring that the annealing schedule for $a$ does not exceed the binary point, as doing so offers no advantage in reaching a better final state and merely increases the chances of being trapped in metastable states [22].

We summarize and compare the 1RSB positions of the two points discussed above in Table 2. Their dependence on the threshold $B$ is illustrated in Fig. 8.

| AIMs | CIM | clipCIM | simCIM | digCIM |
| --- | --- | --- | --- | --- |
| Gap-opening | $a = -0.69$ | $a \in [-0.69, -0.24]$ | $a = -0.24$ | $a = 0$ |
| Binary Point | $a = +\infty$ | $a = B^2 - 0.24$ | $a = -0.24$ | $a = 0$ |

Table 2: The gap-opening and binary points for CIM, clipCIM, simCIM and digCIM in the 1RSB approximation of the SK model.

We observe that the gap-opening point increases in the sequence of CIM, clipCIM, simCIM, and digCIM. The binary point for CIM is theoretically at infinity, indicating that in the thermodynamic limit, the energy

landscape of the CIM cannot become equivalent to that of the Ising model. Consequently, it is impossible for the CIM to achieve a ground state configuration identical to the Ising model at finite values of the gain $a$.

Note that for clipCIM, the gap-opening point decreases with the increasing value of the clipping boundary $B$, resulting in the formation of a gap-opening curve. When $B$ increases to infinity, the effects of the clipping boundary is negligible, and the behavior of the clipCIM approaches that of CIM. Hence, the gap-opening point of clipCIM approaches that of the CIM, i.e., $a = -0.69$. On the other hand, as $B$ approaches 0, the effects of the cubic saturation term of clipCIM is restricted by the narrow range of the spin values, and its behavior approaches that of simCIM. Hence, $a$ converges to the value $-0.24$, identical to that of the simCIM. Detailed illustrations are provided in Fig. 8. For the same reasons, the binary point of clipCIM approaches those of CIM and simCIM, respectively in the limits of large and small values of $B$.

For simCIM, the binary point and the gap-opening point are unaffected by the choice of the clipping boundary. This is due to the fact that the clipping boundary acts merely as a scaling parameter within the simCIM framework and can thus be selected arbitrarily. Additionally, since the full range of the continuous component of the spin distribution of simCIM vanishes simultaneously at the gap-opening point, the simCIM experiences a gapless-binary transition. This implies that as long as the system has not undergone transition to the Ising state, the probability density at $x = 0$ remains nonzero. The existence of spin values in the neghorhood of $x = 0$ indicates that it is easy for spin values to switch sign during the optimization dynamics, facilitating more efficient convergence to the final steady state. In contrast, for other systems exhibiting non-vanishing gap regimes, it is difficult for spins to flip across the gap during the optimization dynamics. In such cases, noise (or finite temperature) is required to facilitate the spin update process. SimCIM, in contrast, does not require such assistance.

For digCIM, our analysis suggests that its decoded energy is independent not only of the clipping boundary but also of the parameter $a$, and digCIM can attain the lowest decoded energy throughout the entire phase space. As the gain approaches $a = 0$ from the negative side, the continuous component of the spin distribution becomes increasingly flat and finally disappear completely at $a = 0$, leaving behind the delta function distributions at the clipping thresholds. This means that binary point and the gap-opening point also overlap in digCIM system, and the transition at $a = 0$ can also be considered as a gapless-binary transition. This also implies that assistance from noise (or temperature) is required across the entire phase space during the annealing process.

So far, we found that the one-step RSB ansatz already successfully predicts the existence of gap-opening and binary points in the AIMs. However, when we consider the full RSB ansatz, we find that an important difference is that the susceptibility approaches zero. This result was already proposed during the formulation of the TAP equations [20]. Subsequent calculations by incrementally increasing the replica symmetry-breaking steps showed that the susceptibility approaches zero in the limit of infinite steps [21].

The vanishing of the susceptibility leads to further refinement of the 1RSB framework. We have explained in a previous section how the vanishing of the susceptibility renders the steady-state solution of the digCIM dynamical equation into the TAP equation. Besides, we recall that the gap-opening point of clipCIM and the gapless-binary transition point of simCIM are given by the condition $a_{\text{eff}} = a + \chi = 0$. Hence, these phase transition points become $a = 0$ in the full RSB scheme, as summarized in Table 3. The importance of this correction will be demonstrated in the next two sections.

| AIMs | CIM | clipCIM | simCIM | digCIM |
| --- | --- | --- | --- | --- |
| Gap-opening | $a = -0.45$ | $a \in [-0.45, 0]$ | $a = 0$ | $a = 0$ |
| Binary Point | $a = +\infty$ | $a = 0$ | $a = 0$ | $a = 0$ |

Table 3: The gap-opening and binary points for CIM, clipCIM, simCIM and digCIM in the full RSB ansatz of the SK model.

For the decoded energy in the binary phase, we show below that, not only in the 1RSB framework but also in the Full RSB case, **an Ising machine constrained to the binary phase shares the same global minimum as the original Ising problem.** This can be rigorously proved. Here, we take clipCIM as an example; other systems are essentially similar. In the binary phase, the analog spin is given by $x_i = B \operatorname{sgn} x_i = B s_i$ where $s_i = \pm 1$.

Directly from the definition of the decoded energy, we have

$$
\begin{aligned}
E_{\text{decoded}} &= -\frac{1}{2N} \left[ \frac{\int d\boldsymbol{x} \exp(-\beta H(\boldsymbol{x})) \operatorname{sgn} \boldsymbol{x}^T \boldsymbol{J} \operatorname{sgn} \boldsymbol{x}}{\int \exp(-\beta H(\boldsymbol{x}'))\, d\boldsymbol{x}'} \right]_J, \\
&= -\frac{1}{2N} \left[ \frac{\operatorname{Tr}_s \exp\left( \frac{\beta a B^2}{2} \sum_i s_i^2 + \frac{\beta B^4}{4} \sum_i s_i^4 - \frac{\beta \xi B^2}{2} \sum_{i \neq j} J_{ij} s_i s_j \right) \boldsymbol{s}^T \boldsymbol{J} \boldsymbol{s}}{\operatorname{Tr}_s \exp\left( \frac{\beta a B^2}{2} \sum_i s_i^2 + \frac{\beta B^4}{4} \sum_i s_i^4 - \frac{\beta \xi B^2}{2} \sum_{i \neq j} J_{ij} s_i s_j \right)} \right]_J \\
&= -\frac{1}{2N} \left[ \frac{\operatorname{Tr}_s \exp(-\beta B^2 \xi H_{\text{sk}}) \boldsymbol{s}^T \boldsymbol{J} \boldsymbol{s}}{\operatorname{Tr}_{s'} \exp\left(-\beta B^2 \xi H_{\text{sk}}(\boldsymbol{s}')\right)} \right]_J.
\end{aligned}
\tag{113}
$$

Hence, in the zero temperature limit,

$$
\lim_{\beta \to \infty} E_{\text{decoded}} \equiv \lim_{\beta \to \infty} E_{\text{Ising}}.
$$

Note that, for other phases, calculating the Full RSB value of the decoded energy is highly nontrivial. Nevertheless, it is possible to demonstrate that the expression of the decoded energy is identical to the ground-state energy of the SK model at all levels of replica symmetry-breaking, and FRSB in particular. Starting from Eq. (51), we set $x_\alpha = B S_\alpha$ where $S_\alpha = \pm 1$. Discarding terms that are even in $S_\alpha$, and therefore reduces to constant factors, we arrive at the expression

$$
E_{\text{decoded}} = -\frac{1}{2} \sum_\alpha \beta J^2 \xi^2 B^2 \left[ \operatorname{Tr}_S \exp\left( \beta^2 J^2 \xi^2 \sum_{\gamma < \delta} q_{\gamma \delta} S_\gamma S_\delta \right) S_\alpha S_1 \right]^2.
\tag{114}
$$

Rescaling the temperature by $\xi B$, we arrive at the energy of the SK model.

As a practical and widely used approximation, we employ the 1RSB theoretical results to illustrate the relative behavior of decoded energy across different systems and parameter settings in the main text and subsequent paragraphs. Numerical results—which naturally reflect the Full RSB case—are also provided for various models and networks as supporting evidence for the validity of our approximate theoretical results. Despite the difficulty of obtaining the Full RSB decoded energy, we are able to present full RSB phase diagrams for clipCIM and simCIM shown as the blue curves in Fig. 8. For digCIM, the phase diagram was obtained in the FRSB framework already, thanks to the equivalence with the TAP equation. A detailed comparison of the decoded energy and transition points for these four systems under the 1RSB framework is illustrated in Fig. 10. We also present a summary of the properties for each system after the gap opens in Table 4.

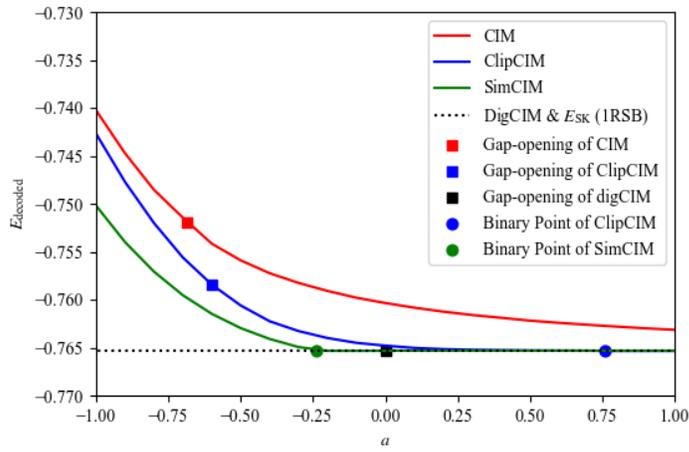

Figure 10: Theoretical decoded energy of CIM, clipCIM, simCIM, and digCIM at zero temperature as a function of the gain $a$ with constant threshold $B = 1$ in clipCIM, simCIM and digCIM. Circles and squares symbolize binary points for clipCIM, simCIM and digCIM, and gap-opening points for CIM and clipCIM under 1RSB assumption respectively. The decoded energy in digCIM consistently coincides with the SK model's ground state across all $a$ ranges, thereby sharing a common dotted line.

| Quantities after the gap opens | CIM | clipCIM | simCIM and digCIM |
|---|---|---|---|
| Spin distribution | The gap appears and widens when $a$ crosses the gap-opening curve. In the large $a$ limit, the peaks become increasingly narrow and asymptotically approaches a dual delta function. | The gap appears and widens when $a$ crosses the gap-opening curve. The spin distribution becomes a dual delta function when $a$ crosses the binary curve. | Dual delta function. |
| Decoded energy | Approaches the ground state energy of the Ising model on increasing $a$, but remains above it at finite values of $a$. | Equals the ground state energy of the Ising model when it crosses the binary curve. | The ground state energy of the Ising model. |
| Order parameter $Q$ | Increases with $a$. Linearly related to $a$ when $a$ is large. | $Q = B^2$, independent of $a$, when $a$ crosses the binary curve. | $Q = B^2$, independent of $a$ when $a$ is positive. |
| Susceptibility $\chi$ | An extra term appears as the gap opens. Decreases with increasing $a$, and converges to zero in the large $a$ limit. | $\chi = 0$, independent of $B$ and $a$ when $a$ crosses the binary curve. | $\chi = 0$, independent of $a$ in the full RSB ansatz. |

Table 4: Summary of the properties of CIN, clipCIM, simCIM and digCIM after the gap opens.

## 12 Dynamical implication: the gapless-binary coexistence region

In previous sections, we employed steady-state tools to analyze target Ising machine systems, focusing on deriving quantities such as the decoded energy. These quantities adhere to the assumptions of steady states, notably requiring sufficiently long evolution times. Consequently, the decoded energy we report represents a theoretical performance upper bound rather than an average value achievable within practical, finite evolution times. This discrepancy highlights a gap between our theoretical predictions and actual experimental outcomes, potentially leading to failures in regions that initially appear promising.

Notwithstanding these challenges, the experimental results closely aligned with the quasi-static state of a finite-time evolved Ising machine, as demonstrated by the congruence between numerical and analytical spin distributions discussed earlier. Therefore, based on our investigation of phase behaviors and phase

transitions, we have gleaned heuristic insights into the system dynamics that enable us to offer more promising recommendations for practical experimental setups.

As detailed in the previous section, the gap-opening point marks the phase transition from the gapless phase to the gap opened phase, where the decoded energy approaches the global minimum of the target Ising model. However, post-transition, the density of analog spins at $x = 0$ decreases to zero, and in the absence of noise, spin flipping ceases. This characteristic persists in a quasi-static system, indicating that in spite of its favorable decoded energy at the steady state, the gap-opening phase will likely exhibit poor dynamics, unable to spontaneously reduce the decoded energy effectively.

The binary point signifies another phase transition, where the analog spins become distinctly Ising-like, a trait also observed in quasi-static systems with finite-time dynamics. Nevertheless, in the binary phase, even though the system shares the same ground state as the Ising model, achieving this state is challenging. The dynamics of the system are nearly frozen, preventing any further evolution of either the spin's sign or amplitude. Even with noise intervention, the energy barriers in this phase are significantly higher compared to those in the gapless phase, necessitating much longer dynamics for effective searching. In fact, when one operates the dynamics deep inside the binary phase, simulations show that the approach to the ground state is hindered by the existence of metastable states [22].

Thus, from the dynamics point of view, the best strategy to attain the ground state of the Ising model exhibited in the binary phase is to operate the dynamics at the point where the spin distribution remains gapless (or approach it in an annealing schedule), such that spin flipping to lower the decoded energy remains feasible. This ideal operation point (or region) is the gapless-binary transition point (or region).

Drawing from the above understanding of phase properties, we propose that the gapless-binary coexistence region – defined as the intersection of the gap-opening and binary curves, or equivalently the overlap of the gapless and binary (and gap-opening, if one takes the point of view that the gap must open during the transition from the gapless phase to the binary phase) phases – represents an optimal location which shares the advantages of both the optimal steady-state decoded energy and the efficient gapless dynamics. Consequently, this region is an ideal parameter endpoint for AIM annealing operations.

A parameter annealing schedule for AIMs that begins in the gapless phase and concludes at the coexistence region can ensure that dynamics spontaneously reduce decoded energy throughout the entire process, thereby preventing the system from becoming prematurely frozen due to the gap.

As a demonstration of its advantage, we apply the notion of the coexistence region to the clipCIM. In the $a$-$B$ parameter space, the gapless-binary transition point is located at $(a, B) = (-0.24, 0)$ (for 1RSB), or $(a, B) = (0, 0)$ (for full RSB), as shown in Fig. 8(a). In practical terms, however, this intersection is challenging to exploit since a clipping threshold of zero results in an undefined spin configuration. The introduction of temperature can extend the coexistence point to the three-dimensional space of $a$, $B$ and

$T$. It is worth noting that gap-opened phases are only clearly defined at $T = 0$. The presence of noise at nonzero temperatures can fill in the gap (creating a pseudogap) and induce a degeneration of the gap-opened phase into the gapless phase. Thus, within the expanded parameter space of $a$, $T$, and $B$, the binary curve in Fig. 8(a) also acts as the gap-opening curve, thus, coexistence curve. Based on this theoretical prediction, we simulate the clipCIM starting at a higher $T$, and gradually anneal it until it reaches the vicinity of the coexistence curve in the phase space. This provides a superior solution in terms of accuracy compared to other parameter schedules aimed at non-coexistence regions as shown in Fig. 11. Additionally, we have applied the full RSB ansatz to the coexistence curve, as indicated by the red dotted line on the graph. It is noteworthy that the experimental decoded energy within the range from the 1RSB to the full RSB coexistence points remains fairly consistent, showing that the 1RSB solution is a reasonable approximation.

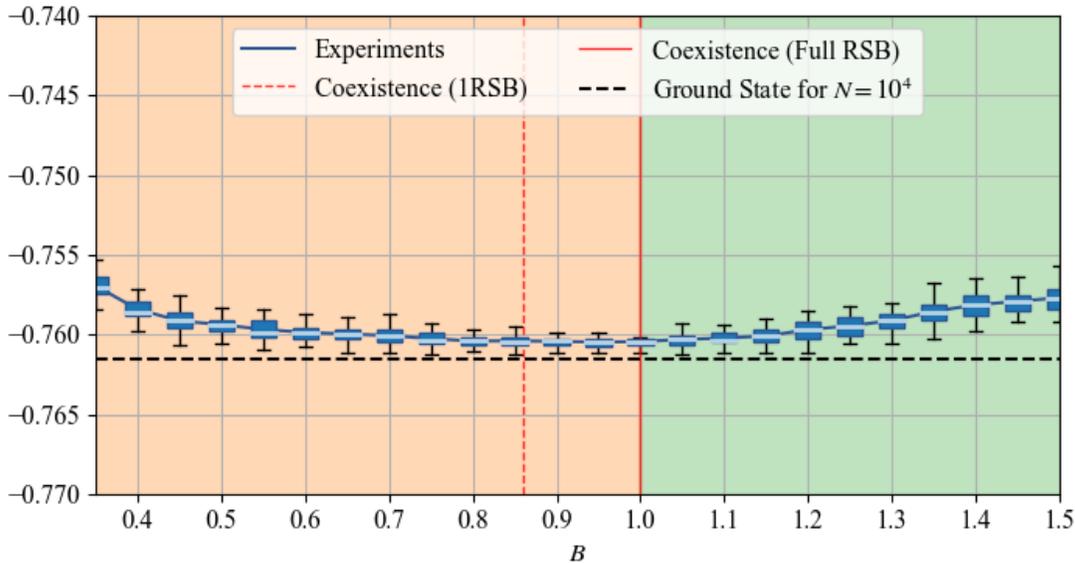

Figure 11: Decoded energy of clipCIM versus clipping threshold at $a = 1$. The blue boxes are the results of numerical experiments on a SK model with 10,000 spins, running dynamics 50 times through the Euler-Maruyama method for 100,000 steps with a step size of 0.1. On the left-hand side (binary phase), the increase in decoded energy is due to poor dynamical properties. On the right-hand side (gap-opening phase), the increase results from lower accuracy. The mean field ground state line corresponds to the full RSB result including finite-size effects. The lowest decoded energy is achieved using parameters within the coexistence region.

SimCIM exhibits a behavior different from clipCIM, where the coexistence region is a single point within the $a$-$B$ parameter space. As discussed in the previous section, the binary curve coincides with the gap-opening curve in the simCIM, resulting in a vertical coexistence curve within the $a$ and $B$ parameter space. Hence, the coexistence region is the entire gapless-binary transition line. A numerical experiment conducted on

the simCIM system aligns with our theoretical predictions. Starting from the same $a$ value in the gapless phase and stopping at various points along the coexistence curve, systems targeting the coexistence curve consistently achieved better results compared to others, as illustrated in Fig. 12.

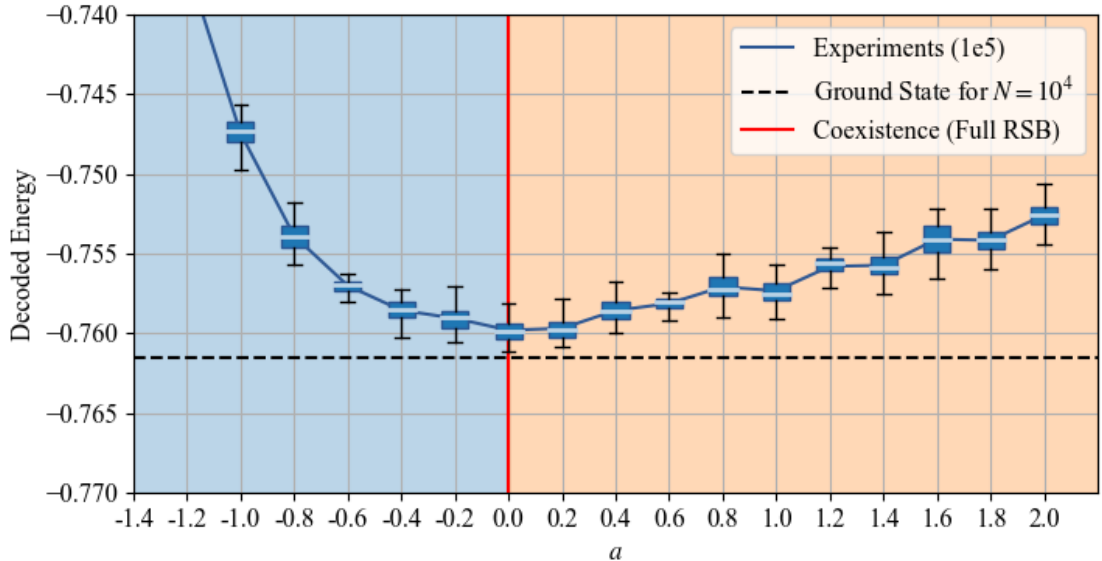

Figure 12: Decoded energy of simCIM versus gain $a$ at $B = 1$. The blue boxes are the results of numerical experiments on a SK model with 10,000 spins, running dynamics 40 times through the Euler-Maruyama method for 100,000 steps with a step size of 0.1. Temperature annealing from 1 to 0 for each data point. The lowest decoded energy is achieved using parameters within the coexistence region. The mean field ground state line corresponds to the full RSB result including finite-size effects.

For digCIM, the binary curve also coincides with the gap-opening curve, similar to simCIM. However, a significant distinction arises because the steady states of the digCIM at negative values of $a$ are isomorphic through rescaling $x_i$ and $B$ by $|a|$. To see this, we multiply both sides of Eq. (106) by a factor $\lambda$,

$$\lambda x_i = \min\left(\frac{|H_i|}{|a|/\lambda}, \lambda B\right) \operatorname{sgn} H_i. \tag{115}$$

This means that if $|a|$ is divided by a factor $\lambda$, then after $B$ is multiplied by the same factor, $\{\lambda x_i\}$ will become the solution of digCIM with the rescaled parameters. The profile of $P(x)$ is the same, except that the variable axis is extended by $\lambda$. Thus, all models with the same value of $B|a|$ are isomorphic. Consequently, this rescaling expands the coexistence curve to encompass the entire area where $a < 0$, forming a comprehensive coexistence region. The entire region exhibits a spin distribution with the dual delta component and dual bands touching at $x = 0$. Experimentally, this configuration provides superior dynamics, leading to better results comparing to other systems. Additionally, due to its wide range of optimal parameter settings, selecting an annealing path for the digCIM is more convenient. Unlike the clipCIM and simCIM, adjusting

36

the parameters $a$ and $B$ in the digCIM does not require meticulous attention, simplifying the parameter selection process.

Comparing the dynamics of digCIM and simCIM, a further advantage of digCIM lies in its band-touching distribution, which vanishes linearly near $a$ as predicted by TAP equations in the coexistence phase. This feature enables soft spins to flip signs more readily during optimization. In contrast, simCIM operating near the coexistence phase concentrates most spins at the boundaries ($\pm B$), leaving only a small fraction of continuous spins available for dynamic sign changes. This fundamental difference in spin distribution explains digCIM's superior exploration capability and faster convergence to optimal solutions.

Numerical results of digCIM dynamics are shown in Fig. 13. We conducted the dynamics starting from the same initial temperature while varying the constant gain $a$ for different data points. The temperature was then gradually annealed to zero following the evolution of the dynamics. The coexistence region consistently yields better results compared to setups within the binary phase.

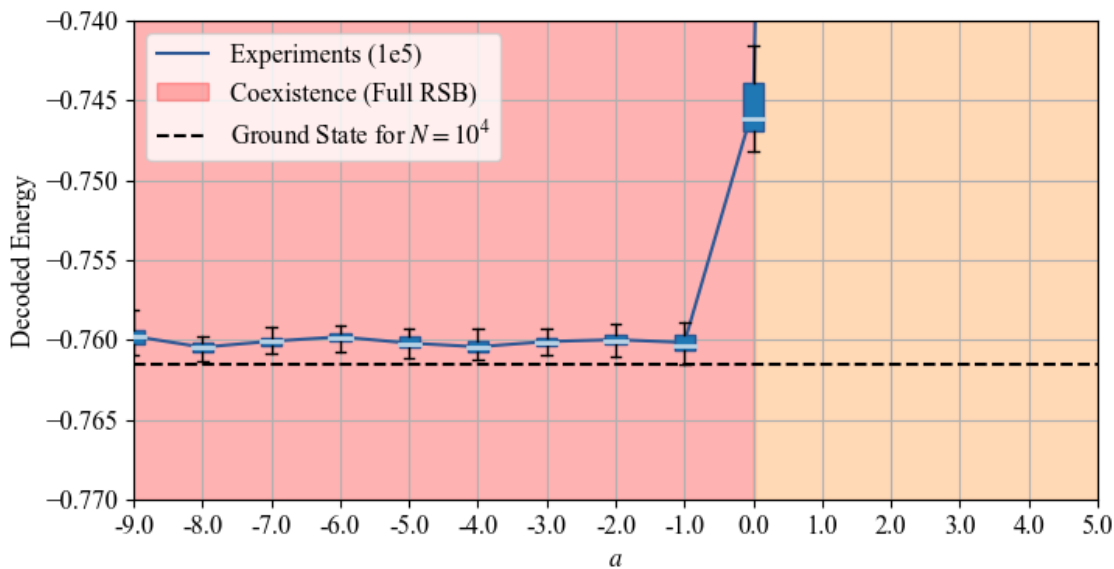

Figure 13: Decoded energy of digCIM versus gain $a$ at $B = 4$. The blue boxes are the results of numerical experiments on a SK model with 10,000 spins, running dynamics 20 times through the Euler-Maruyama method for 100,000 steps with a step size of 0.1. Temperature annealing from 1 to 0 for each data point. It is possible to achieve optimal decoded energy throughout the entire phase space. However, within the orange area (binary phase), the distribution of metastable states hinders the dynamics from freely exploring the solution space, resulting in suboptimal outcomes. The mean field ground state line corresponds corresponds to the full RSB result including finite-size effects.

It is important to note that we intentionally chose a wide range of $a$ in Fig. 13 for demonstration. The

37

distribution of analog spins scale wider by $1/|a|$. With a small $B$ and a small negative $a$, although the distribution still exhibits band touching, the continuous component of the spin distribution is flattened. Consequently, the distribution will resemble a dual delta function, which slows down the dynamics. For a given finite runtime, this leads to higher decoded energy compared to other settings within the same phase. Figures 11–13 provide evidence for the correctness of the coexistence theory and the associated changes in the decoded energy. We further introduce Fig. 14 to illustrate that, given sufficient time, the coexistence phase indeed enables the system to reach the theoretical ground state. Moreover, the increase in decoded energy near $a = 0$ is due to the broadening and concentration of the distribution, which can be mitigated by increasing the runtime.

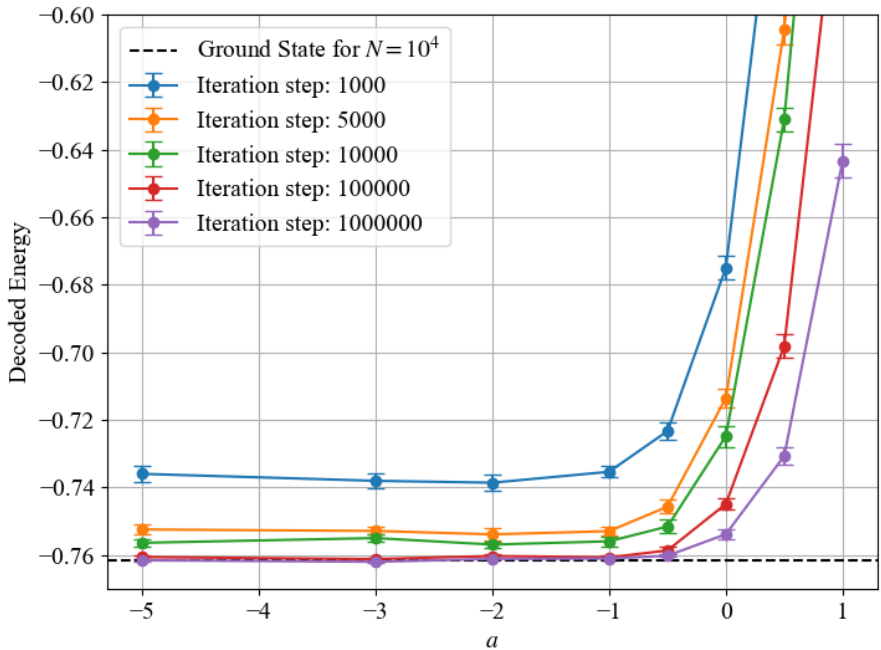

Figure 14: Decoded energy of digCIM as a function of gain $a$ at $B = 4$ for different iteration steps. All numerical experiments are performed on different SK models with $10^4$ spins, running the dynamics 20 times using the Euler-Maruyama method at various iteration steps with a step size of 0.1. The dashed line indicates the mean-field SK ground state energy with finite-size corrections.

The coexistence region for four systems is summarized in Table 5.

| AIMs | CIM | clipCIM | simCIM | digCIM |
|---|---|---|---|---|
| Coexistence region | N/A | $a_{\text{eff}} = B^2$ | $a_{\text{eff}} = 0$ | $a < 0$ |

Table 5: The coexistence regionin CIM, clipCIM, simCIM, and digCIM.

Notably, we use the notation $a_{\text{eff}} = a + \chi$. In the full RSB framework, $\chi = 0$, and all $a_{\text{eff}}$ can be simplified as $a$ in the expressions describing the coexistence regions. We also observed that these exactly calculated coexistence regions are not specific to the SK model but generally applicable to a wide range of models after appropriately rescaling the data. This will be demonstrated and discussed in detail in the latter part of the next section.

# 13 Generalization and the application of full RSB coexistence region

This work employs the SK model as a case study to analyze four machines. However, the underlying physical insights uncovered are not exclusive to the SK model; they represent common features applicable to a broader range of problems. In this section, we extend our experimental findings beyond the SK model, showcasing the relevance and effectiveness of our study in a wider spectrum of applications.

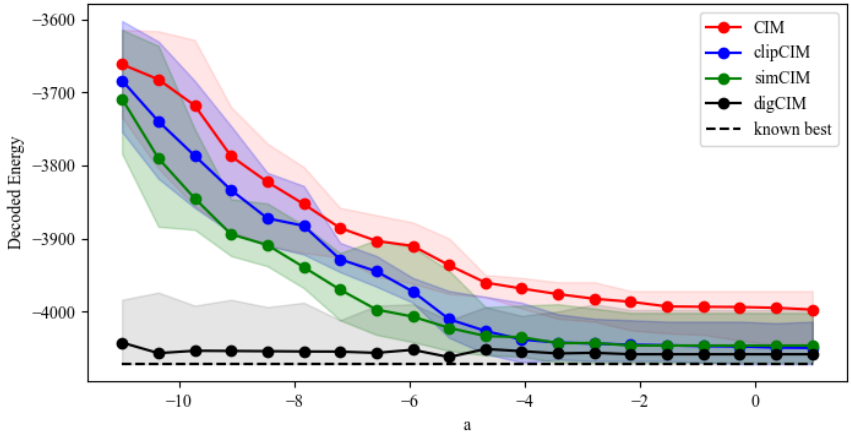

Figure 15: The experimental decoded energy curve for CIM, clipCIM, simCIM, and digCIM applied to dataset G1. The solid lines represent the mean values obtained over 20 trials, while the shaded areas are bounded by the maximum and minimum values observed. In each trial, the parameter $a$ was divided into 20 equidistant ascending steps ranging from -11 to 1. At each step, the temperature was individually annealed from a high value down to 0. The dashed line corresponds to the best-known result for G1, with a value of -4072.

In Fig. 10, we presented a comparison of the decoded energy levels for various systems when applied to the SK model. A corresponding analysis can be conducted for other problem domains as well. Here, we use the MAX-CUT dataset G1 as an example, which comprises a network of 800 nodes and 19,176 edges, with each edge randomly assigned a value of either 1 or 0 (corresponding to $J_{ij} = -1$ or 0, rendering the problem a frustrated network analogous to a spin glass) [26]. The findings are shown in Fig. 15, which displays strong agreement with our theoretical predictions: CIM exhibits the highest decoded energy, followed by clipCIM and simCIM. These latter models are capable of reaching the ground state provided that the parameter $a$ is sufficiently large, and their performance progressively converges to that of digCIM. The mean output of digCIM consistently remains at a low energy level, close to the minimum, demonstrating a persistent capability to achieve the ground state throughout the entire phase space.

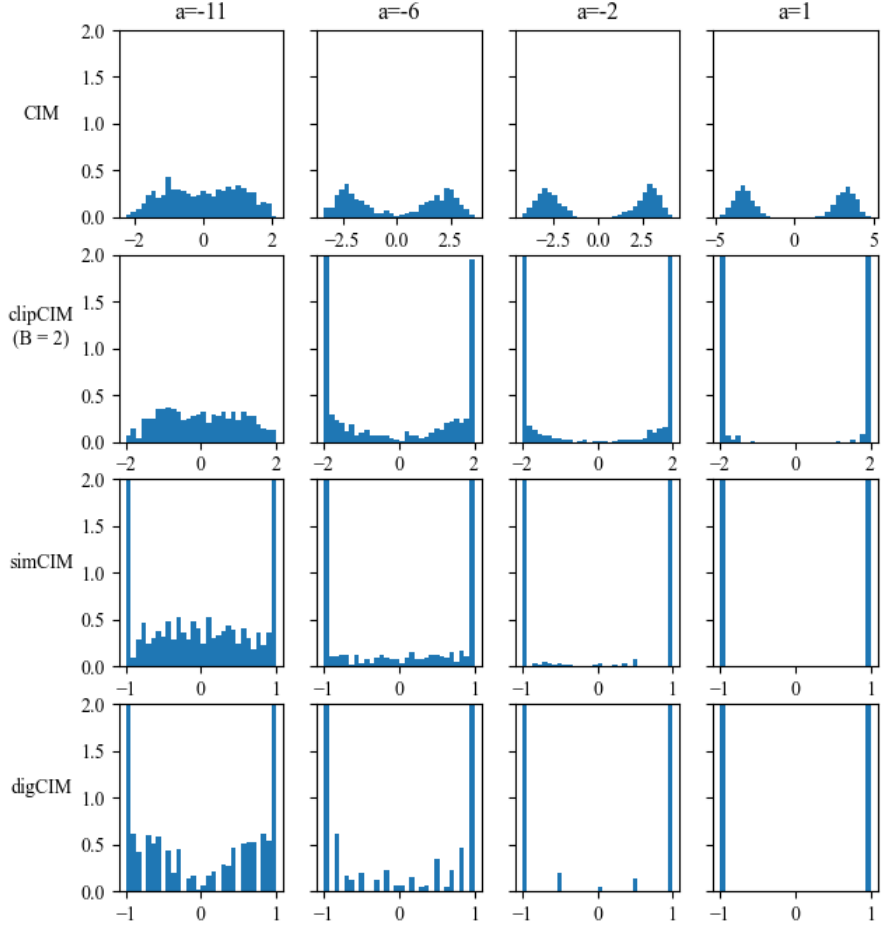

Figure 16: The distribution of analog spins for CIM, clipCIM, simCIM, and digCIM when applied to dataset G1. Each distribution was derived using a 400,000-step Euler–Maruyama method method tailored to the respective dynamics and parameters, with a step size of 0.02.

The distributions exhibit similar patterns across the systems. In Fig. 16, a clear gap-opening phenomenon can be observed in all four systems. As predicted by theory, the gap in the CIM system opens earlier than in the other systems. The gap in the clipCIM system has opened by $a = 1$, yet it has not reached the binary point. The gap-opening and binary points for simCIM and digCIM appear to coincide. Unique among the systems, digCIM exhibits an exceptionally low probability density around $x = 0$ throughout the entire phase. As soon as $a$ becomes positive, the system's distribution transitions into a dual delta function.

As mentioned above, the coexistence region for the SK model significantly enhances the performance of AIMs

in achieving better solutions. This region can be precisely identified using the zero $\chi$ correction from full RSB. For models other than the SK model, if the spins $x$ are fully discretized, the Onsager reaction term $\chi$ is expected to be equal to zero at zero temperature. This is because an infinitesimal magnetic field $h$ applied to a frozen discrete Ising spin should not cause any change in the spin state.

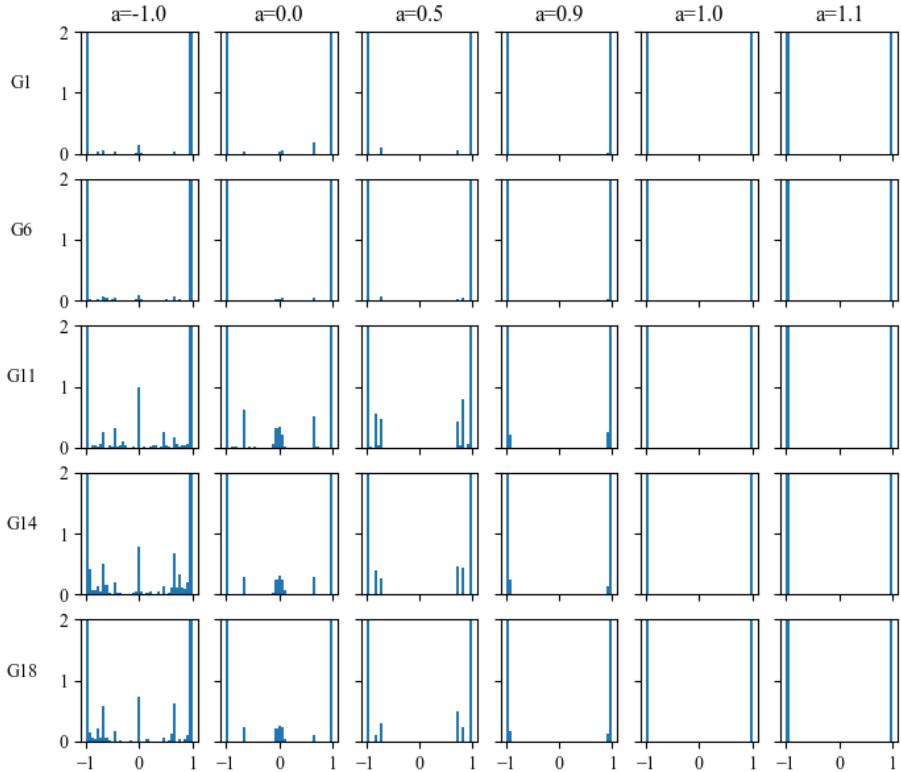

Figure 17: ClipCIM (B = 1) analog spin distributions. Each distribution was derived using a 400,000-step Euler–Maruyama method method tailored to the respective dynamics and parameters, with a step size of 0.02.

Based on this understanding, we argue that the vanishing of $\chi$ and the expression of the coexistence region can be extended from the SK model to other Ising models. This proposal aligns well with experimental results. Here, we use G1, G6, G11, G14, and G18 as representative examples to test our prediction. These examples vary in weight and topology: G1 and G6 are random graphs, G14 and G18 are planar graphs, and G11 is a toroidal graph. Figure 17 shows the final spin distributions for different models under various setups. We expect that after the system enters the coexistence region ($a > B^2$), it will exhibit a dual delta function, whereas before entering ($a < B^2$), it may display continuous components of the spin distribution. All different models align with this prediction. More clipCIM results are presented in Figs. 18-19, where

$B = 0.5$ and $B = 0.1$, Accordingly, the predicted phase transitions are at $a = 0.25$ and $a = 0.01$, respectively, which are consistent with the experimental results.

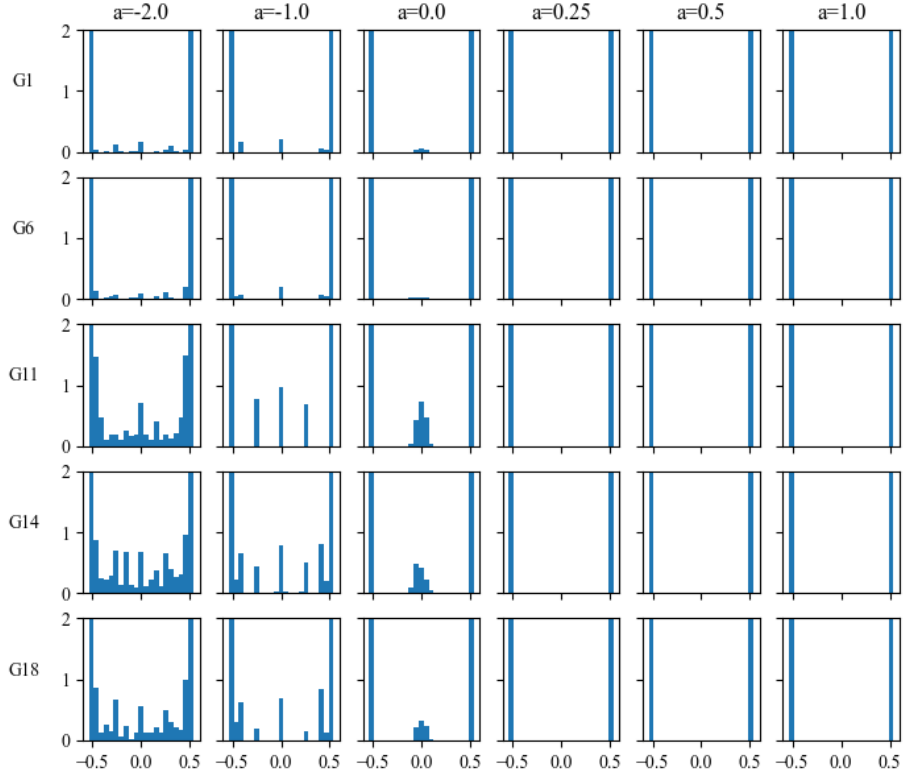

Figure 18: ClipCIM ($B = 0.5$) analog distributions. Each distribution was derived using a 400,000-step Euler–Maruyama method method tailored to the respective dynamics and parameters, with a step size of 0.02.

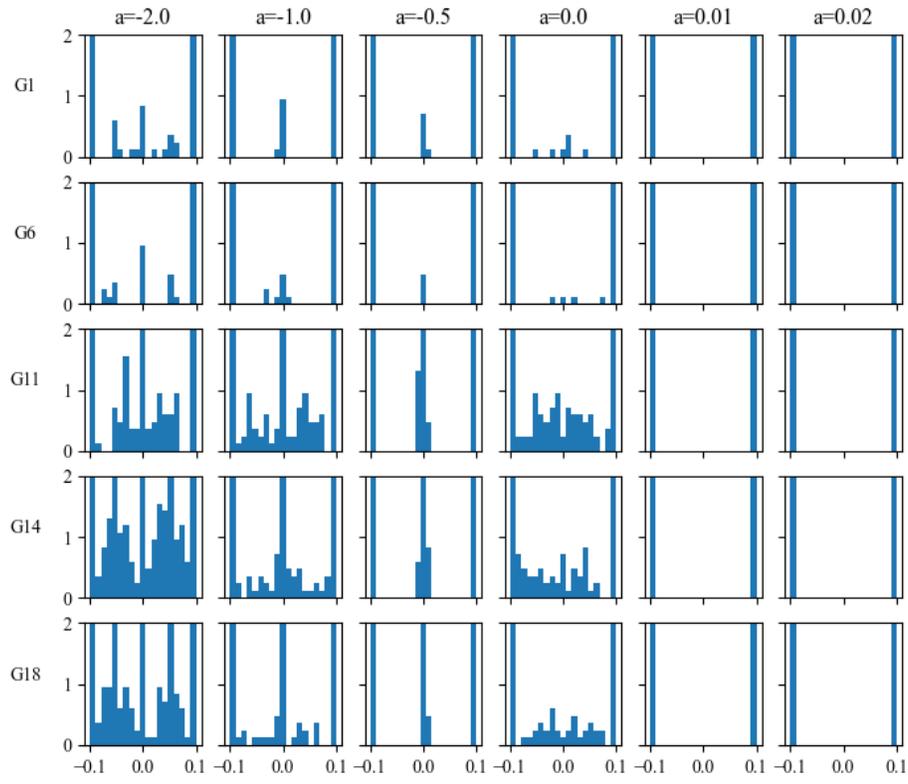

Figure 19: ClipCIM (B = 0.1) analog distributions. Each distribution was derived using a 400,000-step Euler–Maruyama method method tailored to the respective dynamics and parameters, with a step size of 0.02.

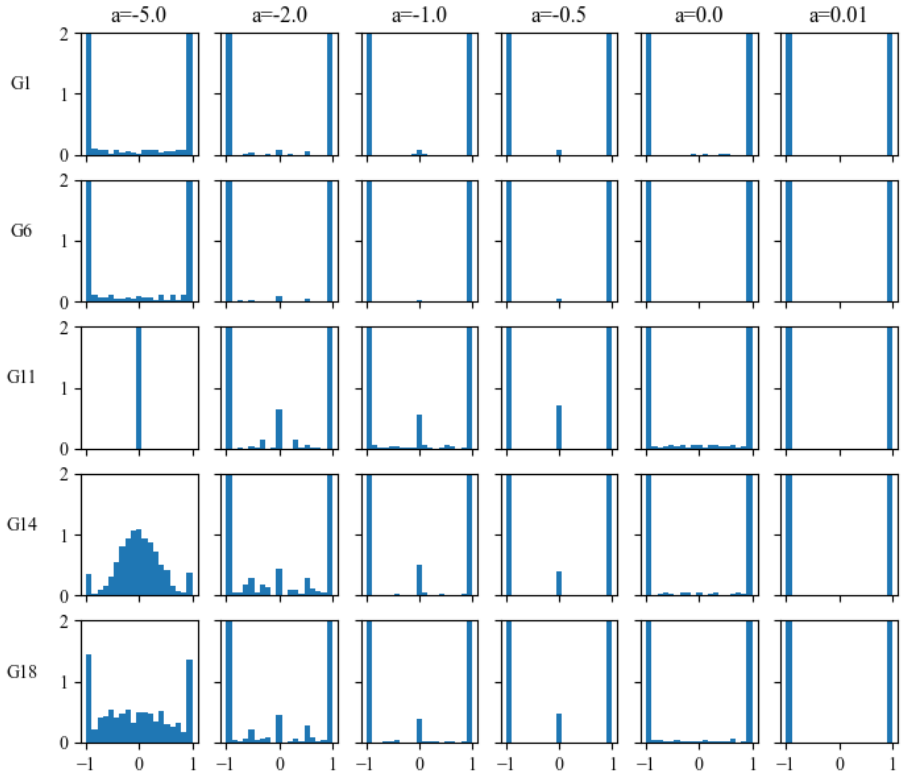

Figure 20: SimCIM (B = 1) analog distributions. Each distribution was derived using a 400,000-step Euler–Maruyama method method tailored to the respective dynamics and parameters, with a step size of 0.02.

For simCIM, one expects the phase transition to always occur at $a = 0$. However, only when $a > 0$ will the distribution take the dual delta form. This is because, at $a = 0$, the energy landscape is flat rather than a dual well (as can be verified by the analysis in the simCIM section), causing the soft spins to be widely spread between the two clipping thresholds. Even for a small negative $a$ the distribution is more concentrated compared with that of $a = 0$. However, this does not affect the fact that $a = 0$ can provide an optimal decoded energy. This property also matches the experimental results for all different models shown in Fig. 20. As $a$ increases to zero, the distribution spreads out, and as it increases slightly further, the distribution starts to concentrate in the neighborhood of the dual delta distribution.

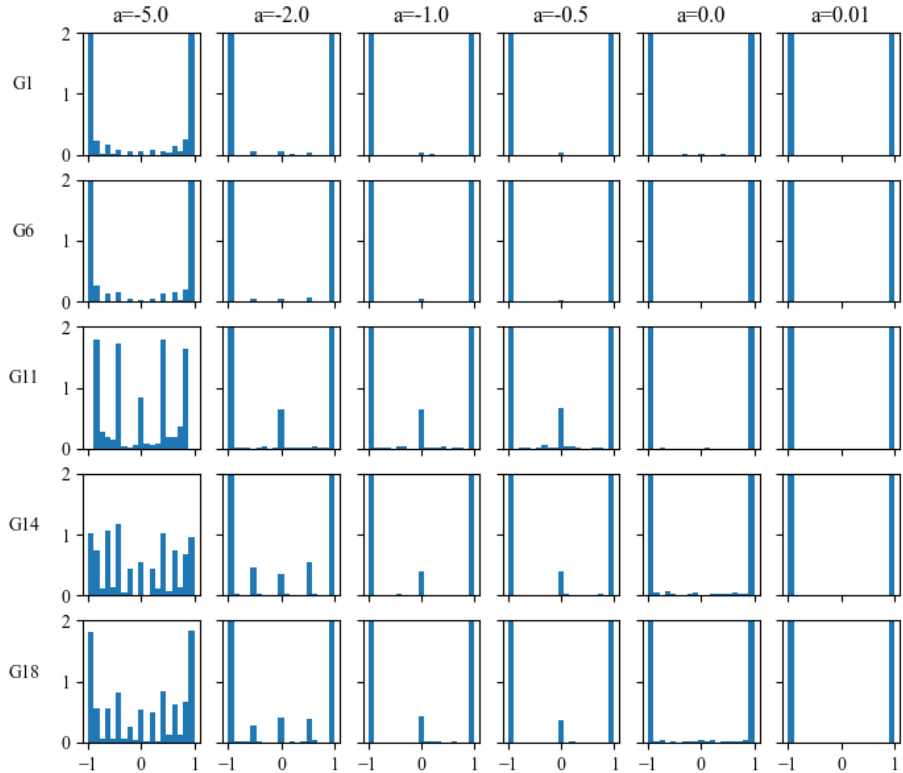

Figure 21: digCIM (B = 1) analog distributions. Each distribution was derived using a 400,000-step Euler–Maruyama method method tailored to the respective dynamics and parameters, with a step size of 0.02.

As shown in Fig. 21, the binary phase of digCIM exists only when $a > 0$.

In the SK model, the clipCIM, simCIM, and digCIM exhibit significantly enhanced performance over the traditional CIM, attributed to the presence of the binary phase and a consistently lower decoded energy. The performance of these AIMs can be further optimized by tailoring setup parameters within the coexistence region. As shown in Tables 6 and 7, experiments conducted in the Gset confirms a performance hierarchy: simCIM and clipCIM (targeted at the coexistence region) outperform their counterparts whose parameters lie outside the coexistence region, and all surpass the conventional CIM, aligning with predictions from the SK model analysis. The following results serve to validate the effectiveness of the coexistence phase and energy hierarchy in the above systems; more extensive computations—including comparisons with state-of-the-art algorithms and parameter selection—will be presented in the final section.

| Data Set | CIM ($\infty$, 1) | clipCIM (2, 1) | simCIM (1, -3) | The Known Best |
| --- | --- | --- | --- | --- |
| G1 | 11565 (11531.65) | 11624 (11617.05) | 11623 (11619.4) | 11624 |
| G2 | 11556 (11533.55) | 11613 (11600.825) | 11616 (11609.05) | 11620 |
| G3 | 11580 (11527.0) | 11621 (11619.125) | 11621 (11619.6) | 11622 |
| G4 | 11594 (11555.7) | 11638 (11638.0) | 11646 (11638.6) | 11646 |
| G5 | 11590 (11540.6) | 11626 (11621.4) | 11631 (11623.625) | 11631 |
| G6 | 2162 (2123.1) | 2174 (2173.05) | 2176 (2172.275) | 2178 |
| G7 | 1993 (1951.0) | 2000 (1994.3) | 2004 (1994.7) | 2006 |
| G8 | 1984 (1959.2) | 1995 (1992.875) | 2004 (1994.15) | 2005 |
| G9 | 2024 (1992.9) | 2046 (2042.575) | 2052 (2044.075) | 2054 |
| G10 | 1986 (1952.3) | 1990 (1988.475) | 1999 (1992.975) | 2000 |
| G11 | 548 (537.5) | 560 (554.05) | 552 (545.65) | 564 |
| G12 | 538 (528.3) | 554 (548.2) | 550 (544.4) | 556 |
| G13 | 568 (558.5) | 580 (573.55) | 578 (568.85) | 582 |
| G14 | 3012 (2924.25) | 3049 (3035.0) | 3031 (3016.05) | 3064 |
| G15 | 2994 (2971.65) | 3035 (3025.5) | 3018 (3002.5) | 3050 |
| G16 | 2995 (2978.1) | 3041 (3029.15) | 3019 (3006.375) | 3052 |
| G17 | 2986 (2972.3) | 3037 (3024.275) | 3013 (3001.875) | 3047 |
| G18 | 946 (918.65) | 971 (967.825) | 980 (974.25) | 992 |
| G19 | 867 (834.89) | 880 (878.25) | 888 (876.25) | 906 |
| G20 | 886 (863.45) | 933 (912.625) | 932 (923.275) | 941 |
| G21 | 898 (859.5) | 914 (912.9) | 916 (907.15) | 931 |
| Error sum | 799 (1457.01) | 190 (322) | 222 (395.925) | - |

Table 6: Performance comparison across three systems on the Gset benchmark outside the coexistence region. For each dataset, we conducted 40 independent trials, with each trial comprising 100,000 iterations. Results are displayed in the format max(mean). Also included are the best-known solutions for reference. Results were obtained using parameter annealing schedules toward the $(B, a)$, as indicated in the column headings, which are outside the coexistence region. Notably, even under these conditions, clipCIM achieved a significant reduction in error—76% (78%) compared with conventional CIM. The final row aggregates the deviations between the current outcomes and the best-known results.

| Data Set | clipCIM (1,1) | simCIM (1,0) | digCIM (1, -10) | The Known Best |
|---|---|---|---|---|
| G1 | 11624 (11621.625) | 11624(11622.425) | 11624(11623.872) | 11624 |
| G2 | 11620 (11610.9) | 11620(11611.15) | 11620(11614.632) | 11620 |
| G3 | 11622 (11621.35) | 11622(11619.725) | 11622(11619.395) | 11622 |
| G4 | 11646 (11639.325) | 11646(11638.5) | 11646(11639.65) | 11646 |
| G5 | 11631 (11624.275) | 11631(11626.15) | 11631(11628.154) | 11631 |
| G6 | 2178 (2174.8) | 2178(2175.0) | 2178(2175.9) | 2178 |
| G7 | 2006 (1998.025) | 2006(1997.575) | 2006(2000.475) | 2006 |
| G8 | 2005 (1996.1) | 2005(1996.275) | 2005(1998.775) | 2005 |
| G9 | 2054 (2047.6) | 2054(2045.975) | 2054(2049.275) | 2054 |
| G10 | 1999 (1994.5) | 1999(1994.725) | 1999(1997.5) | 2000 |
| G11 | 560 (552.8) | 564(554.9) | 564(558.1) | 564 |
| G12 | 554 (547.8) | 554(548.0) | 556(551.1) | 556 |
| G13 | 578 (572.55) | 578(573.6) | 582(578.0) | 582 |
| G14 | 3062 (3048.975) | 3060(3048.3) | 3062(3053.6) | 3064 |
| G15 | 3045 (3036.35) | 3043(3036.0) | 3048(3039.525) | 3050 |
| G16 | 3048 (3036.025) | 3047(3035.375) | 3051(3040.175) | 3052 |
| G17 | 3041 (3033.825) | 3040(3033.075) | 3044(3037.275) | 3047 |
| G18 | 987 (981.725) | 987(982.575) | 988(985.725) | 992 |
| G19 | 904 (893.55) | 903(893.775) | 906(902.2) | 906 |
| G20 | 940 (934.875) | 940(935.175) | 941(935.15) | 941 |
| G21 | 925 (917.525) | 927(916.075) | 928(923.3) | 931 |
| Error sum | 42 (186.5) | 43 (186.65) | 16 (119.223) | - |

Table 7: Performance comparison across three systems on the Gset benchmark, each operating at the coexistence region. This table presents results from 40 independent trials for each dataset, with each trial comprising 100,000 iterations. Results are formatted as max(mean) and include the best-known solutions for reference. The trials utilized the same parameter schedule as Table 6 but targeted at $(B, a)$ as specified in the column name, which are located in the coexistence region. Notably, this parameter setup enabled clipCIM to enhance its error reduction rate from CIM to 95% (87%). Benefiting from its unique mechanism, digCIM outperformed all, achieving a 98% (92%) error reduction from CIM. The final row aggregates the deviations between the current outcomes and the best-known results.

Remarkably, comparing with other Ising machine we discussed, digCIM demonstrates superior performance, benefiting from a broad range of optimal parameters, obviating the need for precise annealing control parameter selection. Additionally, its band-touching distribution, more dynamically friendly compared with the nearly-open gap distribution patterns of other AIMs, further enhances its effectiveness. This is consistent

with the numerical results reported in [19], which show that dSB generally outperforms bSB and aSB. These correspond to the steady-state properties of digCIM, simCIM, and CIM, respectively.

# 14 Potential scaling advantage of digCIM family

This section examines the scaling performance of the digCIM family, including both digCIM and dSB. For the original CIM system, detailed analyses of scaling with problem size are presented in [27], where CIM serves as a benchmark to demonstrate the scaling advantage of the chaotic amplitude control (CAC) algorithm. Here, we adopt their methodology and conduct lighter, yet fair, numerical experiments, showing that digCIM family algorithms achieve significantly improved scaling with problem size compared to the original CIM, and potentially comparable scaling to CAC. We note that our experiments are less comprehensive than those of [27], and thus serve as a preliminary study.

Here, we use the finite-size mean-field results for the SK model's decoded energy and its fluctuations as the theoretical benchmark [28, 29]. Following Fig.2(a,b) in [27], we evaluate the success probability of CIM, digCIM, and dSB as a function of iteration time. The success probability is defined as follows:

$$P_s := P(E_{\text{decode}} \leq E_b); \quad E_b = \langle E_{\text{sk}} \rangle_N + \sigma_N.$$

Figures 22–24 show the success probabilities of CIM, digCIM, and dSB for SK models of varying sizes and iteration steps. Figure 22 reproduces results consistent with Fig.2(a) of [27], despite the use of different benchmarks.[1] Both digCIM and dSB significantly outperform CIM and potentially achieve results comparable to CAC as shown in Fig. 2(b) of [27], underscoring the advantages of their improvement over CIM dynamics. As scaling analysis is not the primary focus here, a rigorous demonstration of the scaling advantage of the digCIM family will be left to more comprehensive and computationally intensive studies in future work.

---

[1] [27] employed large-scale Monte Carlo results for SK models, rather than mean-field results with fluctuations.

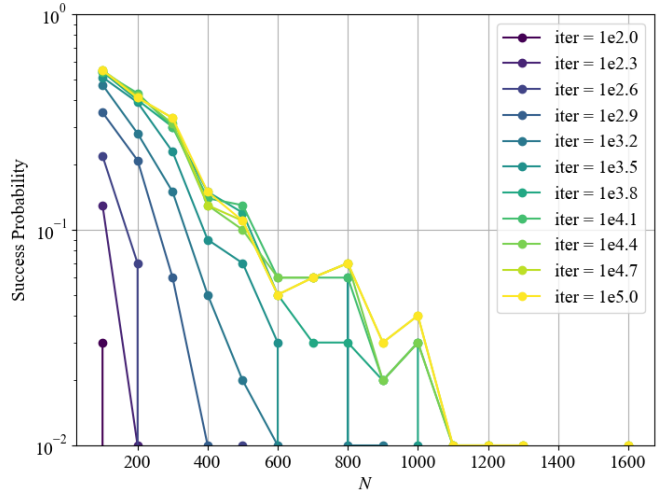

Figure 22: Success probability of CIM as a function of problem size $N$ and runtime. Each point is estimated from 100 independent trials. Dynamics are evolved via the Euler-Maruyama method with step size 0.1. Temperature $T$ and interaction coefficient $\xi$ are fixed at $1 \times 10^{-5}$ and 1, respectively; $a$ is annealed from $-2$ to $2$. [1]

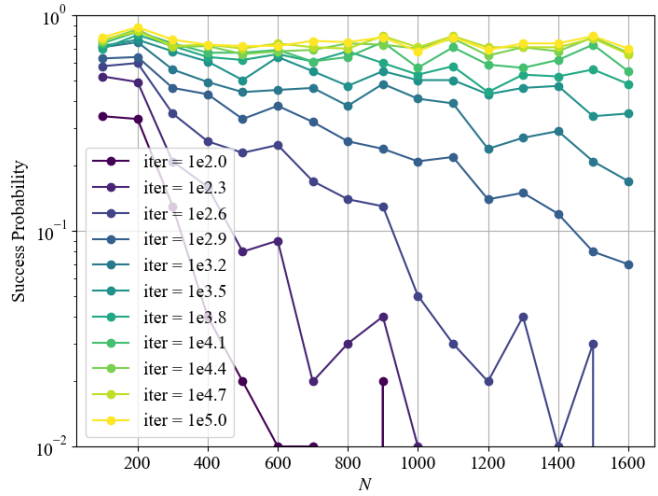

Figure 23: Success probability of digCIM as a function of problem size $N$ and runtime. Each point is estimated from 100 independent trials. Dynamics are evolved via the Euler-Maruyama method with step size 0.1. $a$ and interaction coefficient $\xi$ are fixed at $-5$ and 1, respectively; $T$ is annealed from 0.1 to 0.

---

[1] Here, following the convention of [27], we anneal $a$ in our experiments. Annealing the temperature yields similar results.

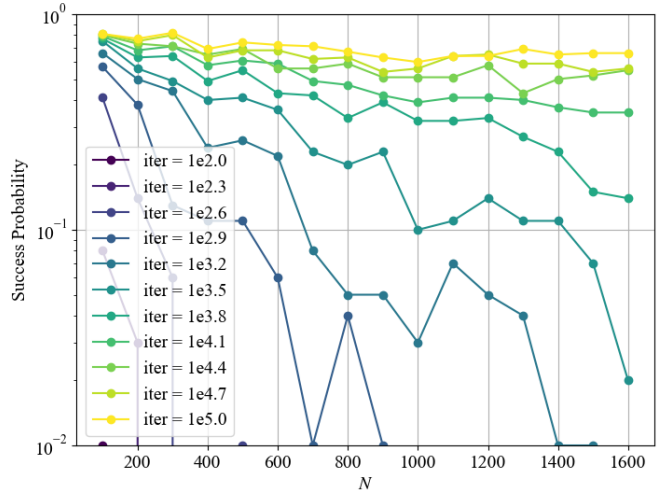

Figure 24: Success probability of dSB as a function of problem size $N$ and runtime. Each point is estimated from 100 independent trials. Dynamics are evolved via the symplectic Euler method with step size 0.1. $a_0$ and interaction coefficient $\xi$ are fixed at 1 and $\frac{1}{2\sqrt{N}\langle J \rangle}$, respectively; $a$ is annealed from 1 to 0. The parameter settings were suggested by [19].

## 15 Performance comparison with state-of-the-art solvers

The performance superiority of digCIM family is demonstrated not only through comparisons with other AIMs but also against popular state-of-the-art (SOTA) commercial solvers. In this section, we evaluate our machine using quadratic unconstrained binary optimization (QUBO) problems from the open benchmark library Quadratic Programming Instances, QPLIB [30]. The problems selected from this library originated from both theoretical studies and real-world applications. We compare our results with other optimization software data sourced from [30], adhering to its format and the constraints pertaining to time and hardware:

"*The simulations were conducted on an AMD Ryzen 9 5900X (12 cores, 128GB RAM) using the 23 unconstrained binary problems from QPLIB. All problems were solved with global solutions being the target. The reported times are elapsed times in seconds with a time limit of 1 hour; computations utilized 12 threads.*"

As shown in Table 8, with appropriate parameter settings, digCIM can maximally solve 19 out of 23 problems. This performance is significantly better than all the listed software. DigCIM fails to solve problems 3650 and 3693, but it still achieves a QUBO value close to the ground truth (-2). Surprisingly, two other problems (5725 and 5875), which digCIM finds challenging, can be easily solved by other solvers. This indicates that digCIM has a complementary relationship with other software; collectively, they can solve nearly all the problems in the benchmark dataset. For reference, simCIM, equipped with a coexistence strategy, also achieves notable performance, solving 8 out of 23 problems and outperforming two other solvers.

| prob# | BARON | SCIP | MCSPARSE | GUROBI | QUBOWL | BIQBIN | SHOT | digCIM | simCIM |
|---|---|---|---|---|---|---|---|---|---|
| Max solved | 12 | 7 | 12 | 12 | 15 | 7 | 12 | 19 | 8 |
| 3506 | 83 | t | 407 | 121 | 102 | t | t | 35.29 | 378.5 |
| 3565 | 2 | 47 | 13 | 2 | 1 | 12 | 4 | 1431.75 | 62.85 |
| 3642 | t | t | t | t | t | f | t | 266.97 | t(-2) |
| 3650 | t | t | t | t | t | t | t | t(-2) | t(-2) |
| 3693 | t | t | t | t | t | f | t | t(-2) | t(-4) |
| 3705 | 5 | 131 | 20 | 5 | 1 | 232 | 3 | 31.23 | 105.73 |
| 3706 | 881 | t | 2418 | 865 | 984 | t | t | 75.89 | 1528.79 |
| 3738 | 78 | t | 137 | 42 | 29 | t | 122 | 905.42 | 1537.50 |
| 3745 | 24 | 706 | 55 | 8 | 12 | t | 11 | 649.50 | 388.50 |
| 3822 | t | t | t | t | t | t | t | 150.95 | 299.61 |
| 3832 | 199 | t | 465 | 342 | 245 | t | 1705 | 1959.66 | t(-4) |
| 3838 | t | t | t | t | t | t | t | 2960.30 | t(-4) |
| 3850 | t | t | t | t | t | f | t | 787.37 | t(-4) |
| 3852 | 1 | 37 | 36 | 2 | 2 | 411 | 2 | 11.70 | 1.13 |
| 3877 | 205 | t | 448 | 457 | 36 | t | 1266 | 829.31 | t(-2) |
| 5721 | t | t | t | 1 | 1 | 2026 | 1 | t | t |
| 5725 | 10 | 63 | 4 | 6 | 1 | 39 | 6 | 263.42 | t |
| 5755 | 2 | 2 | 1 | 1 | 1 | 1317 | 1 | t | t |
| 5875 | t | t | t | t | 3545 | t | t | 15.158 | t |
| 5881 | 441 | 3332 | 75 | t | 29 | t | 81 | 29.56 | t |
| 5882 | t | t | t | t | 607 | t | t | 29.04 | t |
| 5909 | t | t | t | t | t | 1842 | t | 44.33 | t |
| 5922 | t | t | t | t | t | t | t | 491.317 | t |

"t" time exceeded, "f" failed

Table 8: Performance comparison between DigCIM and commercial solvers on the QUBO problems from the QPLIB benchmark dataset, showing the time elapsed in seconds. (Our algorithms were executed on Xeon(R) Gold 6230, 12 cores with 12 threads, whose speed is estimated to be 0.7 times that of the AMD Ryzen hardware used to test other software). The number within the bracket indicates the difference between the best result provided by the Ising machine and the ground truth. This is only added when these two numbers are sufficiently close. [30, 31].

To compare the performance of digCIM with other state-of-the-art Ising machine variants, we report time-to-solution (TTS) results for digCIM on the Gset benchmark. TTS is define as follows:

$$\text{TTS} = t\frac{\log(0.01)}{\log(1-P_s)}, \tag{116}$$

where $t$ is the physical average time for a single trial and $P_s$ is the success rate for achieving the ground state.

Each dataset's TTS was evaluated using 1,000 independent runs. All experiments were implemented in C++ and executed on a single CPU (Intel(R) Xeon(R) Gold 6230, 20 cores) with 40 threads. For a fair comparison, we also provide dSB results obtained on the same hardware, using parameter settings adopted from [19]. The digCIM parameters were selected by grid search over step size $\in 0.03, 0.04, 0.05, 0.06, 0.07, 0.08$ (This can be roughly set as the largest step size that still allows the system to converge), initial temperature $T_{\text{init}} \in 0.1, 0.5, 1, 2, 3$, and number of steps $\in 1000, 3000, 5000, 10000, 50000$, with the boundary $B = 1$, driving term $a = -10$, ending temperature $T_{\text{end}} = 0$, and coupling strength $\xi = 1$ fixed.

| Data Set | Step size | Number of steps | $T_{init}$ | DigCIM TTS (s) | DSB TTS (s) |
|---|---|---|---|---|---|
| G1 | 0.03 | 5000 | 3 | 118.828 | 303.271 |
| G2 | 0.03 | 5000 | 3 | 3969.8 | 3630.45 |
| G3 | 0.03 | 5000 | 3 | 379.503 | 279.748 |
| G4 | 0.03 | 1000 | 3 | 503.952 | 616.024 |
| G5 | 0.03 | 5000 | 3 | 570.282 | 480.614 |
| G6 | 0.07 | 3000 | 2 | 340.854 | 123.259 |
| G7 | 0.07 | 5000 | 2 | 185.569 | 113.626 |
| G8 | 0.07 | 3000 | 1 | 407.9 | 442.672 |
| G9 | 0.05 | 1000 | 1 | 1607.6 | 1038.82 |
| G10 | 0.07 | 1000 | 1 | 1622.63 | 1465.48 |
| G11 | 0.08 | 5000 | 0.5 | 4423.8 | 42.7439 |
| G12 | 0.07 | 3000 | 0.1 | 1000.87 | 129.116 |
| G13 | 0.07 | 3000 | 0.1 | 2004.64 | 404.916 |
| G14 | 0.04 | 50000 | 0.5 | - | - |
| G15 | 0.04 | 50000 | 0.5 | 37074.7 | 7586.37 |
| G16 | 0.04 | 50000 | 0.5 | 46423.5 | 6007.65 |
| G17 | 0.03 | 50000 | 1 | 183455 | 45718.7 |
| G18 | 0.03 | 10000 | 0.5 | - | - |
| G19 | 0.08 | 10000 | 0.5 | 638.523 | 171.501 |
| G20 | 0.07 | 5000 | 0.5 | 160.587 | 179.422 |
| G21 | 0.03 | 10000 | 0.5 | - | - |

Table 9: TTS and parameter settings for digCIM are shown; a dash denotes failure to reach the ground state within 1000 trials[1]

.

From the table, digCIM generally exhibits performance comparable to dSB. For larger instances such as G11 to G16, digCIM is noticeably slower than dSB. This discrepancy is not due to an inherent limitation of digCIM for large-scale problems; our scaling analysis confirms that digCIM has similar scalability. Rather, the slower performance arises from the convergence properties of the numerical method: the standard Euler method used for digCIM converges more slowly than the symplectic Euler method employed for dSB. As a result, digCIM requires a much smaller step size ($\sim 10^{-2}$) compared with dSB ($\sim 1$) to ensure stable numerical dynamics, thereby necessitating more steps and leading to longer physical run times. This issue may be alleviated by employing more advanced numerical algorithms.

---

[1] All three unsolved datasets were successfully solved by both dSB and digCIM in more extensive testing.

In addition, digCIM offers a practical advantage: owing to its gradient descent mechanism, it may be more readily implemented on non-traditional computational hardware compared to dSB—for example, in optical CIM-like Ising machines based on pulsed lasers, where momentum-based approaches like dSB potentially face fundamental hardware constraints [32–34].



\end{document}